\documentclass[prd,tightenlines,nofootinbib,superscriptaddress]{revtex4}

\usepackage{amsfonts,amssymb,amsthm,bbm}
\usepackage{array}
\usepackage{amsmath}
\usepackage{hyperref}
\usepackage{subcaption}
\usepackage{color,psfrag}
\usepackage[dvips]{graphicx}
\usepackage[dvipsnames]{xcolor}

\usepackage{tikz}
\usetikzlibrary{intersections}
\usetikzlibrary{patterns}
\usetikzlibrary{calc}
\usetikzlibrary{decorations.pathmorphing}
\usetikzlibrary{shapes.geometric}
\usetikzlibrary{arrows,decorations.markings}
\usetikzlibrary{arrows.meta, 3d}
\usepackage{tikz-3dplot}

\makeatletter
\def\l@subsubsection#1#2{}
\makeatother

\newcommand{\C}{{\mathbb C}}
\newcommand{\N}{{\mathbb N}}
\newcommand{\R}{{\mathbb R}}
\newcommand{\Z}{{\mathbb Z}}

\newcommand{\cA}{{\mathcal A}}

\newcommand{\cG}{{\mathcal G}}
\newcommand{\cJ}{{\mathcal J}}

\newcommand{\cL}{{\mathcal L}}
\newcommand{\cH}{{\mathcal H}}
\newcommand{\cM}{{\mathcal M}}
\newcommand{\cN}{{\mathcal N}}

\newcommand{\cO}{{\mathcal O}}
\newcommand{\cP}{{\mathcal P}}

\newcommand{\cT}{{\mathcal T}}
\newcommand{\cV}{{\mathcal V}}
\newcommand{\cD}{{\mathcal D}}
\newcommand{\cC}{{\mathcal C}}

\newcommand{\cS}{{\mathcal S}}
\newcommand{\cU}{{\mathcal U}}
\newcommand{\cI}{{\mathcal I}}

\newcommand{\cZ}{{\mathcal Z}}

\newcommand{\SU}{\mathrm{SU}}

\newcommand{\SL}{\mathrm{SL}}
\newcommand{\SO}{\mathrm{SO}}

\newcommand{\U}{\mathrm{U}}

\newcommand{\be}{\begin{equation}}
\newcommand{\ee}{\end{equation}}
\newcommand{\beq}{\begin{eqnarray}}
\newcommand{\eeq}{\end{eqnarray}}
\newcommand{\bes}{\begin{eqnarray}}
\newcommand{\ees}{\end{eqnarray}}

\newcommand{\mat} [2] {\left ( \begin{array}{#1}#2\end{array} \right ) }

\newcommand{\su}{{\mathfrak{su}}}

\renewcommand{\sl}{{\mathfrak{sl}}}

\newcommand{\la}{\langle}
\newcommand{\ra}{\rangle}

\newcommand{\tr}{{\mathrm{Tr}}}
\newcommand{\f}{\frac}

\def\nn{\nonumber}
\def\pp{\partial}
\newcommand{\w}{\wedge}

\def\rd{\mathrm{d}}

\def\vphi{\varphi}
\def\eps{\epsilon}

\newcommand{\id}{\mathbb{I}}

\def\vsigma{\vec{\sigma}}

\def\hH{\hat{H}}
\def\hp{\hat{p}}
\def\tPsi{\widetilde{\Psi}}
\def\tj{\tilde{j}}
\def\tm{\tilde{m}}

\def\tG{\widetilde{G}}
\def\tI{\tilde{I}}

\def\balpha{\bar{\alpha}}
\def\bbeta{\bar{\beta}}
\def\vK{\vec{K}}
\def\vL{\vec{L}}

\definecolor{teallight1}{rgb}{0.9, 0.9, 0.9}
\definecolor{teallight3}{rgb}{0.6, 0.6, 0.6}
\definecolor{teallight2}{rgb}{0.8, 0.8, 0.8}

\def\centerarc[#1](#2)(#3:#4:#5)
    { \draw[#1] ($(#2)+({#5*cos(#3)},{#5*sin(#3)})$) arc (#3:#4:#5); }
    
\def\centerarcnodes[#1](#2)(#3:#4:#5)(#6,#7)
    {\coordinate(#6) at ($(#2)+({#5*cos(#3)},{#5*sin(#3)})$);
    \coordinate(#7) at ($(#2)+({#5*cos(#4)},{#5*sin(#4)})$);
    \draw[#1] ($(#2)+({#5*cos(#3)},{#5*sin(#3)})$) arc (#3:#4:#5); }

\def\angcircle(#1)(#2)(#3:#4) {\coordinate(#1) at ($(#2)+({#4*cos(#3)},{#4*sin(#3)})$); }

\begin{document}

\title{Les Houches lectures on Spinfoam Path Integrals}

\author{{\bf Oleksandra Hrytseniak}}\email{ohrytseniak@perimeterinstitute.ca}
\affiliation{Department of Physics and Astronomy, University of Waterloo, 200 University Avenue West, Waterloo, Ontario, Canada N2L 3G1}
\affiliation{Perimeter Institute for Theoretical Physics, 31 Caroline Street North, Waterloo, Ontario, Canada N2L 2Y5}
\author{{\bf Etera R. Livine}}\email{etera.livine@ens-lyon.fr}
\affiliation{Laboratoire de Physique, Ecole Normale Sup\'erieure de Lyon, CNRS, 46 all\'ee d’Italie, Lyon, 69007 France}
\affiliation{Perimeter Institute for Theoretical Physics, 31 Caroline Street North, Waterloo, Ontario, Canada N2L 2Y5}
\author{{\bf Valentine Maris}}\email{valentine.maris@ens-lyon.fr}
\affiliation{Laboratoire de Physique, Ecole Normale Sup\'erieure de Lyon, CNRS, 46 all\'ee d’Italie, Lyon, 69007 France}

\date{\today}

\begin{abstract}

In these lecture notes for the \href{https://sites.google.com/cstq.org/2025-lqg-blaumann-school}{\nolinkurl{Les} \nolinkurl{Houches} \nolinkurl{School} \nolinkurl{on} \nolinkurl{Loop} \nolinkurl{Quantum} \nolinkurl{Gravity} \nolinkurl{2025}}, which took place in September 2025, we give a pedagogical review of the basics of the spinfoam framework for a quantum gravity path integral.
While  spin network states in loop quantum gravity describe the quantum geometry of the 3d space as dynamical networks of entangled quanta of volumes, spinfoams define transition amplitudes for those spin networks using the reformulation of general relativity as an "almost-topological" field theory and tools from quantum BF theory and topological state-sums.

The lectures were a short format of three times one hour and a half, only allowing to cover the basics and offer a glimpse of more advanced lines of research. We introduce spin foam path integrals for increasing spacetime dimensions starting with 2d BF theory, then build up to 3d quantum gravity with the Ponzano-Regge state-sum and the Turaev-Viro invariant, and finally the quantization of general relativity in four dimensions.

\end{abstract}

\maketitle
\setcounter{secnumdepth}{2}
\setcounter{tocdepth}{2}

\vspace{-6mm}
\tableofcontents

\newpage


Spinfoams are a path-integral formalism designed for quantum gravity, and more broadly applicable to gauge field theories (formulated using a connection 1-form), which is based on striangulated space-times (more generally on discretized geometries), with a specific emphasis on triangulation independence.
Originally introduced as a history formalism for loop quantum gravity in order to produce the transition amplitudes for spin-network states of quantum geometry generated by quantum Einstein equations, it draws heavy inspiration from topological quantum field theory (TQFT), Regge calculus and dynamical triangulations. Based on the exact path-integral quantization of topological BF theory and the definition of a discretization-independent projector onto flat connections, spinfoams can be considered a quantized version of Regge calculus, but, in their non-perturbative formulation in terms of group field theories, they are re-defined as generalized matrix field models and (random) tensor models. 

These lecture notes present path integrals for space-time dimensions from 1 to 4, illustrating the spinfoam logic and how to generate gravity and local degrees of freedom from topological BF theory by introducing a sea of defects.
Starting from the 1d path-integral formulation of quantum mechanics, through 2d gravity and 3d quantum gravity realized as the Ponzano-Regge and Turaev-Viro topological state-sums \cite{PR1968,Turaev:1992hq}, this will lead us to the current standard spinfoam model for 4d  quantum gravity (with Lorentzian spacetime signature), that is the Engle-Pereira-Rovelli-Livine (EPRL in short) model \cite{Engle:2007wy}, which defines what is now called {\it Covariant Loop Quantum Gravity}.
A significant part of these lecture notes is dedicated to three-dimensional spinfoam models, where both the formalism and its applications are far more developed than in four-dimensional spacetime.

Slides telling a brief history of spinfoam models are attached to the present lectures notes, as supplementary material.

\bigskip

A thorough, pedagogical and efficient guide through the basic concepts of quantum gravity to the EPRL spinfoam model for 4d quantum gravity \cite{Engle:2007wy} is the textbook by Rovelli \& Vidotto \cite{Rovelli:2014ssa}.
For a more complete bibliography, here are other relevant introductory reviews for interested readers:
\begin{itemize}
\item a short overview of the main concepts, structures and path integral models by Livine \cite{Livine:2024hhc};
\item a status report with the main results and lines of research developed from the birth of the spinfoam formalism around 1996 to 2010 by Livine \cite{Livine:2010zx};
\item a pair of reviews by Perez with precise mathematical details on spinfoam models up to 2012 \cite{Perez:2012wv,Perez:2012db};
\item a pedagogical review by Engle \& Speziale underlining the geometrical, physical and mathematical logics behind the EPRL spinfoam construction \cite{Engle:2023qsu}.
\end{itemize}
Reviews of more advanced topics can be found:
\begin{itemize}
\item on the group field theory approach \cite{Oriti:2011jm}
\item on refinement, coarse-grainig and the effective framework for the renormalization flow of spinfoams \cite{Asante:2022dnj};
\item on computing spinfoam amplitudes \cite{Dona:2022yyn}.
\end{itemize}
Very clear online lectures, with great material, can be found  on the \href{https://www.youtube.com/@centerspacetimequantum/videos}{\nolinkurl{YouTube} \nolinkurl{CSTQ} \nolinkurl{channel}}~:
\begin{itemize}
\item {\it Spin Foam Model with Cosmological Constant} by Qiaoyin Pan,
available on \href{https://www.youtube.com/watch?v=V-NINFmYEJ8}{\nolinkurl{YouTube}}

\item {\it Covariant Loop Quantum Gravity}  by Francesca Vidotto, 
available on \href{https://www.youtube.com/watch?v=zJ1_rXc8edA}{\nolinkurl{YouTube}} 

\end{itemize}

\section{The Spinfoam Ansatz}

Spinfoams are the discrete path-integral formalism for loop quantum gravity. Quantum states of geometry are defined as spin networks, which represent quantized discrete geometries, and a spinfoam is a history of a spin network evolving in time. Then, a spinfoam model is the assignment of an amplitude to each possible history of quantum states.

\subsection{Loop quantum gravity \& Spin networks}\label{secLQG}

%
Loop Quantum Gravity (LQG) is a canonical formalism for quantum gravity, which relies on a 3+1-dimensional decomposition of space-time and describes the evolution in time of quantum states of 3d (space-like) geometry. This evolution generates a four-dimensional quantum space-time. LQG is based on the Cartan's reformulation of general relativity as a gauge field theory for the Lorentz group in terms of a tetrad field and a Lorentz connection. We will not review the derivation of the phase space and will proceed directly to the description of the quantum states.

\begin{figure}[h!]

\centering

\begin{tikzpicture}[scale=1.5]

\coordinate(A) at (-0.4,0.8) ;
\coordinate(B) at (0.57,1);

\draw (A) node {$\bullet$} node[left]{$s(l)$};
\draw (B) node {$\bullet$} node[above]{$t(l)$};
\draw[->] (-0.4,0.8) arc (140:103:0.7) ;
\draw (-0.4,0.8) arc (140:63:0.8) ;
\node at (0,1.23) {$g_l$};
\end{tikzpicture}
\caption{A spin-network link $l$ with the source node $s(l)$, target node $t(l)$ and an associated $\SU(2)$ group element $g(l)$. 
\label{fig:linktransport}}
\end{figure}
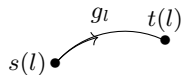

Quantum states realize a sampling of the space manifold's geometry described as transport between points. Let us choose $N$ points with (oriented) links between them forming a graph $\Gamma$. We allow for multiple links between two nodes and for self-links from a node to itself.
Then, we assign an $\SU(2)$ group element $g_{\ell}$ to each link $\ell$, which describes the transport of a 3d frame from the source node $s(\ell)$ to the target node $t(\ell)$ of the link as illustrated on fig.\ref{fig:linktransport}.
Let us keep in mind that those group elements are the integrated holonomies of the Ashtekar-Barbero connection on the canonical hypersurface, which is not simply a 3d spin-connection but it contains information on both 3d intrinsic geometry and extrinsic curvature.
Wave functions in the LQG Hilbert space will be functions of those group elements,
\be
\Psi(\{g_{\ell}\}_{\ell\in\Gamma})
\,.
\ee
The frame field acts on wave functions in the connection polarization as a derivation since it is the conjugate variable to the connection. The 3d metric, being a composite field quadratic in a frame field, will be a higher-order differential operator, as well as areas, volumes and other geometrical observables.

We further require that those wave functions are invariant under a local change of a reference frame, that is under $\SU(2)$ transformations acting at each node:
\be
\label{eqn:SU2inv}
\Psi(\{g_{\ell}\}_{\ell\in\Gamma})
=
\Psi(\{h_{t(\ell)}g_{\ell}h_{s(\ell)}^{-1}\}_{\ell\in\Gamma})
\,,\quad\forall \{h_{n}\}_{n\in\Gamma}
\,.
\ee

\smallskip

Since the frame field, and thus geometry, is conjugate to the connection and transport, it is natural to look for a Fourier basis for such wave functions. This is given by the Peter-Weyl decomposition of functions over $\SU(2)$. According to the Peter--Weyl theorem, matrix elements of irreducible unitary representations of a compact group form a basis of functions over this group. Since representation matrices of $\SU(2)$ (also called Wigner matrices and denoted $D^j_{mn}(g)$) are labeled by a spin $j\in\N/2$ and two magnetic numbers $m,n=-j,-j+1,...,j-1,j$, for a function $\phi:\SU(2)\rightarrow \C$,
\be
\phi(g)\equiv \langle g|\phi\rangle=\sum_{j,m,n} \phi^{j}_{mn}D^{j}_{mn}(g)\,,\label{phi-exp-PW}
\ee
\be
D^{j}_{mn}(g)=\la j,m|g|j,n\ra
\,.
\ee
Alternatively, we can treat the space of functions separately from its dual space $G$,
\begin{equation}
    |\phi\rangle=\sum_{j=0}^\infty\sum_{m,n=-j}^j \phi^j_{mn}D^j_{mn}\equiv \sum_{j=0}^\infty \frac{1}{\sqrt{d_j}}\sum_{m,n=-j}^j \phi^j_{mn}|jmn\rangle,
\end{equation}
which is called a \textit{spin basis} of the LQG wave functions. We denote $\cV^{j}$ the $(2j+1)$-dimensional Hilbert space of the spin-$j$ representation,
\be
\cV^{j}=\bigotimes_{m=-j}^{+j}\C\,|j,m\ra\,,\qquad
d_{j}=\dim \cV^{j}=2j+1
\,.
\ee
An interested reader will find an explanation of the construction of the irreducible representations of $\SU(2)$ (in terms of polynomials of two complex variables) and the computation of the Wigner-matrix elements in the appendix \ref{app:SU2}.
The important point is the orthogonality of the Wigner matrix elements:
\be
\int \rd g\, D^{j_1}_{m_1n_1}(g)\,\overline{D^{j_2}_{m_2n_2}(g)}
\,=\,
\f{\delta_{j_1 j_2}}{d_{j_1}}\,\delta_{m_1 m_2}\delta_{n_1 n_2}
\,,\qquad
\overline{D^{j}_{mn}(g)}={D^{j}_{nm}(g^{-1})}
\,,\label{WignerOrthonorm}
\ee
where the over-bar denotes the usual complex conjugation\footnotemark. The integration measure $\rd g$ is the Haar measure invariant under both left and right group multiplication. It is simply the Lebesgue measure on $\SU(2)$ as the unit 3-sphere embedded in flat Euclidean $\R^{4}$.
\footnotetext{
One should also keep in mind that, unlike higher $\SU(N)$ Lie groups for $N\ge 3$, the complex conjugate representation is isomorphic to the original representation:
\be
|j,m\ra \in \cV^{j}
\,\,\mapsto\,\,
(-1)^{j-m}|j,-m\ra \in \big{(}\cV^{j}\big{)}^{*}\,,
\nn
\ee
leading to the following relation for Wigner matrices:
\be
\overline{D^{j}_{mn}(g)}
=
(-1)^{j-m}(-1)^{j-n}D^{j}_{-m,-n}(g)
\,,\nn
\ee
as can be directly checked for the fundamental representation of spin $\f12$:
\be
g=\mat{cc}{\alpha & -\bbeta \\ \beta & \balpha}\in \SU(2)
\,,\qquad
\overline{D^{\f12}_{\f12,\f12}(g)}
=
\balpha
=
D^{\f12}_{-\f12,-\f12}(g)
\,,\qquad
\overline{D^{\f12}_{\f12,-\f12}(g)}
=
-\beta
=
-D^{\f12}_{-\f12,\f12}(g)
\,.
\nn
\ee
}
The spin coefficients in \eqref{phi-exp-PW} are given by the integration of a function $\phi$ against the Wigner-matrix elements:
\be
\phi^{j}_{mn}= d_{j}\int \rd g\,\phi(g)\overline{D^{j}_{mn}(g)}
\,.
\ee
The Peter-Weyl decomposition is the equivalent of the Fourier transform for $\SU(2)$ seen as a homogeneously curved manifold. It is naturally associated to the Plancherel formula for $\SU(2)$ giving the decomposition of the $\delta$-distribution on the group:
\be
\forall f\in\cC[\SU(2)]\,,\quad
\int \rd g\, f(g)\delta(g)=f(\id)
\,.
\ee
$\delta$-function can be expanded in terms of the characters\footnote{Characters are functions of group elements invariant within conjugacy classes or, said simpler, invariant under an adjoint action onto its argument. Trace is an example of such a function.} $\chi^j$ of a group as
\be
\delta(g)=\sum_{j}d_{j}\chi_{j}(g)
\,,
\quad\quad\chi_{j}(g)=\tr D^{j}(g)=
\f{\sin d_{j}\theta}{\sin\theta}
=
U_{2j}(\cos\theta)
\,,
\ee
where $\theta$ is the class angle of the group element $g$. It is twice the rotation angle of the corresponding 3d rotation, that is $g=\exp(i\theta\hat{u}\cdot\vsigma)$, with $\sigma_{a}$ the three Pauli matrices and $\hat{u}$ a unit 3-vector defining the rotation axis, or it can be determined by $\chi_{\f12}(g)=2\cos\theta$ being the trace of $g$ as a unitary 2$\times$2 matrix. The $U_{n}$ for $n\in\N$ are the Chebyshev polynomials of the 2nd kind.

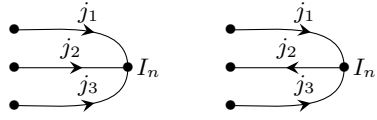
\begin{figure}[h!]

\centering

\begin{tikzpicture}[scale=1]

\draw[in=90,out=0,looseness=1,decoration={markings,mark=at position 0.6 with {\arrow[scale=1.5,>=stealth]{>}}},postaction={decorate}] (0,0.5) to node[above,pos=.5]{$j_{1}$} (1.5,0);
\draw[in=-90,out=0,looseness=1,decoration={markings,mark=at position 0.6 with {\arrow[scale=1.5,>=stealth]{>}}},postaction={decorate}] (0,-0.5) to node[above,pos=.5]{$j_{3}$}(1.5,0);
\draw[decoration={markings,mark=at position 0.6 with {\arrow[scale=1.5,>=stealth]{>}}},postaction={decorate}] (0,0) to node[above,pos=.5]{$j_{2}$} (1.5,0);
\draw (0,0) node {$\bullet$} ;
\draw (1.5,0) node[right] {$I_n$};
\draw (1.5,0) node {$\bullet$} ;
\draw (0,0.5) node {$\bullet$} ;
\draw (0,-0.5) node {$\bullet$} ;

\end{tikzpicture}
\hspace{3ex}
\begin{tikzpicture}
    \draw[in=90,out=0,looseness=1,decoration={markings,mark=at position 0.6 with {\arrow[scale=1.5,>=stealth]{>}}},postaction={decorate}] (0,0.5) to node[above,pos=.5]{$j_{1}$} (1.5,0);
\draw[in=-90,out=0,looseness=1,decoration={markings,mark=at position 0.6 with {\arrow[scale=1.5,>=stealth]{>}}},postaction={decorate}] (0,-0.5) to node[above,pos=.5]{$j_{3}$}(1.5,0);

\draw[decoration={markings,mark=at position 0.6 with {\arrow[scale=1.5,>=stealth]{<}}},postaction={decorate}] (0,0) to node[above,pos=.5]{$j_{2}$} (1.5,0);
\draw (0,0) node {$\bullet$}  ;
\draw (1.5,0) node[right] {$I_n$};
\draw (1.5,0) node {$\bullet$} ;
\draw (0,0.5) node {$\bullet$} ;
\draw (0,-0.5) node {$\bullet$} ;

\end{tikzpicture}
\caption{Graphical representation of interwiners. On the left picture, the node has three incoming links, thus an intertwiner is a map $I_n:\cV^{j_1} \otimes\cV^{j_2} \otimes\cV^{j_3} \to \mathbb{C}$. On the right, the mode has two incoming links and one outgoing link, which makes an intertwiner be a map $I_n:\cV^{j_1} \otimes \cV^{j_3} \to\cV^{j_2}$
\label{fig:interwiner}}
\end{figure}

A function $\Psi(\{g_{\ell}\}_{\ell\in\Gamma})$ is thus decomposed into modes obtained by assigning a spin $j_{\ell}$ to each link $\ell$, which are given by products of Wigner matrices:
\be
\prod_{\ell\in\Gamma}
D^{j_{\ell}}_{m^{t}_{\ell}m^{s}_{\ell}}(g_{\ell})
\,.
\ee
One still has to choose magnetic basis labels $m^{t}_{\ell},m^{s}_{\ell}$ on each link or say how they should be contracted. This is actually fixed by the requirement of local gauge invariance \eqref{eqn:SU2inv}, which means that one must glue the Wigner matrices associated to links with intertwiners living at the graph's nodes. More specifically, we choose an intertwiner at each node sending the incoming spins onto the outgoing spins, as drawn on fig.\ref{fig:interwiner}:
\be
I_{n}:
\bigotimes_{\ell\,|t(\ell)=n}\cV^{j}
\longrightarrow
\bigotimes_{\ell\,|s(\ell)=n}\cV^{j}
\,,\qquad
g\circ I_{n}= I_{n}\circ g\,,\quad \forall g\in\SU(2)
\,,
\ee
where $g$ acts simultaneously on all spins in the tensor products. 3-valent intertwiners are given by the Clebsh-Gordan coefficients, while intertwiners for higher-valent nodes can also be decomposed into 3-valent intertwiners by unfolding such a node into a tree with 3-valent nodes only. An interested reader will find details in \cite{Makinen:2019rou} or appendix \ref{app:SU2}.

Contracting Wigner matrices and interwiners gives spin-network wave functions:
\be
\Psi_{\{j_{\ell}, I_{n}\}}(\{g_{\ell}\})
\,\equiv\,
\sum_{\{m^{s,t}_{\ell}\}}
\prod_{\ell\in\Gamma}
\la j_{\ell},m^{t}_{\ell}|g_{\ell}|j_{\ell},m^{s}_{\ell}\ra
\,
\prod_{n\in\Gamma}
\la \bigotimes_{\ell\,|s(\ell)=n} j_{\ell},m^{s}_{\ell}|
 I_{n}
|\bigotimes_{\ell\,|t(\ell)=n} j_{\ell},m^{t}_{\ell}\ra
\,.
\ee
A scalar product between two wave-functions is the straightforward integration over the $\SU(2)$ Haar measure:
\be
\la \Psi |\tPsi\ra
\equiv
\int \left[\prod_{\ell}\rd g_{\ell}\right]\,
\overline{\Psi}(\{g_{\ell}\})\,\tPsi(\{g_{\ell}\}).
\ee
Since
\be
\la \Psi_{\{j_{\ell}, I_{n}\}} |\Psi_{_{\{\tj_{\ell},\tI_{n}\}}}\ra
=
\prod_{\ell}\f{\delta_{j_{\ell},\tj_{\ell}}}{d_{j_{\ell}}}
\,\prod_{n}\la  I_{n} | \tI_{n} \ra
\,,
\ee
choosing an orthonormal basis of intertwiners at every node yields an orthonormal basis of spin network states. This construction is referred to as the {\it spin-network basis}.

\smallskip

\begin{figure}[h!]

\centering

\begin{tikzpicture}[scale=1.4]

\coordinate(U) at (-0.8,0.2) ;
\coordinate(V) at (.5,1);
\coordinate(W) at (0.1,0);
\coordinate(X) at (.3,-.7);
\coordinate(Y) at (1.2,.6);

\coordinate(a) at (0,1.3);
\coordinate(b) at (0.2,.5);
\coordinate(c) at (-0.1,.3);

\draw[fill=teallight1] (a)--(b)--(c)--(a);
\draw (-0.3,1) node {$a(j_l)$};

\draw (U) node {$\bullet$} ;
\draw (V) node {$\bullet$};
\draw (W) node {$\bullet$} ;
\draw (X) node {$\bullet$} ;
\draw (Y) node {$\bullet$};

\draw (1.8,.6) node {$V(I_n)$};

\draw[decoration={markings,mark=at position 0.6 with {\arrow[scale=1.5,>=stealth]{>}}},postaction={decorate}] (U)--(V);
\draw[decoration={markings,mark=at position 0.6 with {\arrow[scale=1.5,>=stealth]{>}}},postaction={decorate}] (U)--(W);
\draw[decoration={markings,mark=at position 0.6 with {\arrow[scale=1.5,>=stealth]{>}}},postaction={decorate}] (V)--(Y);
\draw[decoration={markings,mark=at position 0.6 with {\arrow[scale=1.5,>=stealth]{>}}},postaction={decorate}] (Y)--(W);
\draw[decoration={markings,mark=at position 0.6 with {\arrow[scale=1.5,>=stealth]{>}}},postaction={decorate}] (U)--(X);
\draw[decoration={markings,mark=at position 0.6 with {\arrow[scale=1.5,>=stealth]{>}}},postaction={decorate}] (X)--(W);
\draw[decoration={markings,mark=at position 0.6 with {\arrow[scale=1.5,>=stealth]{>}}},postaction={decorate}] (X)--(Y);

\coordinate(d) at (1.2,0.8);
\coordinate(e) at (1,0.5);
\coordinate(f) at (1.4,0.3);
\coordinate(g) at (1.5,0.6);

\draw[gray] (d)--(e);
\draw[gray] (d)--(g);
\draw[gray] (d)--(f);
\draw[gray] (e)--(g);
\draw[gray] (g)--(f);
\draw[gray] (e)--(f);

\end{tikzpicture}
\caption{A spin network and its geometrical interpretation. A quantum of area $a(j_l)$ is associated to a link $l$ and a quantum of volume $V(I_n)$ is associated to a node $n$.
\label{fig:spinnetwork}}
\end{figure}
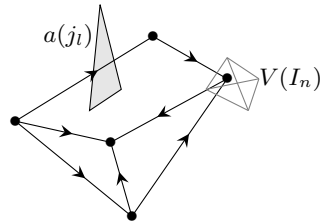

Through the LQG formalism, spin networks are provided with a clear geometrical interpretation as they define discrete quantum geometries. As illustrated on fig.\ref{fig:spinnetwork},
\begin{itemize}
\item Intertwiners $ I_{n}$ represent quanta of volume at each node $n$, with a discrete spectrum that can be a priori computed analytically, but, unfortunately, it does not yet admit a general closed formula \cite{Thiemann:1996au,Brunnemann:2004xi,Flori:2008nw,Bianchi:2012wb};

\item Spins $j_{\ell}$ represent quanta of area dual to links $\ell$, with a discrete spectrum $l_{Planck}^{2}\sqrt{j_{\ell}(j_{\ell}+1)}$ giving the area of a quantum surface interfacing two chunks of spatial volume living at the nodes $s(\ell)$ and $t(\ell)$ \cite{Rovelli:1994ge,Rovelli:1998gg}.

\end{itemize}

Let us underline that, despite this obvious geometrical interpretation, spin-network states are not semi-classical states of geometry. Indeed, they are peaked on areas and volumes, and thus maximally spread on the transport and holonomies, which are their conjugate variables. Semi-classical states of geometry should actually be coherent states in both the gravitational flux (frame field) and holonomy (connection field) variables.
In particular, for a pure spin-network state, intertwiners living at different nodes are completely decoupled and unentangled, and thus do not carry the expected correlations to represent vacuum states of smooth classical geometries. One would require coherent superpositions of spin networks to generate the appropriate large-scale correlations, see e.g. \cite{Bianchi:2023avf}.

\subsection{Spinfoam as history formalism for spin networks}

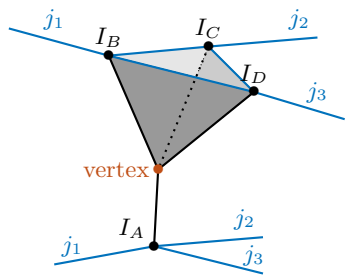
\begin{figure}[h!]

\centering

\begin{tikzpicture}[scale=1.2]

\coordinate(e) at (0.5,0.2);
\coordinate(c) at (-0.6,0.1);
\coordinate(d) at (1,-0.3);
\coordinate(f) at (1.1,-0,3);
\coordinate(g) at (-0.1,-2);

\coordinate(h) at (-1.7,0.4);
\coordinate(i) at (1.7,0.3);
\coordinate(j) at (2,-0.6);
\draw[RoyalBlue] (-1.2,0.5) node {$j_1$} ;
\draw[RoyalBlue] (1.5,0.5) node {$j_2$};
\draw[RoyalBlue] (1.7,-0.3) node {$j_3$};

\coordinate(k) at (-1.2,-2.2);
\coordinate(l) at (1.1,-2.3);
\coordinate(m) at (1.1,-1.9);

\draw[RoyalBlue] (-1,-2) node {$j_1$} ;
\draw[RoyalBlue] (0.9,-1.7) node {$j_2$};
\draw[RoyalBlue] (1,-2.1) node {$j_3$};

\draw[fill=teallight1] (e)--(d)--(c);
\draw[dotted,fill=teallight2] (e)--(f)--(d);
\draw[dotted,fill=teallight3] (f)--(d)--(c);

\draw[thick,RoyalBlue] (c)--(d);
\draw[thick,RoyalBlue] (d)--(e);
\draw[thick,RoyalBlue] (e)--(c);
\draw[thick] (d)--(f);
\draw[thick] (c)--(f);
\draw[thick] (f)--(g);
\draw[thick,RoyalBlue] (c)--(h);
\draw[thick,RoyalBlue] (e)--(i);
\draw[thick,RoyalBlue] (d)--(j);
\draw[thick,dotted] (e)--(f);
\draw[thick,RoyalBlue] (g)--(l);
\draw[thick,RoyalBlue] (g)--(m);
\draw[thick,RoyalBlue] (g)--(k);

\draw (e) node {$\bullet$} node[above]{$I_C$};
\draw[Bittersweet] (f) node {$\bullet$} node [left]{vertex};
\draw (d) node {$\bullet$} node[above]{$I_D$};
\draw (c) node {$\bullet$} node[above]{$I_B$};
\draw (g) node {$\bullet$} ;
\draw (-0.1,-1.8) node[left]{$I_A$};

\end{tikzpicture}
\caption{Spinfoam vertex creating a space-time scattering event between quanta of area, carried by spins propagating on 2d worldsheets (faces), and quanta of volumes, carried by intertwiners propagating on worldlines (edges).
\label{fig:spinfoamvertex}}
\end{figure}

Spinfoams arise as spin networks evolving through time and generating space-time in the process. Spin-network links $\ell$ carrying spins $j_{\ell}$ now sweep spinfoam faces $f$, which thus carry spins $j_{f}$. Spin-network nodes $n$ carrying intertwiners $I_{n}$ now sweep spinfoam edges $e$, which thus carry intertwiners $I_{e}$. A spinfoam is thus a 2-complex, and slicing a spinfoam yields spin networks. We complete the hierarchy of 2d-cells and 1d-cells with 0d-cells, that is spinfoam vertices. At a vertex, the spinfoam edges will branch out, creating new faces or annihilating old faces, as illustrated on fig. \ref{fig:spinfoamvertex}.
Hence, 2d-cells are evolving links. 1d-cells are evolving nodes. And 0d-cells are space-time events where a graph changes and the dynamics of geometry happens. Thus, while spins on spinfoam faces and intertwiners on spinfoam edges can be considered as kinematical data, spinfoam amplitudes live on spinfoam vertices, leading to the {\it local spinfoam ansatz} for the amplitude associated to a given decorated spinfoam 2-complex $\cC$:
\be
\label{eqn:localSFansatz}
Z^{SF}[{\cC}]
=
\sum_{\{j_{f},I_{e}\}}
\prod_{f\in\cC} \cA_{f}[j_{f}]
\prod_{e\in\cC} \cA_{e}[I_{e},\{j_{f}\}_{f\ni e}]
\prod_{v\in\cC} \cA_{v}[\{I_{e}\}_{e\ni v},\{j_{f}\}_{f\ni v}]
\,.
\ee
Here the face amplitude $\cA_{f}$ and the edge amplitude $\cA_{e}$ can be considered as defining the integration measure over quanta of areas and volumes, while the spinfoam vertex  amplitude $\cA_{v}$ truly encodes the dynamics of the theory.

Since a boundary of a spinfoam 2-complex $\cC$ is a spin network (possibly, consisting of multiple disconnected graphs), the discrete path integral \eqref{eqn:localSFansatz} defines a transition amplitude between spin-network states by summing over space-time bulk structures. The hierarchy of structures is summarized in the following table:

\def\arraystretch{1.2}
\setlength{\tabcolsep}{2em}
\noindent
\begin{center}
\begin{tabular}{| c |c|c|c|}
\hline \label{dualitycorrespondence}
spinfoam & algebraic & geometrical & imprint 
\\
bulk & data & interpretation & on boundary
\\\hline\hline
2-complex $\cC$ & $\{j_{f},I_{e}\}$ & space-time & spin network $\Psi_{\Gamma}$
\\\hline
face $f$ & spin $j_{f}$  & area &   link $\ell$
\\\hline
edge $e$ & intertwiner $I_{e}$  & volume  & node $n$
\\\hline
vertex $v$  & amplitude $\cA_{v}$ & 4d event & dynamics
\\\hline
\end{tabular}
\end{center}

The spinfoam 2-complexes can be considered as the (higher-dimensional equivalent of) Feynman diagrams for spin networks, describing the interaction of excited quanta of areas and volumes \cite{Reisenberger:1996pu}. This intuition becomes concrete in the {\it group field theory} formalism, where spinfoam amplitudes are indeed generated as the Feynman diagrams of a non-local field theory living on a group manifold and generalizing tensor models \cite{DePietri:1999bx,Reisenberger:2000zc,Reisenberger:2000fy,Freidel:2005qe}.

\smallskip

The basic spinfoam ansatz is the{ \it BF spinfoam ansatz}, which defines the discretized path integral for BF theory as a topological quantum field theory (TQFT). This is the original ansatz, which gave birth to the spinfoam approach \cite{Rovelli:1993kc,Barrett:1995mg,Baez:1997zt,Baez:1999sr}, and whose main examples are the Ponzano--Regge state-sum for 3d quantum gravity \cite{PR1968,Freidel:2004vi,Freidel:2005bb,Barrett:2008wh,Livine:2021sbf}, the Turaev-Viro model \cite{Turaev:1992hq,Dittrich:2016typ}  for 3d quantum gravity with cosmological constant, and the Ooguri model \cite{Ooguri:1992eb} and Crane--Yetter model for 4d BF theory \cite{Crane:1993if}.

One chooses a (semi-simple and compact) Lie group  $\cG$ and considers its set of irreducible (unitary) representations $\cJ$ (and, more generally, its category of representations equipped with a tensor product of representations, which can be decomposed into direct sum of irreducible representations), and intertwiners $\cI$ between them.
The BF ansatz is given by, up to signs:
\be
\cA_{f}[\cJ]=\dim \cJ
\,,\quad
\cA_{e}[\cI]=1
\,,\quad
\cA_{v}[\{\cJ_{f},\cI_{e}\}_{f,e\ni v}]=\Psi_{\{\cJ_{f},\cI_{e}\}}(\id)\,,
\ee
where $\Psi_{\{\cJ_{f},\cI_{e}\}}(\id)$ is the evaluation of the boundary spin network around the vertex $v$. This boundary spin network is obtained by enclosing the vertex $v$ with an abstract sphere and considering all the faces meeting at $v$ (which become the links of the boundary spin network) and their intersections (which become the nodes of the boundary spin network). Then the evaluation of the boundary spin network is the straightforward contraction of its intertwiners $I_{e}$, up to orientation switches due to the choice of orientation of the links. These orientation switches might generate extra signs, which we did not include in the ansatz above, but they become crucial when one actually computes amplitudes, for instance, to ensure the topological invariance of the spinfoam amplitude (i.e. that the details of the combinatorics of the bulk 2-complex do not matter but only its overall topology does).

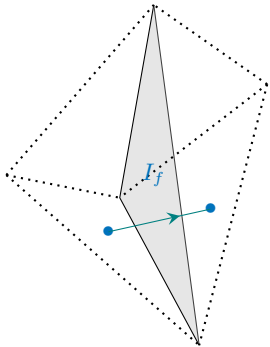
\begin{figure}[h!]

\centering

\begin{tikzpicture}[scale=1.5]

\coordinate(a) at (0,0) ;
\coordinate(c) at (.4,-3);
\coordinate(b) at (-.3,-1.7);
\coordinate(d) at (-1.3,-1.5);
\coordinate(e) at (1,-.7);
\coordinate(X) at (-0.4,-2);
\coordinate(Y) at (0.5,-1.8);

\draw[RoyalBlue] (X) node {$\bullet$};
\draw[RoyalBlue] (Y) node {$\bullet$};

\draw[fill=teallight1] (a)--(b)--(c);
\draw[darkgray] (a)--(c);
\draw[thick,dotted] (d)--(a);
\draw[thick,dotted] (d)--(b);
\draw[thick,dotted] (d)--(c);
\draw[thick,dotted] (e)--(a);
\draw[thick,dotted] (e)--(b);
\draw[thick,dotted] (e)--(c);
\draw[RoyalBlue] (0,-1.5) node {$I_f$};

\draw[decoration={markings,mark=at position 0.7 with {\arrow[scale=1.5,>=stealth]{>}}},postaction={decorate},teal]  (X)--(Y);

\end{tikzpicture}

\caption{Spinfoam edge and intertwiner identification. In dotted line, the two tetrahedra sharing one face (in gray), decorated by the three valent intertwiner $I_f$, associated with the spinfoam edge (in blue).
\label{fig:spinfoamedge}}
\end{figure}

The edge amplitude is set to 1, as long as we work with an orthonormal basis of intertwiners. The key point of this construction is that there is a single intertwiner on each spinfoam edge $e$. More precisely, if an edge $e$ is shared by two spinfoam vertices (the two endpoints of the edge), the same intertwiner $I_{e}$ appears in the boundary spin networks around both of these spinfoam vertices, as illustrated on fig. \ref{fig:spinfoamedge}. This ensures the topological invariance of the spinfoam path integral \cite{Girelli:2001wr,Bahr:2010bs}.

The resulting space-time picture is that of bubbles around each spinfoam vertex, so that vertices can be considered as ``centers'' of the bubbles. The amplitude of each bubble is given by its boundary spin-network  evaluation. Bubbles are glued together through the identification of intertwiners (up to orientation switches) along the (spinfoam) edges linking vertices.
This motivates the idea of a {\it foam of spins}.
Spinfoam models for non-topological theories (such as gravity) are typically constructed from their classical reformulations as a BF theory plus an additional potential or constraints. This leads to a constrained-BF spinfoam ansatz, where instead of summing over all possible labels, we restrain the sets of admissible representations and intertwiners. Adding such constraints can be formalized in terms of topological defects put on all faces and edges of the spinfoam, thus describable as a {\it sea of defects}.

\subsection{Dual picture: space-time triangulations}

It is natural to extend the interpretation of spin networks as discretized geometries to spinfoams. In fact, discretized space-time manifolds can be naturally interpreted as spinfoam 2-complexes.

Indeed, let us consider a four-dimensional triangulation made of 4-simplices glued together through shared tetrahedra. Then, the topological dual of this 4d triangulation is a spinfoam 2-complex or, simply said, a spinfoam, and the topological dual of its boundary 3d triangulation is a spin network.

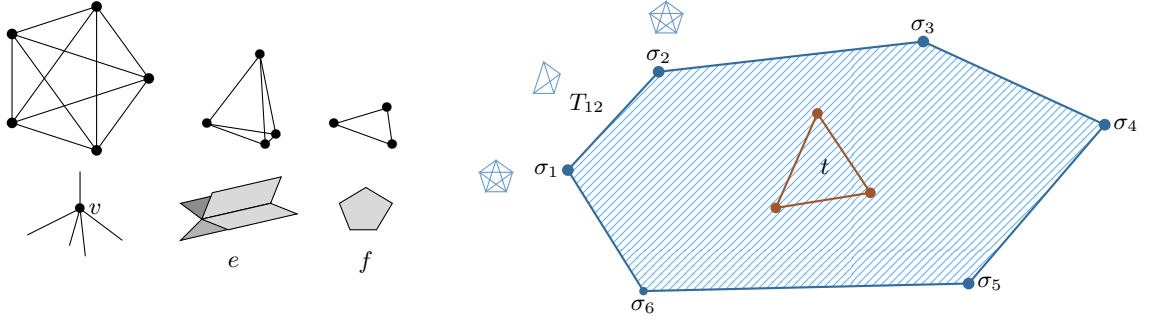
\begin{figure}[h!]
\begin{subfigure}[h]{0.3\linewidth}
  \begin{tikzpicture}[scale=0.70]
        \tikz\foreach \i in {1,...,5}
    \fill (\i*360/5:1) coordinate (n\i) circle(2 pt)
      \ifnum \i>1 foreach \j in {\i,...,1}{(n\i) edge (n\j)} \fi;

      \coordinate(x) at (2,1.9);
    \coordinate(y) at (2.1,0.2);
     \coordinate(z) at (1,0.6);
    \coordinate(w) at (2.3,0.4);
    \draw (x)--(y);
    \draw (y)--(z);
    \draw (w)--(y);
    \draw (z)--(w)--(x)--(z);
    \draw (x) node {$\bullet$} ;
    \draw (y) node {$\bullet$} ;
    \draw (z) node {$\bullet$} ;
    \draw (w) node {$\bullet$} ;

    \coordinate(A) at (4.4,.9);
    \coordinate(B) at (4.5,0.2);
     \coordinate(C) at (3.4,0.6);
     \draw (A)--(B)--(C)--(A);
     \draw (A) node {$\bullet$} ;
    \draw (B) node {$\bullet$} ;
    \draw (C) node {$\bullet$} ;

    \coordinate (s) at (-1.4,-1);
    \coordinate (s1) at (-1.4,-0.3);
    \coordinate (s2) at (-2.4,-1.5);
    \coordinate (s3) at (-0.6,-1.6);    
    \coordinate (s4) at (-1.3,-1.9);
    \coordinate (s5) at (-1.6,-1.7);

    \draw (s)--(s1);
    \draw (s)--(s2);
    \draw (s)--(s3);
    \draw (s)--(s4);
    \draw (s)--(s5);

    \draw (s) node {$\bullet$} node[right]{$v$};

    \coordinate (r1) at (0.9,-1.2);
    \coordinate (r2) at (2.2,-0.9);
    \draw (r1)--(r2);

    \coordinate (r3) at (1.1,-0.7);
    \coordinate (r4) at (2.4,-0.4);
    \coordinate (r5) at (1.4,-1.4);
    \coordinate (r6) at (2.7,-1.1);
    \coordinate (r7) at (0.5,-1.6);
    \coordinate (r8) at (1.8,-1.3);
    \coordinate (r9) at (0.5,-0.9);
    \coordinate (r10) at (1.8,-0.6);

    \filldraw[fill=gray!90] (r1)--(r2)--(r10)--(r9)--(r1);
    \filldraw[fill=gray!30] (r1)--(r2)--(r4)--(r3)--(r1);
    
    \filldraw[fill=gray!60] (r1)--(r2)--(r8)--(r7)--(r1);
    \filldraw[fill=gray!30] (r1)--(r2)--(r6)--(r5)--(r1);
    \coordinate (e) at (1.5,-2);
    \draw (e) node {$e$};

    \coordinate (p3) at (4,-0.6);
    \coordinate (p4) at (3.5, -0.9);
    \coordinate (p5) at (4.5,-0.9);
    \coordinate (p6) at (3.7,-1.4);
    \coordinate (p7) at (4.2,-1.4);

    \filldraw[fill=gray!30] (p3)--(p4)--(p6)--(p7)--(p5)--(p3);
    \coordinate (f) at (4,-2);
    \draw (f) node {$f$};
    \end{tikzpicture}
\end{subfigure}
\hspace*{20mm}
\begin{subfigure}[h]{0.55\linewidth}
\begin{tikzpicture}[scale=1]

\coordinate (G1) at (0,1.4);
\coordinate (G2) at (1.2,2.7);
\coordinate (G3) at (4.7,3.1);
\coordinate (G4) at (7.1,2.0);
\coordinate (G5) at (5.3,-0.1);
\coordinate (G6) at (1.0,-0.2);

\fill[pattern=north east lines,
      pattern color=RoyalBlue!60]
(G1)--(G2)--(G3)--(G4)--(G5)--(G6)--cycle;

\draw[RoyalBlue!70!black, thick]
(G1)--(G2)--(G3)--(G4)--(G5)--(G6)--cycle;

\foreach \P in {G1,G2,G3,G4,G5}{
    \fill[RoyalBlue!70!black] (\P) circle (2.2pt);
}
\fill[RoyalBlue!70!black] (G6) circle (1.6pt);

\node[left]      at (G1) {$\sigma_1$};
\node[above]     at (G2) {$\sigma_2$};
\node[above]     at (G3) {$\sigma_3$};
\node[right]     at (G4) {$\sigma_4$};
\node[right]     at (G5) {$\sigma_5$};
\node[below]     at (G6) {$\sigma_6$};

\node[above left] at ($(G1)!0.5!(G2)$) {$T_{12}$};

\coordinate (A) at (2.75,0.9);
\coordinate (B) at (3.3,2.15);
\coordinate (C) at (4.0,1.1);

\draw[Bittersweet!80!black, thick]
(A)--(B)--(C)--cycle;

\fill[Bittersweet!80!black] (A) circle (2pt);
\fill[Bittersweet!80!black] (B) circle (2pt);
\fill[Bittersweet!80!black] (C) circle (2pt);

\node at (3.4,1.45) {$t$};

\newcommand{\pentagram}[2]{
\begin{scope}[shift={(#1,#2)},scale=0.23]
\draw[RoyalBlue!60]
(90:1)--(306:1)--(162:1)--(18:1)--(234:1)--cycle;
\draw[RoyalBlue!60]
(18:1)--(90:1)--(162:1)--(234:1)--(306:1)--cycle;
\end{scope}
}

\newcommand{\tetrahedron}[2]{
\begin{scope}[shift={(#1,#2)},scale=0.23]
\draw[RoyalBlue!60]
(90:1)--(234:1)--(306:1)--(18:1)--(90:1)--(306:1);
\draw[RoyalBlue!60]
(18:1)--(234:1);
\end{scope}
}

\pentagram{-0.95}{1.3}
\pentagram{1.3}{3.4}

\tetrahedron{-0.32}{2.6}

\end{tikzpicture}

\end{subfigure}
 
\caption{
On the left side, an illustration of the three levels of geometric structure: a 4d simplex and its dual spinfoam vertex, a 3d tetrahedron and its dual spinfoam edge, a 2d triangle and its dual spinfoam face or plaquette.
On the right side, the plaquette dual to a triangle: its boundary loops across all the 4-simplices sharing the triangle.
\label{fig:dualplaquette}}

\end{figure}

Starting with the 4d triangulation, a 4-simplex will be represented as a  spinfoam vertex. Two neighbouring 4-simplices are represented as two spinfoam vertices connected by a spinfoam edge, which represents a shared tetrahedron. The dual of a triangle is a spinfoam face, or plaquette, which is topologically a disk, 1d boundary of which is made of spinfoam vertices dual to the 4-simplices sharing this triangle connected by the spinfoam edges dual to the tetrahedra sharing the same triangle, as illustrated on fig. \ref{fig:dualplaquette}.
This can be summarized in the following table:
\def\arraystretch{1.2}
\setlength{\tabcolsep}{2em}
\noindent
\begin{center}
\begin{tabular}{| c |c|c|}
\hline
4d triangulation $\Delta$& spinfoam 2-complex $\cC$ & algebraic data
\\\hline\hline
4-simplex $\sigma$ & vertex $v$  & vertex amplitude $\cA_{v}$
\\\hline
tetrahedron $T$ &edge $e$ & intertwiner $I_{e}$
\\\hline
triangle $t$ &face $f$ & spin $j_{f}$
\\\hline 
\end{tabular}
\end{center}
\label{duality4drules}

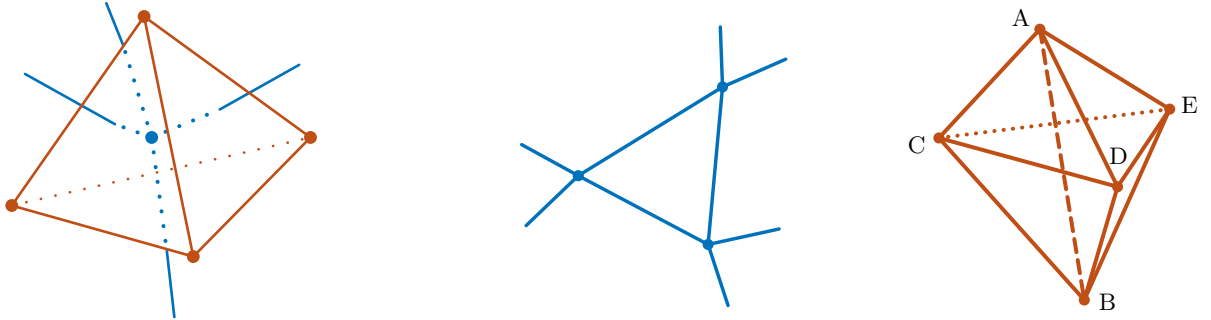
\begin{figure}[h!]

\centering

\begin{subfigure}[h]{0.3\linewidth}

\parbox[c][5cm][c]{5cm}{
\centering 
\begin{tikzpicture}[
  scale=0.5,
  edge/.style   ={Bittersweet,  line width=1pt,   line cap=round},
  hidden/.style={Bittersweet,  line width=1pt, line cap=round, dash pattern=on 0pt off 5.5pt},
  ray/.style    ={RoyalBlue, line width=1pt,   line cap=round},
  hray/.style  ={RoyalBlue, line width=1.5pt, line cap=round, dash pattern=on 0pt off 5.5pt}
]

\coordinate (A) at (4.40,8.70);  
\coordinate (B) at (8.80,5.50);  
\coordinate (C) at (0.90,3.70);  
\coordinate (D) at (5.70,2.35);  
\coordinate (O) at (4.60,5.50);  

\coordinate (p1) at (3.85,7.93);  \coordinate (e1) at (3.40,9.05);
\coordinate (p2) at (3.63,5.85);  \coordinate (e2) at (1.25,7.18);
\coordinate (p3) at (6.4,6.24);  \coordinate (e3) at (8.50,7.42);
\coordinate (p4) at (5.00,2.50);  \coordinate (e4) at (5.2,0.75);
 
\draw[hidden] (C) -- (B);

\foreach \i in {1,2,3,4}{
  \draw[hray] (O) -- (p\i);
  \draw[ray]  (p\i) -- (e\i);
}
 
\draw[edge] (A) -- (C) -- (D) -- (B) -- cycle;
\draw[edge] (A) -- (D);
 
\foreach \v in {A,B,C,D}{\fill[Bittersweet] (\v) circle[radius=0.17];}
\fill[RoyalBlue] (O) circle[radius=0.17];
 
\end{tikzpicture}
}

\caption{A tetrahedron and its dual: a 4-valent node with four dual edges dual to the four triangles.}
\end{subfigure}
\hspace{10mm}
\begin{subfigure}[h]{0.6\linewidth}

\parbox[c][5cm][c]{10cm}{
\centering 
\begin{tikzpicture}[
  x=1pt, y=-1pt,                       
  ln/.style     ={Bittersweet,  line width=1.5pt, line cap=round, line join=round},
  dsh/.style    ={Bittersweet,  line width=1.5pt, line cap=round, dash pattern=on 5pt off 3pt},
  dot/.style    ={Bittersweet,  line width=1.5pt, line cap=round, dash pattern=on 0pt off 3.4pt},
  bl/.style     ={RoyalBlue, line width=1.3pt, line cap=round},
  lab/.style    ={black,  font=\small}, scale=0.835
]

\coordinate (N1) at (109.5, 39);
\coordinate (N2) at ( 44.5, 79);
\coordinate (N3) at (103.0,110);
 
\draw[bl] (N1) -- (N2) -- (N3) -- (N1);
\draw[bl] (N1) -- (108.5, 12);   \draw[bl] (N1) -- (138.0, 26.5);
\draw[bl] (N2) -- ( 19.0, 65);   \draw[bl] (N2) -- ( 21.0,101.5);
\draw[bl] (N3) -- (135.0,103);   \draw[bl] (N3) -- (112.0,137.5);

\foreach \n in {N1,N2,N3}{\fill[RoyalBlue] (\n) circle[radius=2.5pt];}
 
\coordinate (A) at (252.5, 13);   
\coordinate (C) at (207.0, 62);   
\coordinate (E) at (311.0, 49);   
\coordinate (D) at (287.5, 84);   
\coordinate (B) at (272.5,135);   
 
\draw[dot] (C) -- (E);                       
\draw[dsh] (A) -- (B);                       
\draw[ln]  (A) -- (C) -- (B) -- (E) -- cycle;
\draw[ln]  (A) -- (D) -- (B);
\draw[ln]  (C) -- (D) -- (E);
 
\foreach \v in {A,C,E,D,B}{\fill[Bittersweet] (\v) circle[radius=2.5pt];}
 
\node[lab] at (244, 8) {A};
\node[lab] at (197, 65) {C};
\node[lab] at (320, 47) {E};
\node[lab] at (288, 70) {D};
\node[lab] at (283,136) {B};
 
\end{tikzpicture}
}

\caption{An assemblage of three tetrahedra around a shared edge and its dual: three 4-valent nodes, each dual to a tetrahedron (ABCD, ABDE and ABEC), linked together by a triangular loop (in {\color{RoyalBlue} blue}) dual to the bulk shared edge (in {\color{Bittersweet} dash}).}
\end{subfigure}

\caption{
Basic examples of 3d triangulations and their dual graph.
\label{fig:dual4valentnode}
}

\end{figure}

The boundary of a 4d triangulation is, of course, a 3d triangulation made of tetrahedra glued together through shared triangles. This leads to the following duality structure:
\def\arraystretch{1.2}
\setlength{\tabcolsep}{2em}
\noindent
\begin{center}
\begin{tabular}{| c |c|c|}
\hline
3d triangulation $\pp\Delta$& spin network $\Gamma=\pp\cC$ & algebraic data
\\\hline\hline
tetrahedron $T$&node  $n$ & intertwiner $I_{n}$
\\\hline
triangle $t$ &link $\ell$ & spin $j_{\ell}$
\\\hline
\end{tabular}
\end{center}
The resulting spin network is automatically built on a 4-valent graph, with 4-valent nodes dual to tetrahedra as illustrated on fig. \ref{fig:dual4valentnode}.

\smallskip

This provides a natural geometrical interpretation for the evolution of spin networks. Indeed, let us consider a 3d triangulation and its dual 4-valent spin network. Let us now make it evolve by laying a single 4-simplex on top of it. This is the equivalent of a single spinfoam vertex, i.e. a single space-time event. There are different ways to lay a 4-simplex on this initial 3d triangulation. It could rest on a single tetrahedron, or two tetrahedra, or three, or four. The resulting evolution corresponds to the various ways to slice a 4-simplex into an initial and final sets of tetrahedra. Since a 4-simplex is made of five tetrahedra, the possible moves are, as drawn on fig. \ref{fig:4simplexPachnermoves}:
\begin{itemize}
\item the $1\leftrightarrow 4$ Pachner move, mapping one tetrahedron into four tetrahedra by adding a new point inside (or outside) that tetrahedron, and its reverse move;
\item the $2\leftrightarrow 3$ Pachner move, transforming two neighbouring tetrahedra into three tetrahedra by adding a new (triangulation) edge, and its reverse move.
\end{itemize}
Drawing the topological dual leads to the possible evolution moves for (4-valent) spin networks. These moves are the elementary space-time events. And spinfoams are all about prescribing precise probability amplitudes to the dynamics consisting of a combination of the moves resulting from piling up 4-simplices.
\begin{figure}[h!]

\centering

\begin{tikzpicture}[scale=1.2]
  \tikz\foreach \i in {1,...,5}
    \fill (\i*360/5:1) coordinate (n\i) circle(2 pt)
      \ifnum \i>1 foreach \j in {\i,...,1}{(n\i) edge (n\j)} \fi;

      \coordinate(a) at (-1.4,-0.1);
      \coordinate(b) at (0,0.5);
      \draw[thick,teal] (a)--(b);
\end{tikzpicture}
\hspace{30ex}
\begin{tikzpicture}[scale=1.2]

 \coordinate(a) at (0,0);
 \coordinate(b) at (0,0.5);
 \coordinate(c) at (-0.6,-0.2);
 \coordinate(d) at (0.7,0.3);
 \coordinate(e) at (0.6,-0.3);
 \draw (a)--(b);
 \draw (a)--(c);
 \draw (a)--(d);
 \draw (a)--(e);
 \draw (a) node {$\bullet$};
 \draw [->, line width = 1] (1.5,0) -- (2.5,0);

 \coordinate(f) at (4.6,0.2);
 \coordinate(g) at (4.3,-0.2);
 \coordinate(h) at (3.8,0.2);
 \coordinate(i) at (3.5,-0.2);
 \coordinate(j) at (4.8,0.4);
 \coordinate(k) at (4.5,-0.4);
 \coordinate(l) at (3.6,0.4);
 \coordinate(m) at (3.3,-0.5);

 \draw (f)--(g);
 \draw (f)--(h);
 \draw (i)--(g);
 \draw (i)--(h);
 \draw[teal] (f) node {$\bullet$};
 \draw[teal] (g) node {$\bullet$};
 \draw[teal] (h) node {$\bullet$};
 \draw[teal] (i) node {$\bullet$};
 \draw[teal] (f)--(i);
 \draw[teal] (g)--(4.12,-0.05);
 \draw[teal] (h)--(4,0.05);
 \draw (f)--(j);
 \draw (g)--(k);
 \draw (h)--(l);
 \draw (i)--(m);

 \coordinate(x) at (0,-1.1);
 \coordinate(y) at (-0.4,-1.9);
 \coordinate(z) at (0.6,-1.8);
 \coordinate(w) at (0.3,-2);
 \draw (x)--(y);
 \draw (y)--(z);
 \draw (w)--(y);
 \draw (z)--(w)--(x)--(z);

 \draw [->, line width = 1] (1.5,-1.5) -- (2.5,-1.5);
 \coordinate(x2) at (4,-1.1);
 \coordinate(y2) at (3.6,-1.9);
 \coordinate(z2) at (4.6,-1.8);
 \coordinate(w2) at (4.3,-2);
 \coordinate(o) at (4,-1.7);
 \draw (x2)--(y2);
 \draw (y2)--(z2);
 \draw (w2)--(y2);
 \draw (z2)--(w2)--(x2)--(z2);
 \draw[teal] (o)--(x2);
 \draw[teal] (o)--(y2);
 \draw[teal] (o)--(z2);
 \draw[teal] (o)--(w2);
 \draw[teal] (o) node {$\bullet$};
\end{tikzpicture}
\caption{ Slicing a 4-simplex into $1\leftrightarrow 4$ Pachner moves and the resulting spin-network evolution moves.
\label{fig:4simplexPachnermoves}}
\end{figure}
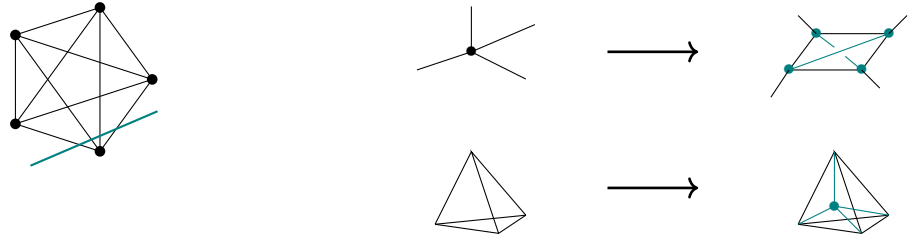

\begin{figure}[h!]

\centering

\begin{tikzpicture}[scale=1.2]
  \tikz\foreach \i in {1,...,5}
    \fill (\i*360/5:1) coordinate (n\i) circle(2 pt)
      \ifnum \i>1 foreach \j in {\i,...,1}{(n\i) edge (n\j)} \fi;

      \coordinate(a) at (-1.8,0.7);
      \coordinate(b) at (-0.2,0.2);
      \draw[thick,teal] (a)--(b);
\end{tikzpicture}
\hspace{30ex}
\begin{tikzpicture}[scale=0.8]

\coordinate(A) at (0.5,3.5);
\coordinate(B) at (-1.3,3.5);
\coordinate(A1) at (1.2,4);
\coordinate(A2) at (1.2,3);
\coordinate(A3) at (1.2,3.5);
\coordinate(B1) at (-2,4);
\coordinate(B2) at (-2,3);
\coordinate(B3) at (-2,3.5);

\draw (A)--(B);
\draw (A)--(A1);
\draw (A)--(A2);
\draw (A)--(A3);
\draw (B)--(B1);
\draw (B)--(B2);
\draw (B)--(B3);
\draw (A) node {$\bullet$} ;
\draw (B) node {$\bullet$} ;

\coordinate(X) at (4.5,3.4);
\coordinate(Y) at (5.9,3.4);
\coordinate(Z) at (5.2,4.2);
\coordinate(X1) at (4,3);
\coordinate(Y1) at (6.3,3.8);
\coordinate(Z1) at (5.6,4.7);
\coordinate(X2) at (4,3.8);
\coordinate(Y2) at (6.3,3);
\coordinate(Z2) at (4.8,4.7);

\draw (X)--(Y)--(Z)--(X);
\draw (X1)--(X);
\draw (X2)--(X);
\draw (Y1)--(Y);
\draw (Y2)--(Y);
\draw (Z1)--(Z);
\draw (Z2)--(Z);

\draw (X) node {$\bullet$} ;
\draw (Y) node {$\bullet$} ;
\draw (Z) node {$\bullet$} ;

 \draw [->, line width = 1] (1.5,3.5) -- (3,3.5);

 \coordinate(a2) at (0.1,2.5) ;
\coordinate(c2) at (0.3,-0.3);
\coordinate(b2) at (-0.3,.5);
\coordinate(d2) at (-1.1,0.7);
\coordinate(e2) at (1,1.3);

\draw (a2)--(b2)--(c2);
\draw (d2)--(a2);
\draw[darkgray] (d2)--(b2);
\draw (d2)--(c2);
\draw (e2)--(a2);
\draw[darkgray] (e2)--(b2);
\draw (e2)--(c2);

\draw[dotted] (e2)--(d2);

 \draw [->, line width = 1] (1.5,1.2) -- (3,1.2);
 \coordinate(a) at (5.1,2.5) ;
\coordinate(c) at (5.3,-0.3);
\coordinate(b) at (4.7,.5);
\coordinate(d) at (3.9,0.7);
\coordinate(e) at (6,1.3);

\draw (a)--(b)--(c);
\draw[thick,teal] (a)--(c);
\draw (d)--(a);
\draw[darkgray] (d)--(b);
\draw (d)--(c);
\draw (e)--(a);
\draw[darkgray] (e)--(b);
\draw (e)--(c);

\draw[dotted] (e)--(d);
\end{tikzpicture}
\caption{ Slicing a 4-simplex into $2\leftrightarrow 3$ Pachner moves and the resulting spin-network evolution moves.
\label{fig:4simplexPachnermoves}}
\end{figure}
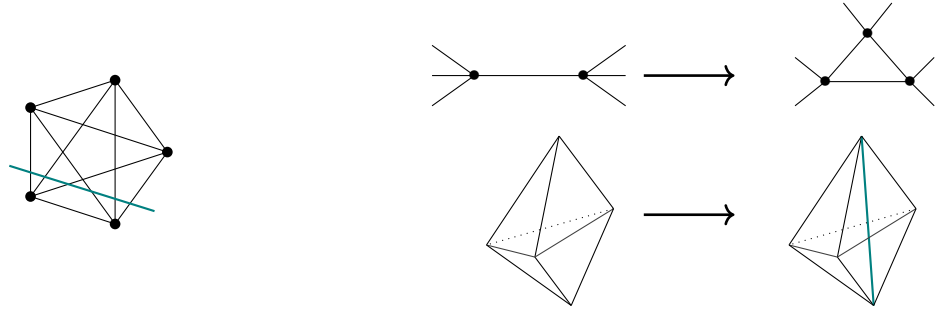
\section{1d Spinfoams: Quantum Mechanics}

Now, we would like to start with one-dimensional systems to illustrate the basic logic of path integrals and topological invariance.
Let us consider a simple mechanical system and quantize it using a path integral. The action reads:
\be
S[t,q(t)]=\f m{2}\int_{t_{i}}^{t_{f}} \rd t\,\left[
 \dot{q}^{2}-V(q)\right]
 \,,\qquad
 \cS_{H}[t,q(t),p(t)]=
\int \rd t\,\Big{[}
p\dot{q}-\f{p^{2}}{2m}-V(q)
\Big{]}
\,.
\ee
For the sake of simplicity, we focus on the case of a free particle. Then, $V=0$,
\be
S[t,q(t)]=\f m{2}\int_{t_{i}}^{t_{f}} \rd t\,
 \dot{q}^{2}
 \,,\qquad
 \cS_{H}[t,q(t),p(t)]=
\int \rd t\,\Big{[}
p\dot{q}-\f{p^{2}}{2m}
\Big{]}
\,.
\ee
Let us consider the path-integral definition of a transition amplitude,
\be
\cA\big{[}(t_{i},t_{f}),(q_{i},q_{f})\big{]}
=
\int_{q(t_{i,f)}=q_{i,f}} [\cD q(t)]\, e^{\f{im}{2\hbar}\int_{t_{i}}^{t_{f}}\rd t\, \dot{q}^{2}}
\,.
\ee
We  discretize this path integral by introducing intermediate times:
\be
t_{i}=t_{0}<t_{1}<t_{2}<..<t_{N-1}<t_{N}=t_{f}
\ee
and integrating over the position $q_{1},..,q_{N-1}$ at those intermediate times:
\be
\cA_{N}=
\cN_{N}\int \prod_{k=1}^{N-1}\rd q_{k}\,
e^{\f{im}{2\hbar}\sum_{k=0}^{N-1}(t_{k+1}-t_{k})\left(\f{q_{k+1}-q_{k}}{t_{k+1}-t_{k}}\right)^{2}}\,,\label{1damplitude}
\ee
where $\cN_{N}$ is a to-be-stipulated renormalization factor. The challenge of a path integral is to understand the $N\rightarrow +\infty$ limit of this multi-dimensional integral and  how it depends on the choice of intermediate times.

The factor $(t_{k+1}-t_{k})$ gives the length of each interval and leads to the more usual scaling factor of $\f1N$ in the equidistant case, thereby ensuring the correct normalization of the Riemann sum.
We show here that we do not need to take equal time intervals and that the path integral is actually independent of the discretization truncation level $N$ and the values of the times $t_k$.

\medskip

In order to exactly perform a Gaussian integral over each intermediate position, we notice that the correct normalization factor of the discretized path integral is
\be
\cN_{N}=\prod_{k=0}^{N-1}\sqrt{\f{\lambda}{\pi(t_{k+1}-t_{k})}} 
\qquad\textrm{with}\quad
\lambda=\f{m}{2i\hbar}
\,.
\ee
Such factor warrants the 1d topological invariance of the path integral.

Each intermediate position appears in two terms of the sum in the exponent in \eqref{1damplitude}. For instance, integrating over $q_{1}$, we get:
\be
\sqrt{\f{\lambda}{\pi(t_{1}-t_{0})}}\sqrt{\f{\lambda}{\pi(t_{2}-t_{1})}}
\int_{\R} \rd q_{1}\,e^{-\lambda\f{(q_{1}-q_{0})^{2}}{(t_{1}-t_{0})}}
e^{-\lambda\f{(q_{2}-q_{1})^{2}}{(t_{2}-t_{1})}}
=
\sqrt{\f{\lambda}{\pi(t_{2}-t_{0})}}e^{-\lambda\f{(q_{2}-q_{0})^{2}}{(t_{2}-t_{0})}}
\,.
\ee
The dependence on the intermediate $t_{1}$ totally disappears and we have a weight which is exactly invariant under coarse-graining and refinement of the time axis.
In summary, this leads to the following transition amplitude:
\be
\cA_{N}
=
\sqrt{\f{\lambda}{\pi\,T}}\,e^{-\lambda\f{(\Delta q)^{2}}{T}}
=
\sqrt{\f{m}{2i\pi  \hbar\,T}}\,e^{\f{im}{2\hbar}\f{(\Delta q)^{2}}{T}}\,.
\ee
We have an exact discretization of the path integral, without the need to take the limit $N\rightarrow +\infty$. This property, and its higher-dimensional generalisation, is often referred to as a {\it perfect discretization} \cite{Bahr:2011uj}. It is a key property of topological field theories and a powerful tool for computing quantum corrections to almost topological field theories such as gravity \cite{Asante:2021blx}.

Moreover, this discretization-independent path integral matches exactly the quantum mechanics computation! Indeed, the quantum transition amplitude is simply given by
\be
\cA\big{[}(t_{i},t_{f}),(q_{i},q_{f})\big{]}
=
\la q_{f}|e^{-i\hbar^{-1}T\hH}|q_{i}\ra
\,,\qquad
\hH=\f1{2m}{\hp^{2}}\,,
\ee
with $T=(t_{f}-t_{i})$.
Switching to the momentum basis by a Fourier transform, we compute
\beq
\cA\big{[}(t_{i},t_{f}),(q_{i},q_{f})\big{]}
&=&
\f1{2\pi \hbar}\int \rd p_{i}\rd p_{f}
\,
e^{i\hbar^{-1}(q_{f}p_{f}-q_{i}p_{i})}e^{-i\hbar^{-1}T\f{p_{i}^{2}}{2m}}\delta(p_{f}-p_{i})
\\
&=&
\f1{2\pi \hbar}\int \rd p\,e^{ip\f{(q_{f}-q_{i})}{\hbar}}e^{-ip^{2}\f{T}{2m\hbar}}
\nn\\
&=& 
\sqrt{\f{m}{2i\pi \hbar\,T}}\,e^{\f{im}{2\hbar}\f{(\Delta q)^{2}}{T}}
\,,
\eeq
which leads back to the previous path integral result as expected.

Although this a priori seems to be a feature of a free particle (for instance, the discretization might not be as perfect when a potential is switched on\footnotemark{}), this is actually a natural property of the composition of unitary operators,
\beq
\la q_{N}|e^{-i\hbar^{-1}(\tau_{N}-\tau_{0})\hH}|q_{0}\ra
&=&
\la q_{N}|e^{-i\hbar^{-1}(\tau_{N}-\tau_{N-1})\hH}..e^{i(\tau_{2}-\tau_{1})\hH}e^{i(\tau_{1}-\tau_{0})\hH}|q_{0}\ra
\nn\\
&=&
\int [\rd q_{k}]_{k=1..N-1}\,
\prod_{n=0}^{N-1}\la q_{n+1}|e^{-i\hbar^{-1}(\tau_{N}-\tau_{0})\hH}|q_{n}\ra
\,,
\eeq
\footnotetext{When a potential admitting a convergent Taylor series in a finite neighborhood of, say, $q=0$, is turned on, it can be included in a straightforward manner. However, terms with power higher than quadratic will not translate into Gaussian terms and will have to be expanded into integrals of the type $\int q^{n} e^{-\gamma q^{2}}$, which are nevertheless easily evaluated.}
and is therefore an inherent property of path integrals in quantum mechanics.
This makes quantum mechanics (for a time-independent Hamiltonian) a basic example of a 1d TQFT, and its path integral -- a toy-model archetype for spinfoam models in higher dimensions.
%

\section{2d Spinfoams}

Let us move one dimension up and look into 2d spinfoams. We consider 2d BF theory and we'll build the discretized path integral for two cases: the $\U(1)$ gauge group and the $\SU(2)$ gauge group. The two constructions will be almost identical, up to details due to $\SU(2)$ being non-abelian. In both cases, the key point will be the topological invariance of BF theory. Topological invariance means that the field theory does not have any local physical degrees of freedom, leaving only topological (global) and boundary degrees of freedom.
Not only does this yield intrinsically holographic quantum theories, but it also implies that discretization does not result in a loss of degrees of freedom or truncation of the theory. Thus, we are able to define a discretized path integral in such a way that it does not depend on the details of a discretization, just like the 1d ``spinfoam'' path integral described above.

\subsection{2d BF theory for the $\U(1)$ group}

\subsubsection{Classical 2d BF theory}

We start with a 2d BF theory for the $\U(1)$ group defined as a classical field theory with the following Lagrangian:
\be
S[A,B]
=
\int \rd^{2} x\,  B F[A]
=
\int \rd t \rd x  \,B \big{(}\partial_t A_x - \partial_x A_t\big{)}
\,.
\ee
Let us give a brief overview of the classical theory.
The Lagrangian is invariant under local gauge transformations parametrized by a field $\phi(t,x)$:
\be
\left|
\begin{array}{lcl}
A&\mapsto& A+\rd\phi
\,,\\
B&\mapsto& B
\,.
\end{array}
\right.
\ee
This is a direct consequence of the gauge invariance of the curvature:
\be
F=\partial_t A_x - \partial_x A_t
\,\,\mapsto\,\,
\partial_t (A_x+\pp_{x}\phi) - \partial_x (A_t+\pp_{t}\phi))
=\partial_t A_x - \partial_x A_t=F
\,.
\ee
The Euler-Lagrange equation of this classical action imposes a flat connection $A$ and a constant scalar field $B$:
\be
\f{\delta S}{\delta B}=0
\Leftrightarrow
F[A]=0
\,,\qquad
\f{\delta S}{\delta A}=0
\Leftrightarrow
\rd B=0
\,.
\ee
To clarify the phase space of the theory, we can perform a Hamiltonian analysis for a foliation by constant-$t$ slices $X_t$\footnotemark.
\footnotetext{
We can perform a similar canonical analysis for an arbitrary foliation of the 2d space-time using the covariant phase space formalism, e.g. \cite{Girelli:2021zmt,Geiller:2020edh}.  This allows for a cleaner analysis clarifying the role of boundary terms.
}
Let us identify the terms with a time derivative in the Lagrangian:
\be
S[A,B]
=
\int \rd t \int \rd x  \, \Big{[}
B \partial_t A_x + A_t\partial_x B
\Big{]}
+\int \rd t\, BA_{t}|_{\pp X_t}
\,.
\ee
This leads to the canonical Poisson bracket on a constant-$t$ slice:
\be
\{A_{x}(t,x),B(t,y))\}=\delta(x-y)\,,
\ee
with the field $B$ being the momentum canonically conjugate to the connection $A_{x}$.
The Hamiltonian is:
\be
H=-\int \rd x\,A_t\partial_x B
\,.
\ee
The field $A_{t}$ does not have a canonical momentum, it is not a dynamical field and plays the role of a Lagrange multiplier enforcing the Hamiltonian constraint $\cH\equiv \partial_x B=0$. We recognize the spatial projector onto the constant-$B$ solutions to the equations of motion. The key point is to further realize that the dynamics generated by this Hamiltonian (constraint) is given purely by a gauge transformation:
\be
\left|
\begin{array}{lclcl}
\delta_{\lambda}A_{x}
&=&
\{A_{x},\int \lambda\pp_{x} B \}
&=&
-\pp_{x} \lambda
\,,\\
\delta_{\lambda}B
&=&
\{B,\int \lambda\pp_{x} B \}
&=&
0
\,.
\end{array}
\right.
\ee
Setting $\lambda=-A_{t}$, we of course recover the equations of motion:
\be
\pp_{t} A_{x}=-\delta_{A_{t}}A_{x}=\pp_{x} A_{t}
\,,\qquad
\pp_{t}B=\delta_{A_{t}}B=0\,,
\ee
with the last piece of the equations of motion being simply the Hamiltonian constraint, $\partial_x B=0$, itself.
The implementation of dynamics as a gauge transformation is a key aspect of topological theories that needs to be preserved by the path-integral formalism.

\subsubsection{Holonomies: non-local observables and gauge-covariance.}

In order to discretize the path integral, we naturally turn towards non-local observables. These are the $\U(1)$ holonomies of the connection $A$, which define the integrated version of the connection along curves.
More precisely, for a curve $\gamma:s\in[0,1]\mapsto \gamma(s)\in\cM_{2d}$, we define the abelian phase:
\be
g[\gamma]
\equiv
e^{i\int_{0}^{1}\rd s\, \pp_{s}\gamma^{a}(s)\, A_{a}[\gamma(s)]}
\quad\in\U(1)
\,.
\ee
Let's look at the action of  a gauge transformation $A\mapsto A+\rd\phi$ on this integral along the curve $\gamma$:
\be
\Theta[\gamma]=\int_{0}^{1}\rd s\, \pp_{s}\gamma^{a}(s)\, A_{a}[\gamma(s)]
\quad\longmapsto\quad
\int_{0}^{1}\rd s\, \pp_{s}\gamma^{a}(s)\, (A_{a}+\pp_{a}\phi)
=
\Theta[\gamma]+\phi(\gamma(1))-\phi(\gamma(0))
\,.
\ee
Gauge transformations thus act in a simple way on holonomies through $\U(1)$ transformations  at the endpoints of the considered path:
\be
g[\gamma]
\,\,\mapsto\,\,
h_{t(\gamma)}\,g[\gamma]\,h_{s(\gamma)}^{-1}
\,,\qquad
h_{p}\equiv e^{i\phi(p)}\,.
\ee
A direct consequence is that the holonomy along an open curve is not gauge-invariant, but gauge-covariant. Nonetheless, a holonomy around a closed loop is gauge-invariant, just like the curvature $F$. This is called a {\it Wilson loop} observable. It defines a finite equivalent of the curvature.

\subsubsection{The discretized path integral for 2d BF theory}

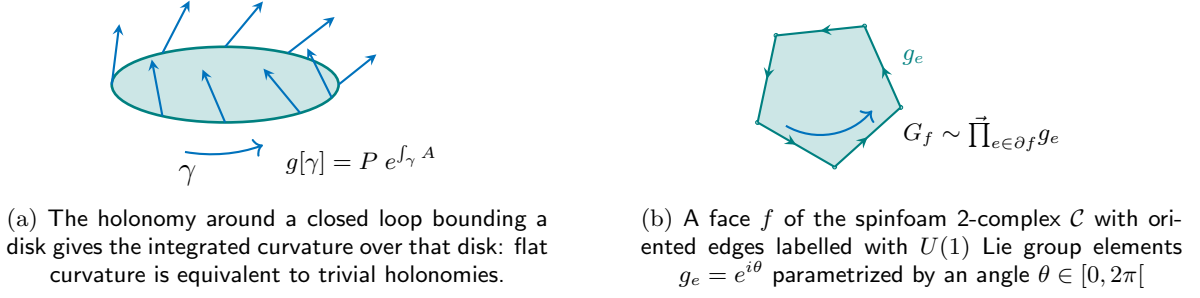
\begin{figure}[h!]
\begin{subfigure}[h]{0.4\linewidth}
\parbox[c][2cm][c]{7cm}{
\begin{tikzpicture}[
  tangent/.style={
    decoration={
      markings,
      mark=
      at position #1
      with
      {
        \coordinate (tangent point-\pgfkeysvalueof{/pgf/decoration/mark info/sequence number}) at (0pt,0pt);
        \coordinate (tangent unit vector-\pgfkeysvalueof{/pgf/decoration/mark info/sequence number}) at (1,0pt);
        \coordinate (tangent orthogonal unit vector-\pgfkeysvalueof{/pgf/decoration/mark info/sequence number}) at (0pt,1);
      }
    },
    postaction=decorate
  },
  use tangent/.style={
    shift=(tangent point-#1),
    x=(tangent unit vector-#1),
    y=(tangent orthogonal unit vector-#1)
  },
  use tangent/.default=1
  ]

  \draw[ thick, RoyalBlue, <-] (3,-0.8) arc (-70:-100:2);
  \node at (2,-1.2) {{\large $\gamma$}};
  \node at (4.3,-1) { $g[\gamma] = P \;e^{\int_\gamma A}$};

  \filldraw[
  tangent=0,
  tangent=0.125,
  tangent=0.25,
  tangent=0.375,
  tangent=0.5,
  tangent=0.625,
  tangent=0.75,
  tangent=0.9,
  tangent=0.97,line width=1pt,color=teal,fill=teal!20] (2.5,0) ellipse (1.5 and 0.5);
  
  \draw [RoyalBlue, thick, use tangent, -stealth] (0,0) -- (0.4,-0.5) coordinate (A);
  \draw [RoyalBlue, thick, use tangent=2, -stealth] (0,0) -- (-0.5,-0.6);
  \draw [RoyalBlue, thick, use tangent=3, -stealth] (0,0) -- (-0.5,-0.6);
  \draw [RoyalBlue, thick, use tangent=4, -stealth] (0,0) -- (-0.5,-0.6);
  \draw [RoyalBlue, thick, use tangent=5, -stealth] (0,0) -- (-0.8,0.1);
  \draw [RoyalBlue, thick, use tangent=6, -stealth] (0,0) -- (-0.3,0.7);
  \draw [RoyalBlue, thick, use tangent=7, -stealth] (0,0) -- (-0.3,0.7);
  \draw [RoyalBlue, thick, use tangent=8, -stealth] (0,0) coordinate (a) -- (-0.3,0.7) coordinate (b);
  \draw [RoyalBlue, thick, use tangent=9, -stealth] (0,0) -- (0.2,0.7);

\end{tikzpicture}}
\caption{The holonomy around a closed loop bounding a disk gives the integrated curvature over that disk: flat curvature is equivalent to trivial holonomies.}
\label{fig:closedloopholo}
\end{subfigure}
\hspace{1 cm}
\begin{subfigure}[h]{0.4\linewidth}

\parbox[c][2cm][c]{7cm}{
\begin{tikzpicture}[
    scale=0.5,decoration={markings,
    mark= at position 0.5 with {\arrow{stealth}},
    mark= at position 2cm with {\arrow{stealth}}}
]
  \def\r{2}
  \def\rotation{-30}

  \foreach \i in {1,...,5} {
    \coordinate (A\i) at ({90 + 72*(\i-1) + \rotation}:\r);
  }

  \foreach \i in {1,...,5} {
    \filldraw[color=teal,fill=teal!20] (A\i) circle (1.5pt);
  }
  \fill[teal!20] (A1) -- (A2) -- (A3) -- (A4) -- (A5) -- (A1);
  \draw [thick,color=teal,postaction={decorate}] (A1) -- (A2);
  \draw [thick,color=teal,postaction={decorate}] (A2) -- (A3);
  \draw [thick,color=teal,postaction={decorate}] (A3) -- (A4);
  \draw [thick,color=teal,postaction={decorate}] (A4) -- (A5);
  \draw [thick,color=teal,postaction={decorate}] (A5) -- (A1);

    \node at (4.1,-1.1) {$G_f \sim \vec{\prod}_{e \in \partial f} g_e$};
    \node[teal] at (2.3,0.9) {$g_e$};
    \draw[ thick, RoyalBlue, <-] (1.2,-0.5) arc (-40:-120:1.8);
    
\end{tikzpicture}}
\caption{A face $f$ of the spinfoam 2-complex $\cC$ with oriented edges labelled with $U(1)$ Lie group elements $g_{e} = e^{i \theta}$ parametrized by an angle $\theta \in [0,2\pi[$}
\label{fig:faceholonomy2d}
\end{subfigure}
\caption{On the left side a continuous description of an holonomy, on the right side its discrete version}
\end{figure}

Let us now discretize the path integral and derive a quantum theory.
The goal is to define and compute the path integral over the fields $A$ and $B$:
\be
\cZ_{BF}^{(2d)}
=
\int \cD A \cD B \; e^{i \int_{\cM_{2d}} B F[A]}
\propto
\int \cD A \; \delta(F[A])
\,,
\ee
where we have formally integrated over the field $B$, which straightforwardly enforces the flatness of the connection. To discretize this expression, we replace the curvature $F[A]$ by holonomies around closed loops. Indeed, if we consider a surface with the topology of a disk, the integral of the curvature on that surface is exactly equal, by Stokes theorem, to the integral of the connection on the circle bounding that surface:
\be
\iint_{\textrm{Disk}} \rd^{2}x\,F
=
\oint_{\pp\textrm{Disk}} \rd s\, \pp_{s}\gamma^{a}(s)\, A_{a}[\gamma(s)]
\,,
\ee
for the curve $\gamma$ parameterizing the boundary circumference $\pp\textrm{Disk}$, as illustrated on fig.\ref{fig:closedloopholo}.
This means that imposing the flatness of the connection, $F[A]=0$, is exactly equivalent to imposing that the holonomies around all contractible loops are trivial, $g[\gamma]=1$.

Having this mind, discretization of the path integral $\cZ_{BF}^{(2d)}$ simply amounts to the following algorithm:
\begin{itemize}
\item first, choose a 2-complex $\cC$ dual to a 2d triangulated manifold; this is the spinfoam 2-complex;
\item second, discretize the connection $A$ as the holonomies $g_{e}$ on the edges of the spinfoam 2-complex $\cC$;
\item third, define the path integral as the product of $\delta$-functions over $\U(1)$ imposing that the holonomies around every face $f$ are flat;
\item fourth, observe that this path integral is topological, in the sense that it does not depend on the details of the 2-complex but only on its overall topology and its boundary data;
\item finally, recognize the quantum version of the $B$-field as the Fourier modes  dual to the $\U(1)$ holonomies.
\end{itemize}

\medskip
Let us thus start by considering a 2d triangulation\footnotemark{} $\Delta_{2d}$ and its dual 2-complex $\cC$, as summarized below.
\footnotetext{
One can consider more generally an arbitrary cellular decomposition of a 2d manifold in terms of polygons with arbitrary number of edges. Here we stick to 2d triangulations simply to keep notations as light as possible and attempt to avoid confusion between geometrical structures on the triangulations and on the spinfoam complex, e.g. between triangulation edges and their topological dual, the spinfoam edges.
}
\bigskip
\def\arraystretch{1.2}
\setlength{\tabcolsep}{2em}
\noindent
\begin{center}
\begin{tabular}{| c |c|c|}
\hline
2d triangulation $\Delta$& spinfoam 2-complex $\cC$ &  data
\\\hline\hline
triangle (2-simplex) & spinfoam vertex $v$  & 
\\\hline
triangulation edge & spinfoam edge $e$ & holonomy $g_{e}\in\U(1)$
\\\hline
point & spinfoam face $f$ & connection flatness $\delta(\prod_{e\in \pp f} g_{e})$
\\\hline
\end{tabular}
\end{center}
\bigskip
The configuration variables for the discretized path integral are the group elements $g_e\in\U(1)$ along the spinfoam edges.
A spinfoam edge $e$ represents a map between the two triangles dual to its source and target spinfoam vertices $s(e)$ and $t(e)$. The holonomy $g_{e}$ then represents the integrated connection, i.e. the $\U(1)$ transport, between those two triangles.

The curvature of each spinfoam face $f$ is then computed as a product of holonomies along the edges $e \in \partial f$, which form a closed loop around this face.
The discretized path integral formulation of the BF theory is then given by:
\be
Z[\cC]
=
\int [\rd g_e]_{e\in\cC} \,
\prod_{f\in\cC}\delta\big(\overrightarrow{\prod_{e \in \partial f}}g_e\big)
\,.
\label{2dSFpathintegral}
\ee
The $\delta$-distribution $\delta(g)$ on $\U(1)$ simply constrains the group element to be the unit element $g=1$: if
\be
g=e^{i\theta}\in\U(1) \textrm{ with }\theta\in[0,2\pi[,
\ee
\be
\delta(g)=2\pi\delta(\theta),
\ee
so that
\be
\int \rd g\,f(g)\delta(g)
=
\int \f{\rd \theta}{2\pi}\,f(e^{i\theta})\delta(\theta)
=
f(1)
\,
\ee
for any continuous function $f$ on $\U(1)$.

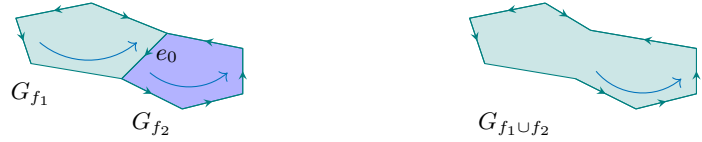
\begin{figure}[h!]
\begin{tikzpicture}[scale=2,decoration={markings,
    mark= at position 0.5 with {\arrow{stealth}},
    mark= at position 2cm with {\arrow{stealth}}}
]

    \coordinate (p3) at (4,-0.6);
    \coordinate (p4) at (3.5, -0.7);
    \coordinate (p5) at (4.5,-0.8);
    \coordinate (p6) at (3.6,-1);
    \coordinate (p7) at (4.2,-1.1);
    
    \coordinate (p1) at (4.6,-1.3);
    \coordinate (p2) at (5,-1.2);
    \coordinate (p8) at (5,-0.9);

    \filldraw[color=teal,fill=teal!20] (p3)--(p4)--(p6)--(p7)--(p5)--(p3);

    \filldraw[color=teal,fill=blue!30] (p5)--(p7)--(p1)--(p2)--(p8)--(p5);
    
    \draw [color=teal,postaction={decorate}] (p3) -- (p4);
    \draw [color=teal,postaction={decorate}] (p4) -- (p6);
    \draw [color=teal,postaction={decorate}] (p5) -- (p7);
    \draw [color=teal,postaction={decorate}] (p3) -- (p5);
    \draw [color=teal,postaction={decorate}] (p7) -- (p1);
    \draw [color=teal,postaction={decorate}] (p1) -- (p2);
    \draw [color=teal,postaction={decorate}] (p2) -- (p8);
    \draw [color=teal,postaction={decorate}] (p8) -- (p5);
    
    \coordinate (q3) at (7,-0.6);
    \coordinate (q4) at (6.5, -0.7);
    \coordinate (q5) at (7.3,-0.78);
    \coordinate (q6) at (6.6,-1);
    \coordinate (q7) at (7.2,-1.1);
    
    \coordinate (q1) at (7.6,-1.3);
    \coordinate (q2) at (8,-1.2);
    \coordinate (q8) at (8,-0.9);

    \filldraw[color=teal,fill=teal!20] (q3)--(q4)--(q6)--(q7)--(q1)--(q2)--(q8)--(q5)--(q3);
    \draw [color=teal,postaction={decorate}] (q3) -- (q4);
    \draw [color=teal,postaction={decorate}] (q4) -- (q6);
    \draw [color=teal,postaction={decorate}] (q3) -- (q5);
    \draw [color=teal,postaction={decorate}] (q7) -- (q1);
    \draw [color=teal,postaction={decorate}] (q1) -- (q2);
    \draw [color=teal,postaction={decorate}] (q2) -- (q8);
    \draw [color=teal,postaction={decorate}] (q8) -- (q5);

    \draw[RoyalBlue, <-] (4.3,-0.86) arc (-50:-130:0.5);
    \draw[RoyalBlue, <-] (4.9,-1.06) arc (-50:-130:0.4);
    \draw[RoyalBlue, <-] (7.9,-1.1) arc (-50:-140:0.4);
    
    \node at (4.5,-0.95) {{\small $e_0$}};
    \node at (3.6,-1.2) {{\small $G_{f_1}$}};
    \node at (4.4,-1.4) {{\small $G_{f_2}$}};
    \node at (6.8,-1.4) {{\small $G_{f_1\cup f_2}$}};
\end{tikzpicture}
\caption{
Merging two faces $f_1$ and $f_2$ sharing a common edge $e_0$ into the larger face $f_{1}\# f_{2}$:
choosing the same root vertex for the two faces, the two holonomies $G_{f_1} = \overrightarrow{\prod_{e \in \partial f_1 }} g_e=G_{f_{1}\setminus e_{0}} g_{e_0}^{-1} $ and $G_{f_2}= \overrightarrow{\prod_{e \in \partial f_2 }} g_e=g_{e_0} G_{f_{1}\setminus e_{0}}$ are merged into $G_{f_{1}\# f_{2}}=G_{f_{1}}G_{f_2}=G_{f_{1}\setminus e_{0}}G_{f_{2}\setminus e_{0}}$.
}
\label{fig:facemerge2d}
\end{figure}

This path integral is topological and does not depend on the details of the 2-complex. Indeed, we can merge two neighbouring faces without affecting the overall path integral. Let us show this by merging explicitly faces $f_{1}$ and $f_{2}$ sharing an edge $e_{0}$ into a single face $f_{12}=f_{1}\# f_{2}$ obtained by erasing $e_{0}$, as drawn on fig. \ref{fig:facemerge2d}.
The holonomy $g_{e_{0}}$ only appears in the holonomies around $f_{1}$ and $f_{2}$. Integration over it leads to a mere convolution of the $\delta$-functions associated to those two faces:
\be
\int \rd g_{e_{0}}\,
\delta\bigg{(}\overrightarrow{\prod_{e \in \partial f_{1}}}g_e\bigg{)}
\delta\bigg{(}\overrightarrow{\prod_{e \in \partial f_{2}}}g_e\bigg{)}
=
\int \rd g_{e_{0}}\,
\delta\Big{(}g_{e_{0}}^{-1}G_{f_{1}\setminus e_{0}}\Big{)}
\delta\Big{(}g_{e_{0}}G_{f_{2}\setminus e_{0}}\Big{)}
=
\delta\big{(}G_{f_{1}\setminus e_{0}}G_{f_{2}\setminus e_{0}}\big{)}
=
\delta\big{(}G_{f_{1}\# f_{2}}\big{)}
\,.
\ee
One can thus merge or refine faces without affecting the overall path integral as long as one does not change the boundary data, i.e. the boundary edges and the boundary holonomies carried by those edges. This is the 2d equivalent of the invariance of the 1d path integral under the choice of decomposition of the overall time interval shown in the previous section.

\medskip

Since we discretized the path integral after integrating out the field $B$, it is expressed in terms of the discrete counterpart of the connection $A$ -- holonomies. However, it is natural to wonder what is the discretized equvalent containing the data of the field $B$. This is crucial for higher-dimensional spinfoams in 3d and 4d, where it is $B$ that carries geometric observables such as lengths, areas and volumes. Moreover, in four space-time dimensions, one has to impose constraints on the quantum $B$-field so that it faithfully represents metric data, which allows one to recover 4d gravity from 4d BF theory.

The mechanism is simple. At the level of classical continuous fields,  $B$ is canonically conjugate to the connection $A$. It is therefore expected that $B$ is the conjugate momentum to $A$ and that it re-appears as the Fourier modes. Moreover, since $B$ is a scalar field, it is naturally associated to space-time points, i.e. to spinfoam faces by duality. Let us see how this works more concretely.

Whenever we Fourier-transform a function on $\U(1)$, we swap the angle $\theta$ parametrizing the group element for an integer $n\in\Z$,
\be
f(e^{i\theta})=\sum_{n\in\Z} f_{n}e^{in\theta}
\,,\qquad
\int \f{\rd \theta}{2\pi}\,\overline{e^{in\theta}}\,e^{i\tilde{n}\theta}=\delta_{n,\tilde{n}}
\,,\qquad
\delta(g)=\sum_{n\in\Z}e^{in\theta}
\,.
\ee
This means that we will have one integer $n_{e}$ for each spinfoam edge $e$, dual to the associated group element $g_{e}$ in the path integral.
Now, we can decompose the $\delta$-distributions entering the path integral. This gives for each face $f$:
\be
\delta\big(\overrightarrow{\prod_{e \in \partial f}}g_e\big)
=
\sum_{n_{f}\in\Z}e^{in_{f}\sum_{e\in \pp f}\eps^{(f)}_{e}\theta_{e}}\,,
\ee
where $g_{e}=e^{i\theta_{e}}$ and the sign $\eps^{(f)}_{e}=\pm$ registers a relative orientation of the edge $e$ with respect to the chosen orientation of the contour of the face $f$.
This effectively identifies the integer numbers $n_{e}$ for all the edges belonging to the same face. This directly leads to re-writing of the path integral in terms of Fourier modes as:
\be
Z[\cC]
=
\int [\rd g_{e}]_{e\in\cC}\sum_{\{n_{f}\}_{f\in\cC}}
\prod_{f}e^{in_{f}\sum_{e\in \pp f}\eps^{(f)}_{e}\theta_{e}}
=
\int [\rd g_{e}]_{e\in\cC}\sum_{\{n_{f}\}_{f\in\cC}}
e^{i\sum_{f}\sum_{e\in \pp f}\eps^{(f)}_{e}n_{f}\theta_{e}}
\,.
\ee
This is simply the path integral of a discretized action, where we recognize the angles $\theta_{e}$ as the {\it discretized connection} along the edges $e$, and the integers $n_{f}$ as the {\it discretized $B$-field} at each face $f$.

We can now integrate over the angles $\theta_{e}$ and look at the resulting path integral it induces for the integers $n_{f}$, that is for the quantized $B$-field. Since an edge $e$ is only shared by two faces, the source and target faces $s(e)$ and $t(e)$, integrating over $\theta_{e}$ simply identifies the integers of those two faces, $\delta_{n_{s(e)},n_{t(e)}}$. As a result, all the integers $n_{f}$ on the whole 2-complex $\cC$ (assuming it is connected) are equal. This is consistent with the classical field equation $\rd B=0$.

\subsubsection{Computing the BF path integral: evolution by gauge transformation}

Let us compute this path integral in particular cases. As the first example, we consider a 2d disk with a 1d circle as a boundary. Let the boundary circle be made of $N$ edges, all of which are oriented in the same direction and carry group elements $g_{1},..,g_{N}$. Then, regardless of the cellular decomposition of the disk, the integrand reduces to an overall $\delta$-function imposing the flatness of the bulk curvature through the triviality of the boundary holonomy (since one can integrate out the bulk group elements described above):
\be
Z[\textrm{Disk}]
=
\delta\big{(}g_{N}..g_{1}\big{)}
=
\sum_{n\in\Z} e^{in(\theta_{1}+..+\theta_{N})}
=
\sum_{n_{1},..n_{N}} e^{i\sum_{k}^{N}n_{k}\theta_{k}}\,\sum_{n\in\Z}\prod_{k}^{N}\delta_{n,n_{k}}
\,.
\ee
This simple building block allows to express the partition function for a 2d BF theory on manifolds with all possible topologies and boundaries.

\begin{figure}[h!]
\begin{tikzpicture}[scale=1.4]

\tikzset{
    midarrow/.style={
        postaction={
            decorate,
            decoration={
                markings,
                mark=at position 0.52 with {\arrow[scale=1.6]{stealth}}
            }
        }
    }
}

\draw[midarrow, blue!80, thick] (-1,-0.7) -- (1,-0.7);  
\draw[midarrow, blue!80, thick] (-1, 0.7) -- (1, 0.7);  
\draw[midarrow,teal, thick] (-1,-0.7) -- (-1,0.7);  
\draw[midarrow, teal, thick] ( 1,-0.7) -- ( 1,0.7); 

\node at (0,0.9) {$g_f$};
\node at (0,-0.9) {$g_i$};
\node at (-1.25,0) {$h_L$};
\node at ( 1.25,0) {$h_R$};

\coordinate (A1) at (-1,0.6);
\coordinate (A2) at (-0.6,0.3);
\coordinate (A3) at (-0.1,0.3);
\coordinate (A5) at (0.3,0.7);
\coordinate (A6) at (0.1,-0.1);
\coordinate (A7) at (-0.1,-0.5);
\coordinate (A8) at (0.6,-0.1);
\coordinate (A9) at (0.9,0.2);
\coordinate (A10) at (1,0.2);
\coordinate (A11) at (0.8,0.6);
\coordinate (A12) at (1,0.7);
\coordinate (A13) at (-0.6,-0.5);
\coordinate (A14) at (-0.8,0);
\coordinate (A15) at (-1,0);
\coordinate (A16) at (-1,-0.7);
\coordinate (A17) at (0.1,-0.7);
\coordinate (A18) at (0.8,-0.4);
\coordinate (A19) at (0.7,-0.7);
\coordinate (A20) at (0.9,-0.4);

\draw[dashed] (A1)--(A2);
\draw[dashed] (A3)--(A2);
\draw[dashed] (A3)--(A5);
\draw[dashed] (A3)--(A6);
\draw[dashed] (A7)--(A6);
\draw[dashed] (A8)--(A6);
\draw[dashed] (A8)--(A9);
\draw[dashed] (A10)--(A9);
\draw[dashed] (A11)--(A9);
\draw[dashed] (A11)--(A5);
\draw[dashed] (A11)--(A12);
\draw[dashed] (A7)--(A13);
\draw[dashed] (A14)--(A13);
\draw[dashed] (A14)--(A2);
\draw[dashed] (A14)--(A15);
\draw[dashed] (A16)--(A13);
\draw[dashed] (A17)--(A7);
\draw[dashed] (A18)--(A8);
\draw[dashed] (A18)--(A8);
\draw[dashed] (A18)--(A19);
\draw[dashed] (A18)--(A20);

\end{tikzpicture}

\caption{Evolution of Lie group element $g_i$ and $g_f$ depending only on the border elements $h_L$ and $h_R$ after internal faces have been merged}
\label{fig:2dSFevolve}
\end{figure}
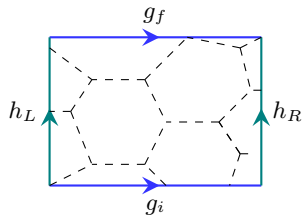

For instance, the evolution of a single edge with an associated group element from an initial to a final segment is given by the case of a discretized circle approximated by a rectangle, as drawn on fig.\ref{fig:2dSFevolve}.
The boundary circumference is made of four edges: an initial holonomy $g_{i}$, a final holonomy $g_{f}$ and two ``time-like'' boundary edges carrying boundary holonomies $h_{L}$ and $h_{R}$, with $L$ and $R$ standing for left and right correspondingly. The partition function is simply given by
\be
Z\big{[}g_{i}\underset{h_{L,R}}\longrightarrow g_{f}\big{]}
=
\delta(h_{L}^{-1}g_{f}^{-1}h_{R}g_{i})\,.
\ee
This describes the basic evolution of a holonomy, from the initial group element $g_{i}$ to the final group element $g_{f}=h_{R}g_{i}h_{L}^{-1}$. As expected, this evolution is  a gauge transformation generated by the data on the ``time-like'' boundary. 

In the $B$-polarization, obtained by a Fourier transform, states of the initial and final slices are simply labeled by integers $n_{i}$ and $n_{f}$. The BF dynamics is simple: it enforces that $n_{i}=n_{f}$ and the boundary condition that $n_{L}=n_{R}=n_{i}$ on the ``time-like'' boundaries. This is similar to the conservation of the momentum of a free particle, in both classical and quantum theories.

The abelian group $\U(1)$ does not lead to particularly enlightening amplitudes. We move on to the non-abelian gauge group $\SU(2)$ to illustrate the physical content of the 2d BF theory and how it is related to topological invariants for 2d surfaces.

\subsection{2d BF theory for the $\SU(2)$ group}

\subsubsection{Classical 2d BF theory}

We now move up to 2d BF theory for the $\SU(2)$ Lie group:
\be
S[A,B]
=
\int \rd^{2} x\,  \delta_{ij} B^{i} F^{j}[A]
\,,\qquad\textrm{with}\quad
F^{i}= \partial_t A^{i}_x - \partial_x A_t^{i}+\eps^{ijk}A_{t}^{j}A_{x}^{k}
\,.\label{2dsu(2)bf}
\ee
The indices $i,j,k\in\{1,2,3\}$ label the basis of the $\su(2)$ Lie algebra. They are raised and lowered with the flat Euclidean metric on $\su(2)$. The field $B=B^i\f{\sigma_{i}}2$ is a $\su(2)$-valued scalar field, while $A=A^{i}\f{\sigma_{i}}2$ is an $\su(2)$-valued connection 1-form. The curvature $F=\rd A+\f12[A,A]$ is an $\su(2)$-valued 2-form. The $\sigma_{i}$'s are the 2$\times$2 Pauli matrices, which we normalize so that $\sigma_{i}\sigma_{j}=\delta_{ij}\id+i\eps^{ijk}\sigma_{k}$.

For a constant field $B=1$, this simple theory gives 2d gravity.  With an additional a potential $V(B)$, it can also be related to Jackiw-Teitelboim gravity and, in general, the 2d dilatonic gravity. \cite{Jackiw:1984je} \cite{Jackiw:1992bw}\cite{Teitelboim:1983ux}

\medskip

The action \eqref{2dsu(2)bf} is invariant under local gauge transformations parametrized by a field $h(t,x)\in\SU(2)$ such that
\be
\left|
\begin{array}{lcl}
A&\mapsto& h^{-1}Ah+h^{-1}\rd h
\,,\\
B&\mapsto& h^{-1}Bh
\,.
\end{array}
\right.
\ee
This is a direct consequence of the gauge-covariance of curvature:
\be
F\,\,\mapsto\,\,
h^{-1}Fh
\,.
\ee
The Euler-Lagrange equations of this action imposes a flat connection $A$ and a trivial transport of the scalar field $B$:
\be
\f{\delta S}{\delta B}=0
\quad \Leftrightarrow\quad 
F^{i}[A]=0
\,,\qquad
\f{\delta S}{\delta A}=0\quad 
\Leftrightarrow \quad
\rd_{A} B^{i}_{a}= \pp_{a} B^{i}+\eps^{ijk}A_{a}^{j}B^{k}=0
\,.
\ee
The action \eqref{2dsu(2)bf} is already in its Hamiltonian form:
\be
S[A,B]=\int \rd t\,\rd x\, \left[ B^{i}  \partial_t A^{i}_x+ A_t^{i}\big{(}
\partial_x B^{i}+ \eps^{ijk} A_{x}^{j}B^{k}
\big{)}\right]
\ee
up to an integration by parts and a corresponding boundary term.
We can read off a canonical Poisson bracket and a Hamiltonian on a constant-$t$ slice:
\be
\{A^{i}_{x}(t,x),B(^{j}t,y))\}=\delta^{ij}\delta(x-y)
\,,\qquad
H=-\int \rd x\,A_t^{i}\cH_{i}
\quad\textrm{with}\quad
\cH_{i}
=
\delta_{ij}\big{(}\rd_{A}B^{j}\big{)}_{x}
=
\partial_x B_{i}+ \eps_{ijk} A_{x}^{j}B^{k}
\,.
\ee
The field $A_{t}$ does not have a canonical momentum, it is a Lagrange multiplier enforcing the Hamiltonian constraints $\cH_{i}$. These constraints generate $\SU(2)$ gauge transformations:
\be
\left|
\begin{array}{lclcl}
\delta_{\lambda}A_{x}^{i}
&=&
\{A_{x}^{i},\int \lambda^{k}\cH_{k} \}
&=&
-\big{(}
\pp_{x} \lambda^{i}+\eps^{ijk}A_{x}^{j}\lambda^{k}
\big{)}
\,,\\
\delta_{\lambda}B^{i}
&=&
\{B^{i},\int \lambda^{k}\cH_{k}\}
&=&
-\eps^{ijk}B^{j}\lambda^{k}
\,.
\end{array}
\right.
\ee
Setting $\lambda=-A_{t}$, we of course recover the equations of motion:
\be
\pp_{t} A_{x}^{i}
=
-\delta_{A_{t}}A_{x}^{i}
=
\pp_{x} A_{t}^{i}+\eps^{ijk}A_{x}^{j}A_{t}^{k}
\,,\qquad
\pp_{t}B^{i}
=\delta_{A_{t}}B^{i}
=
\eps^{ijk}B^{j}A_{t}^{k}
\,.
\ee
Once again, evolution is given by a gauge transformation with the time component of the connection $A_{t}^{i}$ as a gauge parameter.

\subsubsection{Holonomies: non-local observables}
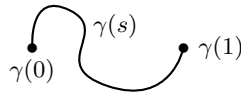
\begin{figure}[h!]
\begin{tikzpicture}
    \draw [thick,xshift=2cm] plot [smooth, tension=2] coordinates { (0,0) (0.5,0.5) (1,-0.5) (2,0)};
\node at (4.5,0) {$\gamma(1)$};
\node at (2,-0.3) {$\gamma(0)$};
\node at (3.1,0.3) {$\gamma(s)$};
\node at (4,0) {$\bullet$};
\node at (2,0) {$\bullet$};
\end{tikzpicture}
\caption{A curve $\gamma:s\in[0,1]\mapsto \gamma(s)\in\cM_{2d}$, on which path-ordered integration will be defined.}
\label{curve}
\end{figure}
In order to discretize the path integral, we follow the same method as for the $\U(1)$ gauge group and define $\SU(2)$ holonomies along curves. Since $\SU(2)$ is a non-abelian Lie group, we now have to care about the order of group multiplication and use a path-ordering for the integration.
More precisely, writing $A_{a}=A^{i}_{a}J_{i}\in\su(2)$ with the Lie-algebra generators $J_{i}$ (expressed via the Pauli matrices in the fundamental representation as $J_i=\sigma_i/2$), we define the path-ordered integration for a curve $\gamma:s\in[0,1]\mapsto \gamma(s)\in\cM_{2d}$ (see figure \ref{curve}) as follows:
\beq
\forall \tau\in[0,1]\,,\quad
g(\tau)
&\equiv&
\cP e^{i\int_{0}^{\tau}\rd s\, \pp_{s}\gamma^{a}(s)\, A_{a}[\gamma(s)]}
\\
&=&
1+i \int_{0}^{\tau}\rd s\, \pp_{s}\gamma^{a}\, A_{a}[\gamma(s)]
+
{i^{2}} \int_{0}^{\tau}\rd s_{2}\int_{0}^{s_{2}}\rd s_{1}\, \pp_{s}\gamma^{a_{1}}\pp_{s}\gamma^{a_{2}}\, 
A_{a_{2}}[\gamma(s_{2})]A_{a_{1}}[\gamma(s_{1})]
+\dots
\,,\nn
\eeq
with ordered partition $0<s_{1}<..<s_{n}<\tau$ at order $n$,
satisfying the differential equation
\be
\pp_{\tau}g(\tau)=i\pp_{\tau}\gamma^{a}(\tau)A_{a}[\gamma(\tau)]g(\tau)
\,,\qquad\textrm{with the initial condition}\quad
g[0]=1
\,.
\ee
When taking $\tau=1$ and thus the final point of the holonomy being the target of the curve, $\gamma(\tau)=t(\gamma)$, we get the holonomy along the path $\gamma$:
\be
\textrm{and for $\tau=1$, we have:}\quad
g[\gamma]\equiv \cP e^{i\int_\gamma\, A}\in\SU(2)
\,.
\ee
It is straightforward to check that for an abelian group such as $\U(1)$ the path-ordered exponentiation simplifies to the standard exponentiation.

Moreover, using the defining differential equation, it is direct to see that $\SU(2)$ gauge transformations $A\mapsto h^{-1}Ah+h^{-1}\rd h$ act in a simple way on holonomies by group multiplication at the endpoints of the considered path:
\be
g[\gamma]
\,\,\mapsto\,\,
h_{t(\gamma)}\,g[\gamma]\,h_{s(\gamma)}^{-1}
\,,\qquad
h_{p}\in\SU(2)\,.
\ee
A holonomy around a closed loop is gauge-covariant since it transforms by a conjugation by a gauge group element at its root (i.e. its start and endpoint). Therefore, the trace of a holonomy $\tr\, g[\gamma]$ around a loop $\gamma$ is gauge-invariant. This is another example of the {\it Wilson loop} observable, the finite equivalent of the curvature $F[A]$.

One can indeed show that requiring the connection be flat, by imposing $F[A]=0$, is equivalent to requiring that holonomies around all the contractible closed loops are trivial, $g[\gamma]=\id$. The requirement of contractibility of the loops is crucial and reflects how the topology of the manifold affects the (moduli space of) flat connections.

\subsubsection{The discrete path integral: 2d spinfoams}

Analogously to the $\U(1)$ case, we work with a 2-complex $\cC$ dual to a 2d triangulation (or any other cellular decomposition of a 2d manifold). We assign group elements $g_{e}\in\SU(2)$ to the (oriented) edges of the spinfoam complex, $e\in\cC$. They represent holonomies of the $\SU(2)$ connection relating the frames of neighbouring triangles connected by these edges. In order to define a path integral, we express a holonomy around each face of the spinfoam complex as a path-ordered product of edge holonomies at its boundary:
\be
Z[\cC]
=
\int [\rd g_e]_{e\in\cC} \,
\prod_{f\in\cC}\delta\big(\overrightarrow{\prod_{e \in \partial f}}g_e\big)
\,.\label{2dSFpartfunc}
\ee
The topological invariance of this path integral is once again ensured by the convolution property of a $\delta$-distribution:
\be
\int_{\SU(2)} \rd g\,\delta (gG)\delta (g^{-1}\tG)=\delta(G\tG)\,.
\ee
This allows merging of neighbouring faces, which shows that the path integral does not depend on the details of the bulk triangulation.
Performing a Fourier transform, one can now switch to the spin basis (see Section \ref{secLQG} by decomposing the $\delta-$distribution as a sum over the characters of the irreducible representations of $\SU(2)$:
\beq
\delta(g_{N}..g_{1})
&=&
\sum_{j\in\f\N2}
d_{j}\chi_{j}(g_{N}..g_{1})
\\
&=&
\sum_{j\in\f\N2}
d_{j}\sum_{\{m_{k}\}}
D^{j}_{m_{1}m_{N}}(g_{N})..D^{j}_{m_{3}m_{2}}(g_{2})D^{j}_{m_{2}m_{1}}(g_{1})
\,.\nn
\eeq
This introduces a spin label per each $\delta$-function, i.e. a spin $j_{f}$ for each face $f$. The path integral then reads:
\be
Z[\cC]
=
\int [\rd g_e]_{e\in\cC} 
\sum_{\{j_{f}\}_{f\in\cC}}
\prod_{f}d_{j_{f}}\chi_{j_{f}}\big(\overrightarrow{\prod_{e \in \partial f}}g_e\big)
\,.
\ee
Since the character can be written in terms of exponentials,
\be
\chi_{j}(g)
=
\f{\sin d_{j}\theta}{\sin \theta}
=
\f{ e^{+id_{j}\theta}-e^{-id_{j}\theta}}{2i\sin \theta}
\,,
\ee
where $\theta$ is the class angle of the group element such that $\chi_{\f12}(g)=2\cos\theta$, one can interpret the spin $j_{f}$ as the quantum version of the discretized $B$-field. This is actually more subtle since $B^{i}$ is a vector with 3 components, while a spin $j_{f}$ is a single number. The correct correspondence is established by interpreting a coherent state\footnotemark{} in $\cV^{j}$ as the quantization of the discretized $B$-field.
\footnotetext{
Using Perelomov coherent states of spin $j$ labeled by a unit 2-vector $\hat{u}$, one has:
\be
\id_{\cV^{j}}=d_{j}\int_{ \cS^{2}}\rd^{2}\hat{u}\,|j,\hat{u}\ra\la j,\hat{u}|
\,,\quad
\chi_{j}(g)=d_{j}\int_{ \cS^{2}}\rd^{2}\hat{u}\,\la j,\hat{u}|g|j,\hat{u}\ra=d_{j}\int_{\cS^{2}}\rd^{2}\hat{u}\,\la \hat{u}|g|\hat{u}\ra^{2j}
\,,\quad
\delta(g)=\int_{\cS^{2}}\sum_{j\in\f\N2}d_{j}^{2}\rd^{2}\hat{u}\,e^{2j\ln\la \hat{u}|g|\hat{u}\ra}\,,
\nn
\ee
where $\la \hat{u}|g|\hat{u}\ra$ is computed in the fundamental 2$\times$2 representation. This allows to recognize the coherent states $|j,\hat{u}\ra$ as the quantum counterpart of the discretized $B$-field \cite{Livine:2007vk}. The measure $\sum_{j\in\f\N2}\int_{\cS^{2}}d_{j}^{2}\rd^{2}\hat{u}$ corresponds to the expected integration measure for 3d vectors with quantized norm given by $d_{j}$. 
This correspondence becomes even clearer in terms of the coherent states labeled by spinors $z\in\C^{2}$ \cite{Dupuis:2010iq,Dupuis:2011fx,Dupuis:2011fz,Dupuis:2011wy}.
\\
Restricting spins to integer values only is equivalent to working with the group $\SO(3)\sim\SU(2)/\Z_{2}$ instead of $\SU(2)$. In this case, the formula above simplifies to \cite{Freidel:2004vi,Freidel:2005me}:
\be
\sum_{j\in\N}d_{j}\chi_{j}(g)=\delta_{\SU(2)}(g)+\delta_{\SU(2)}(-g)=\delta_{\SO(3)}(g)
\propto
\int_{\R^{3}} \rd^{3}\vec{X} \,e^{\tr (\vec{X}\cdot\vsigma)g}\,,
\ee
where the trace in the exponent is that of 2$\times$2 matrices. The vector $\vec{X}$ can be straightforwardly recognized as a discretized $B$-field. This is not entirely satisfactory since we are missing a half of the $\SU(2)$ group and all of its half-integer representations.
However, lifting this formula from $\SO(3)$ to $\SU(2)$ without coherent states is rather cumbersome \cite{Freidel:2005ec,Livine:2008sw}.
\\
The proper identification of the quantized field $B$ is instrumental in the implementation of the simplicity constraints, which promote 4d BF theory to 4d gravity. Coherent states are one of the key techniques for this in the spinfoam framework \cite{Freidel:2007py,Livine:2007ya,Engle:2007wy,Rovelli:2014ssa}.
}

To obtain a formulation entirely in terms of the quantum $B$-field, i.e. only in terms of the spins, one should integrate over the group elements $g_{e}$ in the path integral.
As each edge is shared by only two faces, each group element $g_{e}$ only appears twice and can be integrated over using the orthonormality of the Wigner matrices \eqref{WignerOrthonorm}.

First, this implies that the spins on the two faces, $f$ and $\tilde{f}$ sharing an edge $e$ are necessary equal. This propagates to the entire 2-complex (as long as it is connected), which implies that we are dealing with a unique spin. Second, identification of the magnetic indices $a,b,...$ simply amounts to merging two faces into a fused face $f\#\tilde{f}$ by erasing their shared edge. Third, we shall keep in mind the factor $1/d_{j}$ produced by such an integration.

Finally, consider a spinfoam vertex $v$. Integration over all the group elements $g_{e}$ associated to the edges around $v$ leads to a merge of all faces around this vertex. Then, a careful analysis of the identification of the magnetic indices yields a factor:
\be
\sum_{m}\delta_{mm}=\chi_{j}(\id)=d_{j}.
\ee
This means that we can write down the discretized path integral in the form of the spinfoam ansatz \eqref{eqn:localSFansatz} with an amplitude assigned to each face, edge and vertex:
\be
Z[\cC]=\sum_{\{j_{f}\}}\prod_{f}d_{j_{f}}\prod_{e}\f{\delta_{j_{f_1^{e}}j_{f_2^{e}}}}{d_{j_{f_1^{e}}}}\prod_{v}\cA_{v},
\qquad\textrm{with}\quad
\cA_{v}=\sum_{j_{v}}d_{j_{v}}\prod_{f\ni v}\delta_{j_{v},j_{f}}.
\ee
The amplitude of an edge $e$ identifies spins carried by the two faces $j_{f_1^{e}}$ and $j_{f_2^{e}}$ sharing this edge and produces a factor of $d_{j_f^e}^{-1}$. The amplitude of a vertex $v$ identifies the spins carried by all the faces $f$ sharing this vertex and produces a factor $d_{j_v}$.

Assuming that the 2-manifold has no boundary, it is straightforward to see that this whole sum reduces to a mere counting of faces, edges and vertices of the 2-complex:
\be
Z[\cC]=\sum_{j\in\f\N2}d_{j}^{F-E+V}
\,.
\ee
In the exponent, we recognize the Euler characteristic 
\be
\chi=F-E+V\,.
\ee
This is a topological invariant, which depends on the genus $g$ of the surface, 
\be
\chi=2-2g\,.
\ee 
For instance, $\chi=2$ for a 2-sphere, $\chi=0$ for a 2-torus, and $\chi=-2$ for a double-torus or Bolza surface.
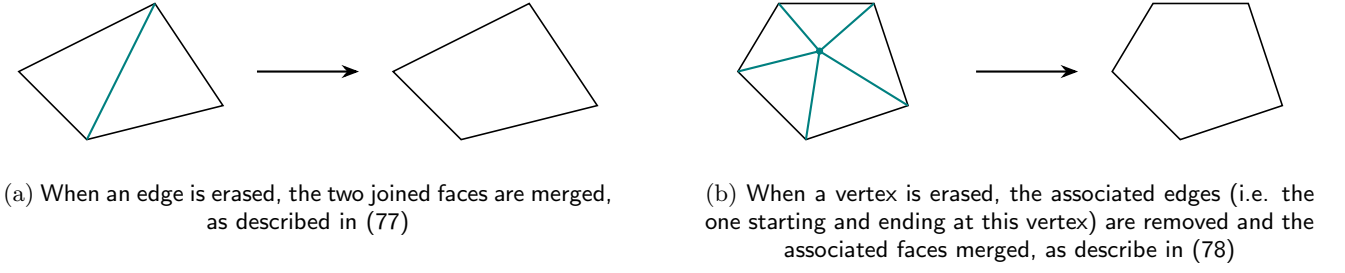
\begin{figure}[t]
\raggedright
\begin{subfigure}[t]{0.45\linewidth}

\parbox[c][2cm][c]{8cm}{
\begin{tikzpicture}[scale=0.9,  >=Stealth]

\coordinate (A) at (0,0);
\coordinate (B) at (2,1);
\coordinate (C) at (3,-0.5);
\coordinate (D) at (1,-1);

\draw[semithick] (A)--(B)--(C)--(D)--cycle;
\draw[thick, teal] (B)--(D);

\draw[->, thick] (3.5,0) -- (5,0);

\begin{scope}[xshift=5.5cm]
\coordinate (A2) at (0,0);
\coordinate (B2) at (2,1);
\coordinate (C2) at (3,-0.5);
\coordinate (D2) at (1,-1);

\draw[semithick] (A2)--(B2)--(C2)--(D2)--cycle;
\end{scope}

\end{tikzpicture}}

\caption{When an edge is erased, the two joined faces are merged, as described in (\ref{erasedEdge})}
\end{subfigure}
\hspace{1 cm}
\begin{subfigure}[t]{0.45\linewidth}

\parbox[c][2cm][c]{8cm}{
\centering
\begin{tikzpicture}[scale=0.9,  >=Stealth]

\coordinate (A) at (0,0);
\coordinate (E) at (0.6,1);
\coordinate (B) at (2,1);
\coordinate (C) at (2.5,-0.5);
\coordinate (D) at (1,-1);
\coordinate (O) at (1.2,0.3);

\draw[semithick] (A)--(E)--(B)--(C)--(D)--cycle;

\draw[thick, teal] (O)--(A);
\draw[thick, teal] (O)--(E);
\draw[thick, teal] (O)--(B);
\draw[thick, teal] (O)--(C);
\draw[thick, teal] (O)--(D);

\fill[teal] (O) circle (1.5pt);

\draw[->, thick] (3.5,0) -- (5,0);

\begin{scope}[xshift=5.5cm]
\coordinate (A2) at (0,0);
\coordinate (E2) at (0.6,1);
\coordinate (B2) at (2,1);
\coordinate (C2) at (2.5,-0.5);
\coordinate (D2) at (1,-1);
\coordinate (O2) at (1.2,0.3);

\draw[semithick] (A2)--(E2)--(B2)--(C2)--(D2)--cycle;
\end{scope}

\end{tikzpicture}}

\caption{When a vertex is erased, the associated edges (i.e. the one starting and ending at this vertex) are removed and the associated faces merged, as describe in (\ref{erasedVertex})}

\end{subfigure}

\caption{Face merge moves}
\label{fig:2derasing}
\end{figure}

The simplest way to see that $\chi$ is a topological invariant is a direct counting in the context of the face merge moves. Indeed, as drawn on fig.\ref{fig:2derasing}, erasing an edge shared by two faces leads to:
\be
F-E+V\mapsto (F-1)-(E-1)+V=F-E+V\,.
\label{erasedEdge}
\ee
Erasing all the edges around a vertex, merging all faces separated by these edges, and thus erasing the vertex itself gives:
\be
F-E+V\mapsto (F-n+1)-(E-n)+(V-1)=F-E+V\,.
\label{erasedVertex}
\ee
It is sometimes more convenient to work with the 2d triangulation dual to a 2-complex. According to the duality rules, triangles are dual to spinfoam vertices, edges are dual to spinfoam edges and points are dual to spinfoam faces. Hence, it has $T=V$ triangles, $E$ edges and $P=F$ points, and thus the same Euler characteristic as the 2-complex, $T-E+P=F-E+V=\chi$. A mathematical theorem states that two triangulations with the same topology are related by a finite sequence of Pachner moves. These moves can be of two types: $1\leftrightarrow 3$, which creates a point within a triangle or erases it, or $2\leftrightarrow 2$, which changes the position of an edge shared by two triangles.  As illustrated on fig.\ref{fig:Pachner2-2}, these moves result from evolving a 2d triangulation by attaching a tetrahedron to it. An easy counting shows that the Euler characteristic is invariant under these two Pachner moves and is therefore a topological invariant:
\be
\left|\begin{array}{llcl}
1\rightarrow 3 \,: \quad &T-E+P &\mapsto & (T+2)-(E+3)+(P+1)=T-E+P\,,
\\
2\leftrightarrow 2 \,: \quad &T-E+P &\mapsto & T-E+P\,.
\end{array}\right.
\ee
\begin{figure}[b]

\begin{subfigure}[t]{0.35\textwidth}

\parbox[c][2cm][c]{5cm}{
\begin{tikzpicture}[scale=1,  >=Stealth]

\coordinate (A) at (0,0);
\coordinate (B) at (2,1);
\coordinate (C) at (3,-0.5);
\coordinate (D) at (1,-1);

\fill[BurntOrange!10] (B)--(C)--(D)--cycle;
\draw[semithick] (B)--(C)--(D)--cycle;

\draw[<->, thick] (3.5,0) -- (5,0);

\begin{scope}[xshift=4cm]
\coordinate (A2) at (0,0);
\coordinate (B2) at (2,1);
\coordinate (C2) at (3,-0.5);
\coordinate (D2) at (1,-1);
\coordinate (O) at (2.1,0);

\fill[teal!10] (B2)--(C2)--(D2)--cycle;
\draw[semithick] (B2)--(C2)--(D2)--cycle;

\draw[thick, teal] (O)--(B2);
\draw[thick, teal] (O)--(C2);
\draw[thick, teal] (O)--(D2);

\end{scope}

\end{tikzpicture}}
\caption{Pachner move $1\rightarrow3$ as a subdivision of a triangle}
\end{subfigure}
\hspace{1 cm}
\begin{subfigure}[t]{0.35\textwidth}
\parbox[c][2cm][c]{5cm}{
    \begin{tikzpicture}[scale=1,  >=Stealth]

\coordinate (A) at (0,0);
\coordinate (B) at (1.7,1.5);
\coordinate (C) at (3,0);
\coordinate (D) at (1,-1);

\fill[BurntOrange!10, blend mode=multiply] (A)--(C)--(D)--cycle;
\fill[teal!10, blend mode=multiply] (A)--(B)--(D)--cycle;
\fill[teal!10, blend mode=multiply] (A)--(C)--(B)--cycle;
\fill[teal!10, blend mode=multiply] (B)--(C)--(D)--cycle;

\draw[semithick] (A)--(D);
\draw[semithick] (C)--(D);
\draw[semithick, dashed] (A)--(C);
\draw[thick, teal] (B)--(D);
\draw[thick, teal] (B)--(A);
\draw[thick, teal] (B)--(C);

\coordinate (mid) at ($(A)!0.5!(B)!0.5!(C)!0.333!(D)$);
\coordinate (arrowStart) at ($(mid)+(0.0,-0.18)$);
\draw[semithick, black, <->, >=Stealth] (arrowStart) -- ($(arrowStart) + (0,1)$);

\end{tikzpicture}}

\caption{Pachner move $1\rightarrow3$ as an evolution of one triangle into three}
\end{subfigure}

\caption{Pachner move $1\leftrightarrow3$ adds or removes a vertex in a triangle}
\label{fig:Pachner1-3}
\end{figure}
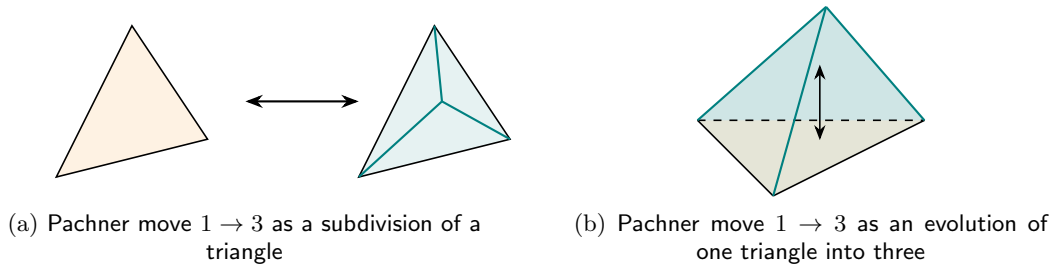
\begin{figure}[h!]

\begin{subfigure}[t]{0.4\linewidth}
\parbox[c][2.6cm][c]{7.6cm}{
\begin{tikzpicture}[scale=0.9,  >=Stealth]

\coordinate (A) at (0,0);
\coordinate (B) at (1.7,1);
\coordinate (C) at (2.8,0);
\coordinate (D) at (1,-1);

\fill[BurntOrange!10] (A)--(B)--(C)--(D)--cycle;
\draw[semithick] (A)--(B)--(C)--(D)--cycle;
\draw[thick, BurntOrange] (A)--(C);

\draw[<->, thick] (3.5,0) -- (5,0);

\begin{scope}[xshift=5.5cm]
\coordinate (A2) at (0,0);
\coordinate (B2) at (1.7,1);
\coordinate (C2) at (2.8,0);
\coordinate (D2) at (1,-1);

\fill[teal!10] (A2)--(B2)--(C2)--(D2)--cycle;
\draw[semithick] (A2)--(B2)--(C2)--(D2)--cycle;
\draw[thick, teal] (B2)--(D2);

\end{scope}

\end{tikzpicture}}

\caption{Pachner move $2\rightarrow2$ as a change of an edge dividing two triangles}
    
\end{subfigure}
\hspace{1cm}
\begin{subfigure}[t]{0.3\linewidth}
\parbox[c][2.6cm][c]{3.1cm}{
\begin{tikzpicture}[scale=1,  >=Stealth]

\coordinate (A) at (0,0);
\coordinate (B) at (1.7,1.5);
\coordinate (C) at (3,0);
\coordinate (D) at (1,-1);

\fill[teal!10, blend mode=multiply] (B)--(C)--(D)--cycle;
\fill[teal!10, blend mode=multiply] (A)--(B)--(C)--cycle;
\fill[BurntOrange!10, blend mode=multiply] (A)--(B)--(D)--cycle;
\fill[BurntOrange!10, blend mode=multiply] (A)--(C)--(D)--cycle;

\draw[thick, BurntOrange] (A)--(D);
\draw[semithick] (C)--(D);
\draw[semithick, dashed] (A)--(C);
\draw[semithick] (B)--(D);
\draw[semithick] (B)--(A);
\draw[thick, teal] (B)--(C);

\coordinate (midTeal)   at ($(A)!0.5!(D)$);
\coordinate (midOrange) at ($(B)!0.5!(C)$);

\draw[semithick, black, ->, >=Stealth, shorten >=10pt, shorten <=10pt] 
    (midTeal) -- (midOrange);

\end{tikzpicture}}

\caption{Pachner move $2\rightarrow2$ as an evolution of a pair of triangles into a different pair}
    
\end{subfigure}

    \caption{Pachner move $2\leftrightarrow2$ changes the orientation of a shared edge between two triangles}
    \label{fig:Pachner2-2}
\end{figure}
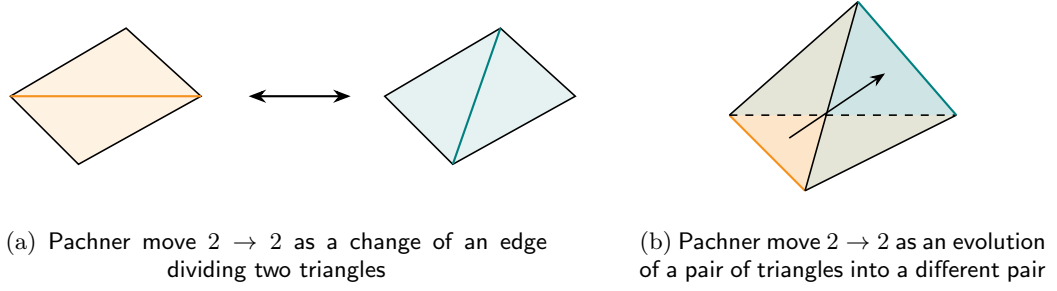

\medskip

Let us now see how to compute the path integral directly in terms of group elements, derive the dynamics of the holonomies and recover the Euler characteristic.

\subsubsection{Computing the BF path integral: evolution and Euler characteristic}

\begin{figure}[b]
\begin{subfigure}[t]{0.23\linewidth}
\raggedright
\parbox[c][2.5cm][c]{2.5cm}{
\begin{tikzpicture}[scale=1.4]

\tikzset{
    midarrow/.style={
        postaction={
            decorate,
            decoration={
                markings,
                mark=at position 0.52 with {\arrow[scale=1.6]{stealth}}
            }
        }
    }
}

\draw[midarrow, Bittersweet, thick] (-1,-0.7) -- (1,-0.7);   
\draw[midarrow, Bittersweet, thick] (-1, 0.7) -- (1, 0.7);   
\draw[midarrow,RoyalBlue, thick] (-1,-0.7) -- (-1,0.7);   
\draw[midarrow, RoyalBlue, thick] ( 1,-0.7) -- ( 1,0.7);   

\node at (0,0.9) {$g_f$};
\node at (0,-0.9) {$g_i$};
\node at (-1.25,0) {$h_L$};
\node at ( 1.25,0) {$h_R$};

\end{tikzpicture}}

\caption{A rectangle representing the evolution of an edge}
\label{fig:rectangle}
\end{subfigure}
\hspace{0.01\linewidth}
\begin{subfigure}[t]{0.23\linewidth}

\parbox[c][2.5cm][c]{2.5cm}{
\begin{tikzpicture}[scale=1, >=stealth]

\tikzset{
    midarrow/.style={
        postaction={
            decorate,
            decoration={
                markings,
                mark=at position 0.6 with {\arrow[scale=1.6]{stealth}}
            }
        }
    }
}

\def\r{1}     
\def\h{2}     
\def\e{0.35}  

\draw[semithick] (-\r,0) -- (-\r,-\h);
\draw[semithick] (\r,0) -- (\r,-\h);
\draw[midarrow,RoyalBlue, thick] (-\r+0.5,-\h-0.3) -- (-\r+0.5,-0.3);

\draw[midarrow,Bittersweet, thick] (-\r,0) arc[start angle=180, end angle=360, x radius=\r, y radius=\e];
\draw[Bittersweet, thick] (-\r,0) arc[start angle=180, end angle=0, x radius=\r, y radius=\e];

\draw[midarrow, Bittersweet, thick] (-\r,-\h) arc[start angle=180, end angle=360, x radius=\r, y radius=\e];
\draw[dashed, Bittersweet, thick] (-\r,-\h) arc[start angle=180, end angle=0, x radius=\r, y radius=\e, dashed];

\node at (0.2,-0.6) {$g_f$};
\node at (0.2,-\h-0.6) {$g_i$};
\node at (-0.25,-\h/2-0.2) {$h$};

\end{tikzpicture}}

\caption{A 2-cylinder is a rectangle with ``time-like" edges identified}
\label{fig:2dcyl}
\end{subfigure}
\hspace{0.01\linewidth}
\begin{subfigure}[t]{0.23\linewidth}

\parbox[c][2.5cm][c]{2.5cm}{
\begin{tikzpicture}[scale=0.8, >=stealth]

\tikzset{
    midarrow/.style={
        postaction={
            decorate,
            decoration={
                markings,
                mark=at position 0.55 with {\arrow[scale=1.6]{stealth}}
            }
        }
    }
}

\def\r{1.5}     
\def\e{0.45}    

\draw[semithick] (0,0) circle (\r);

\draw[semithick, dashed]
    (-\r,0) arc[start angle=180, end angle=0, x radius=\r, y radius=\e];
\draw[semithick]
    (-\r,0) arc[start angle=180, end angle=360, x radius=\r, y radius=\e];


\draw[RoyalBlue, thick, midarrow]
    (0,-\r) arc[start angle=270, end angle=90, x radius=\e, y radius=\r];

\fill[Bittersweet] (0,\r) circle (2pt);
\fill[Bittersweet] (0,-\r) circle (2pt);

\node at (0,\r+0.4) {$g_f$};
\node at (0,-\r-0.4) {$g_i$};
\node at (-0.1,0) {$h$};

\end{tikzpicture}}

\caption{A 2-sphere is a 2-cylinder with initial and final boundaries shrunk to a point (and thus trivial corresponding holonomies)}
\label{fig:2dsphere}
\end{subfigure}
\hspace{0.01\linewidth}
\begin{subfigure}[t]{0.23\linewidth}

\parbox[c][2.5cm][c]{2.5cm}{
\begin{tikzpicture}[scale=0.87]

\tikzset{
    midarrow/.style={
        postaction={
            decorate,
            decoration={
                markings,
                mark=at position 0.8 with {\arrow[scale=1.6]{stealth}}
            }
        }
    }
}

\def\R{1.5}     
\def\r{0.5}     
\def\rh{1.2} 
\def\e{0.25}   

\draw[semithick] (0,0) circle (\R);

\draw[semithick] (0,0) circle (\r);

\draw[midarrow, RoyalBlue, thick] (-\rh,0) arc[start angle=180,end angle=90,x radius=\rh,y radius=\rh];
\draw[thick, RoyalBlue] (0,0) circle (\rh);

\draw[midarrow, Bittersweet, thick] (-\R,0) arc[start angle=-180,end angle=0,x radius=(\R-\r)/2,y radius=\e];
\draw[Bittersweet, thick, dashed] (-\R,0) arc[start angle=180,end angle=0,x radius=(\R-\r)/2,y radius=\e];

\node at (-0.6,-\r) {$g_i$};
\node at (-0.4,0.8) {$h$};

\end{tikzpicture}}

\caption{A 2-torus is a 2-cylinder with ``space-like" boundaries identified}
\label{fig:2dtorus}
\end{subfigure}

\caption{Illustration of transition amplitudes as 2d geometries. Red edges represent initial and final states (``space-like" boundaries). Blue edges correspond to gauge transformations at boundary nodes (``time-like" boundaries).}
\label{fig:2dgeometries}
\end{figure}
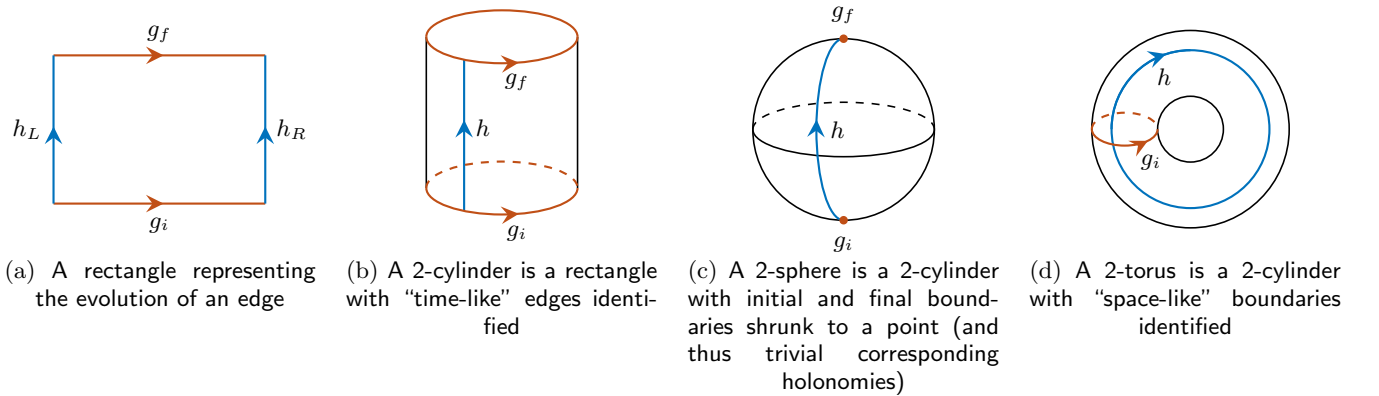

Like for the $\U(1)$ group case, we start from a disk with a boundary circumference made of $N$ edges carrying group elements $g_{1},..,g_{N}$. Regardless of the details of a cellular decomposition of the disk, one can integrate over all bulk holonomies and obtain a path integral, which amounts to requiring that the holonomy around the disk is flat.  Assuming that all edges are oriented in the same way, this gives:
\be
Z[\textrm{Disk}]
=
\delta\big{(}g_{N}..g_{1}\big{)}
\,.
\ee
Then, choosing a rectangular boundary configuration, as drawn on fig.\ref{fig:rectangle} with an initial group element $g_{i}$ evolving into a final group element $g_{f}$ and group elements $h_{L}$ and $h_{R}$ along the time-like boundaries, one gets:
\be
Z\big{[}g_{i}\underset{h_{L,R}}\longrightarrow g_{f}\big{]}
=
\delta(h_{L}^{-1}g_{f}^{-1}h_{R}g_{i})
\,,
\ee
as earlier for the $\U(1)$ case, showing that the evolution from one slice to another is given simply by a gauge transformation at the two endpoints of the segment, just like in the continuum theory.

One can then identify the ``time-like'' boundaries to transform a rectangle into a cylinder with no ``time-like'' boundaries and only the initial and final circles as ``space-like'' boundaries. As one can see in \ref{fig:2dcyl}, integrating over the bulk holonomy $h=h_{L}=h_{R}$ gives:
\be
Z[\textrm{Cyl}]
=
\int \rd h\,\delta(h^{-1}g_{f}^{-1}hg_{i})
=
\sum_{j\in\f\N2}\chi_{j}(g_{f})\chi_{j}(g_{i})
\,,
\ee
implying that the evolution along the cylinder conserves the spin.

 We can get a 2-sphere, with no boundary by closing the initial and final holes with disks, an initial cap and a final cap, as illustrated on fig.\ref{fig:2dsphere}, yielding:
\be
Z[\cS_{2}]
=
\int \rd g_{i}\,\rd g_{f}\,
\int \rd h\,\delta(h^{-1}g_{f}^{-1}hg_{i})\delta(g_{i})\delta(g_{f})
=
\delta(\id)
=
\sum_{j\in\f\N2}
d_{j}^{2}
\,.
\ee
This amplitude is obviously divergent, but the degree of divergence is the physically relevant information here. Indeed, it diverges as $d_{j}^{2}$, with the power 2 being the Euler characteristic of a 2-sphere. A standard way to gauge-fix the divergence is simply to gauge-fix the spin, say to a fixed value $j_{0}$, which corresponds to fixing the value of the $B$-field (which in the continuum is pure gauge!).

A 2-torus is obtained by identifying the initial and final circumferences of a cylinder, as illustrated on fig.\ref{fig:2dtorus}, yielding:
\beq
Z[\cT_{2}]
&=&
\int \rd g_{i}\,\rd g_{f}\,
\int \rd h\,\delta(h^{-1}g_{f}^{-1}hg_{i})\delta(g_{f}^{-1}g_{i})
=
\int \rd g\,\rd h\,\delta(h^{-1}g^{-1}hg)
\\
&=&
\sum_{j\in\f\N2}d_{j}
\sum_{a,b,c,d}\int \rd g\,D^{j}_{bc}(g^{-1})D^{j}_{da}(g) \, \int\rd h\,D^{j}_{ab}(h^{-1})D^{j}_{cd}(h)
\nn\\
&=&
\sum_{j} d_{j}\f{1}{d_{j}^{2}}
\sum_{a,b,c,d}\delta_{ab}\delta_{cd}\delta_{ad}\delta_{bc}
=
\sum_{j\in\f\N2}
d_{j}^{0}
\,,\nn
\eeq
with the exponent 0 matching the expected Euler characteristic of a 2-torus.

\begin{figure}[h!]

\begin{tikzpicture}[scale=0.7, >=Stealth]

\def\n{10}      
\def\radius{2} 

\foreach \i in {0,...,9} {
    \coordinate (P\i) at (\i*360/\n:\radius);
}

\draw[very thick, Bittersweet] (P0) -- (P1);
\draw[very thick, Goldenrod] (P1) -- (P2);
\draw[very thick, RoyalBlue] (P2) -- (P3);
\draw[thick,dotted] (P3) -- (P4);
\draw[very thick,BurntOrange] (P4) -- (P5); 
\draw[very thick,Bittersweet] (P5) -- (P6);
\draw[very thick,Goldenrod] (P6) -- (P7);
\draw[very thick,RoyalBlue] (P7) -- (P8);
\draw[thick,dotted] (P8) -- (P9);  
\draw[very thick, BurntOrange] (P9) -- (P0);

\draw[thick, RoyalBlue, <->, shorten >=2pt, shorten <=2pt] 
    ($(P2)!0.5!(P3)$) -- ($(P7)!0.5!(P8)$) 
    node[midway, left, black] {\small{identify}};

\end{tikzpicture}

\caption{Arbitrary-genus surface can be created from a polygonal disk via identification of its opposite sides}
\label{fig:2dgenus}
\end{figure}
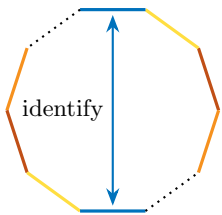

Finally, an (orientable) 2-surface of genus $g$ can be obtained from a disk with a  polygon with $4g$ edges as a boundary by identifying opposite edges, as illustrated on fig.\ref{fig:2dgenus}. The identification creates $N=2g$ non-contractible cycles of the surface. This leads to the path integral:
\be
Z[\cS^{2d}_{g}]
=
\int \prod_{p=1}^{N}\rd g_{p}\,
\delta(g_{N}^{-1}g_{N-1}^{-1}..g_{1}^{-1}g_{N}..g_{1})
\,.
\ee
Upon expansion over the Wigner matrices, integration over the group elements, and a proper identification of magnetic indices, one gets:
\be
Z[\cS^{2d}_{g}]
=
\sum_{j}d_{j}^{2-N}
\,.
\ee
This is indeed the Euler characteristic of the chosen 2d discrete manifold with a single face (the disk), $N=2g$ edges (due to the pair-wise identification) and a single vertex (due to the exhaustive identification of all opposite boundary edges).

\begin{figure}[h!]
    \centering
    \includegraphics[scale=0.37]{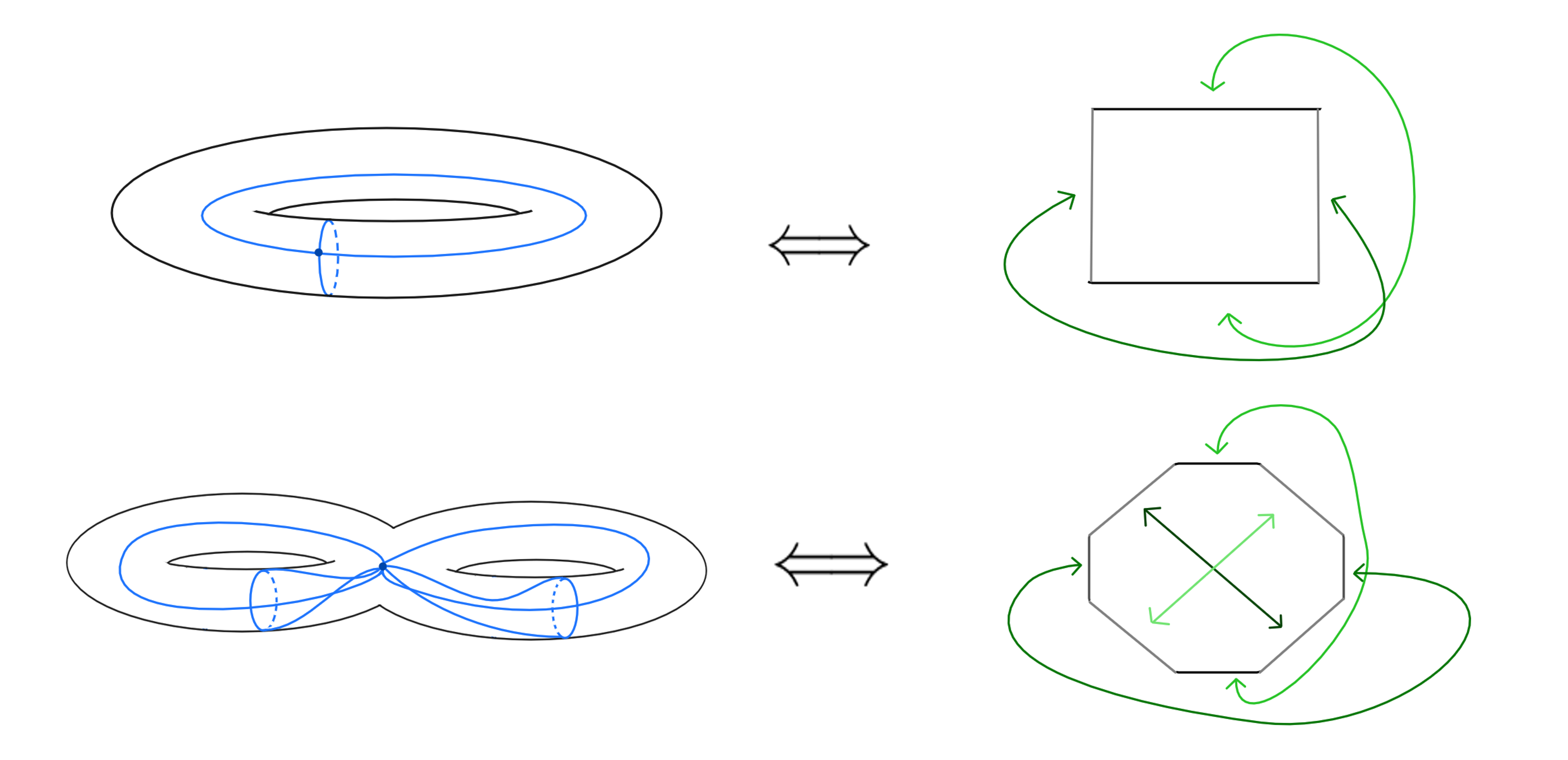}
    \caption{Example of the torus and the double torus. On the left are 1 and 2-genus surfaces, with cycles depicted in blue. On the right is their polygonal disk counterpart, with edges identification depicted in green. }
    \label{fig:my_label}
\end{figure}

\subsection{2d Group field theory and Matrix models}
\label{sec:2dGFT}

Spinfoams are structures encoding the probability amplitude of an initial spin-network state evolving into a final spin-network state as a combination of elementary interaction events between quanta of geometry. The full transition amplitude is then obtained by summing over all the dressed spinfoam complexes existing between the initial and final states. In that sense, spinfoams are to (loop) quantum gravity exactly what Feynman diagrams are to quantum field theory: a mathematical expansion of the full path integral amplitudes, which also have a direct physical interpretation in terms of histories of scattering events. This is, actually, the intuition behind the creation of spinfoams \cite{Reisenberger:1996pu}.

More precisely, for each (local) spinfoam model given by the ansatz \eqref{eqn:localSFansatz}, there exists an auxiliary field theory on a group manifold, dubbed simply {\it group field theory}, such that its Feynamn diagrams are exactly spinfoam 2-complexes (dual to space-time triangulations of the appropriate dimension) and that the evaluation of those Feynman diagrams are exactly the corresponding spinfoam amplitudes.
This logic, providing a systematic non-perturbative definition of spinfoam models, was formalized for 4d quantum gravity in \cite{Reisenberger:2000fy,Reisenberger:2000zc}. For 2d quantum gravity, it simply reduces to matrix models.

\medskip

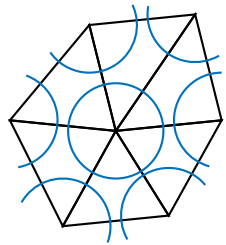
\begin{figure}[h!]
\begin{tikzpicture}[scale=0.7]

\coordinate (A) at (0,0);
\coordinate (B) at (2,0.2);
\coordinate (C) at (1,1.8);
\coordinate (D) at (3,2);
\coordinate (E) at (2.5,4);
\coordinate (F) at (0.5,3.8);
\coordinate (G) at (-1,2);

\draw[thick] (A) -- (B) -- (C) -- cycle;
\draw[thick] (B) -- (C) -- (D) -- cycle;
\draw[thick] (C) -- (D) -- (E) -- cycle;
\draw[thick] (C) -- (E) -- (F) -- cycle;
\draw[thick] (C) -- (F) -- (G) -- cycle;
\draw[thick] (A) -- (C) -- (G) -- cycle;

\draw[RoyalBlue, thick] (C) circle (0.9);

\draw[RoyalBlue, thick] 
    ($(A)+(-20:0.9)$) arc[start angle=-20, end angle=150, radius=0.9];
\draw[RoyalBlue, thick] 
    ($(B)+(40:0.9)$) arc[start angle=40, end angle=210, radius=0.9];
\draw[RoyalBlue, thick] 
    ($(D)+(90:0.9)$) arc[start angle=90, end angle=255, radius=0.9];
\draw[RoyalBlue, thick] 
    ($(E)+(170:0.9)$) arc[start angle=170, end angle=300, radius=0.9];
\draw[RoyalBlue, thick] 
    ($(F)+(215:0.9)$) arc[start angle=215, end angle=385, radius=0.9];
\draw[RoyalBlue, thick] 
    ($(G)+(-80:0.9)$) arc[start angle=-80, end angle=70, radius=0.9];

\end{tikzpicture}
\caption{Feynman diagram as a spinfoam complex (in blue) dual to a 2d triangulation (in black): the two-line propagator combine to form the spinfoam faces, which loop around the points of the 2d triangulation.}
\label{fig:2dGFT}
\end{figure}

Let us take a field $\varphi(h_{1},h_{2})$ living on $\SU(2)^{\times 2}$, which satisfies a reality condition:
\be
\overline{\varphi(h_{1},h_{2})}
\,=\,
\varphi(h_{2},h_{1})
\,,
\ee
and a gauge invariance,
\be
\varphi(h_{1}g,h_{2}g)
\,=\,
\varphi(h_{1},h_{2})
\,,\quad\forall g\in\SU(2)
\,.
\ee
We consider the straightforward field theory action with a trivial propagator and a cubic interaction term:
\be
S^{GFT}_{2d}[\vphi]
\,=\,
\f12\int[\rd h]^{2}\,
\varphi(h_{1},h_{2})\overline{\varphi(h_{1},h_{2})}
+
\f{\lambda}{3}\int[\rd h]^{3}\,
\varphi(h_{1},h_{2})\varphi(h_{2},h_{3})\varphi(h_{3},h_{1})
\,.
\ee
The path integral $\int \cD\vphi\,\exp[iS^{GFT}_{2d}]$ can be expanded in Feynman diagrams using the standard logic and method of quantum field theory.
The resulting Feynman diagrams are obviously made of cubic interaction terms, which are the spinfoam vertices dual to triangles. Those are glued together by trivial propagators, which are the spinfoam edges dual to triangulation edges shared by triangles, as illustrated on fig.\ref{fig:2dGFT}.
\begin{figure}[h!]
\begin{subfigure}[t]{0.4\linewidth}
\vspace{-2cm}
\begin{tikzpicture}[line cap=round, thick]

  \def\ycenter{0.75}
  \def\arrowsep{0.15} 
  \def\arrowheadlength{0.5}
  \def\arrowheadangle{40} 

  \draw[RoyalBlue] (-2,\ycenter+\arrowsep) -- (-1,\ycenter+\arrowsep);
  \draw[RoyalBlue] (-2,\ycenter-\arrowsep) -- (-1,\ycenter-\arrowsep);

  \draw[RoyalBlue] 
    (-1.3, \ycenter) -- ++({-\arrowheadlength*cos(\arrowheadangle)}, {\arrowheadlength*sin(\arrowheadangle)});
  \draw[RoyalBlue] 
    (-1.3, \ycenter) -- ++({-\arrowheadlength*cos(\arrowheadangle)}, {-\arrowheadlength*sin(\arrowheadangle)});

  \draw[RoyalBlue] (-1,0.45) rectangle (0,1.05);
  \node[Bittersweet] at (-0.45,\ycenter) {\textbf{$g_e$}};

  \draw[RoyalBlue] (0,\ycenter+\arrowsep) -- (1,\ycenter+\arrowsep);
  \draw[RoyalBlue] (0,\ycenter-\arrowsep) -- (1,\ycenter-\arrowsep);

  \draw[RoyalBlue] 
    (0.7, \ycenter) -- ++({-\arrowheadlength*cos(\arrowheadangle)}, {\arrowheadlength*sin(\arrowheadangle)});
  \draw[RoyalBlue] 
    (0.7, \ycenter) -- ++({-\arrowheadlength*cos(\arrowheadangle)}, {-\arrowheadlength*sin(\arrowheadangle)});

  \node at (-2.3,1.1) {$h_1^s$};
  \node at (-2.3,0.4) {$h_2^s$};

  \node at (1.3,1.1) {$h_2^t$};
  \node at (1.3,0.4) {$h_1^t$};

\end{tikzpicture}
\vspace{0.75cm}
\caption{Feynman propagator for the GFT edge $e$:
$\int \rd g_e\,\delta(h^{t(e)}_{e,1}{}^{-1}g_{e}h^{s(e)}_{e,2})\delta(h^{t(e)}_{e,2}{}^{-1}g_{e}h^{s(e)}_{e,1})$
}
\label{fig:2dGFTedge}
\end{subfigure}
\hspace*{10mm}
\begin{subfigure}[t]{0.4\linewidth}
\begin{tikzpicture}[scale=0.5]

\coordinate (A) at (90:3);    
\coordinate (B) at (210:3);   
\coordinate (C) at (330:3);   

\coordinate (ABstart) at ($(A)!0.93!(B)$);
\coordinate (ABend)   at ($(B)!0.93!(A)$);

\coordinate (BCstart) at ($(B)!0.93!(C)$);
\coordinate (BCend)   at ($(C)!0.93!(B)$);

\coordinate (CAstart) at ($(C)!0.93!(A)$);
\coordinate (CAend)   at ($(A)!0.93!(C)$);

\coordinate (BCmid) at ($(B)!0.5!(C)$);
\coordinate (BCctrl) at ($(BCmid)+(0,1.0)$); 

\draw[thick, RoyalBlue] (ABstart) .. controls ($(A)!0.5!(B)+(0.87,-0.5)$) .. (ABend);
\draw[thick, RoyalBlue] (BCstart) .. controls (BCctrl) .. (BCend); 
\draw[thick, RoyalBlue] (CAstart) .. controls ($(C)!0.5!(A)+(-0.87,-0.5)$) .. (CAend);

\node at ($(A)+(-1,-0.1)$) {$h_{c,2}$};
\node at ($(A)+(1,-0.1)$) {$h_{c,1}$};

\node at ($(B)+(-0.4,0.9)$) {$h_{b,2}$};
\node at ($(B)+(0.58,-0.8)$) {$h_{b,1}$};

\node at ($(C)+(-0.58,-0.82)$) {$h_{a,1}$};
\node at ($(C)+(0.4,0.9)$) {$h_{a,2}$};

\node[Bittersweet] at (0:0) {\textbf{$v$}};

\end{tikzpicture}
\caption{Feynman interaction for the GFT vertex $v$:
$\delta\left(h^{v}_{e_{a},1}{}^{-1}h^{v}_{e_{b},2}\right)\delta\left(h^{v}_{e_{b},1}{}^{-1}h^{v}_{e_{c},2}\right)\delta\left(h^{v}_{e_{c},1}{}^{-1}h^{v}_{e_{a},2}\right)$}
\label{fig:2dGFTvertex}
\end{subfigure}
\caption{Feynman rules for 2d group field theory.}
\label{fig:2dGFTrules}
\end{figure}
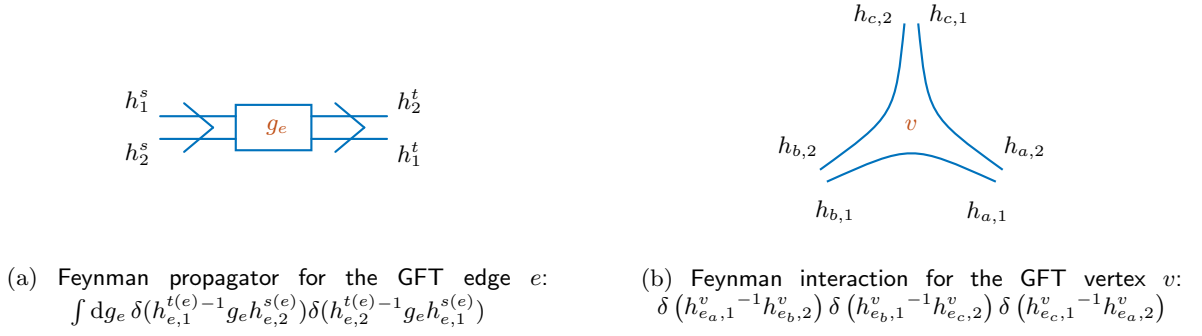

The amplitude of one Feynman diagram, i.e. one spinfoam complex $\cC$, is given by identifying the group elements along the lines of the Feynman diagram, which form loops around the spinfoam faces:
\beq
Z[\cC]
&=&
\lambda^{V}\int \prod_{e}\rd g_{e}\prod_{v}\prod_{e\ni v}\rd h^{v}_{e,1}\rd h^{v}_{e,2}
\,
\prod_{e}\delta\left(h^{t(e)}_{e,1}{}^{-1}g_{e}h^{s(e)}_{e,2}\right)\delta\left(h^{t(e)}_{e,2}{}^{-1}g_{e}h^{s(e)}_{e,1}\right)
\prod_{v}\cA_{v}[\{h^{v}_{e,1},h^{v}_{e,2}\}_{e\ni v}]
\nn\\
&=&
\lambda^{V}\int \prod_{e}\rd g_{e}
\prod_{f}\delta\left(\overrightarrow{\prod_{e\in f}}g_{e}\right)\,,
\eeq
Here, $V$ is the number of interaction vertices in the Feynman diagram, i.e. the number of spinfoam vertices in the 2-complex $\cC$.
The vertex amplitude $\cA_{v}$ is given by the Feynman rule for the interaction vertex, explicitated in fig.\ref{fig:2dGFTrules}.
The group variables $g_{e}$ on the edges $e$ come from the gauge invariance of the field $\vphi$.
This reproduces exactly the expected spinfoam amplitude for BF theory.

\medskip

The link with matrix models comes from the direct decomposition for the group field $\vphi$ in terms of $\SU(2)$ representations. First, we noticed that the required gauge invariance means that the field is actually a function of a single group element $h_{2}h_{1}^{-1}$:
\be
\vphi(h_{1},h_{2})=\phi(h_{2}h_{1}^{-1})
\,,\qquad
\overline{\phi(H)}=\phi(H^{-1})
\,.
\ee
We can then rewrite the group field theory action in terms of this new field $\phi$ and get:
\be
S^{GFT}_{2d}[\vphi]
\,=\,
\f12\int\rd H\,
\phi(H)\phi(H^{-1})
+
\f{\lambda}{3}\int[\rd H]^{3}\,
\delta(H_{3}H_{2}H_{1})\,
\phi(H_{1})\phi(H_{2})\phi(H_{3})
\,.
\ee
One notices that the group elements $H$'s play a role similar to momentum variables with a momentum conservation around every interaction vertex and along every propagator. This insight is behind the intuition of non-commutative geometry driving the propagation of matter coupled to spinfoam models \cite{Freidel:2005bb,Freidel:2005me}.

Second, we can decompose the field $\phi$ as a sum over spins:
\be
\phi(H)=\sum_{j,m,n}d_{j}\Phi^{j}_{mn}D^{j}_{mn}(H)
\,,\qquad
\Phi^{j}=(\Phi^{j})^{\dagger}
\,.
\ee
The Fourier transform of the action then reads:
\be
S^{GFT}_{2d}[\vphi]
\,=\,
\sum_{j}d_{j}\left[
\f12\tr(\Phi^{j})^{2}
+
\f{\lambda}{3}\tr(\Phi^{j})^{3}
\right]
\,,
\ee
where we recover an infinite tower of matrix models \cite{Livine:2001sc,Livine:2003ux}. Directly applying the Feynman rules for matrix models, we see that the Feynman amplitude of a given (closed) spinfoam complex is
\be
Z[\cC]=\sum_{j}\lambda^{V}d_{j}^{V-E+F}\,,
\ee
where each vertex obviously contributes a factor $\lambda d_{j}$, while each edge contributes a propagator factor $d_{j}^{-1}$, and finally each face gives a factor $d_{j}$ due to the size of the matrices. This allows to recover the expected Euler characteristic $\chi_{\cC}=V-E+F$.

We are now ready to move up to  three-dimensional spinfoams.

\section{3d Spinfoams}

Let us move one dimension up and look into 3d spinfoams. We consider 3d gravity as a topological  BF theory. Focussing on an Euclidean space-time signature, we show how to discretize the path integral while preserving the topological invariance of the theory, yielding the Ponzano-Regge model for 3d quantum gravity. It assigns an amplitude to every 3d triangulations (and more generally to every 3d cellular complexes), such that it does not depend on the details of the bulk. In this framework, geometry is represented in terms of algebraic data from the representation theory of the Lie group $\SU(2)$: edges are dressed with irreducible representations - spins, triangles are dressed with intertwiners - singlet states, and tetrahedra are dressed with the $\{6j\}$ symbol, which is interpreted in the semi-classical regime at large spins as the quantum version of the exponential of the Regge action. The topological invariance of spinfoam amplitudes is then ensured by the Biedenharn-Eliott identity satisfied by the 6j-symbol.  This leads the way to a body of results relating the topological invariance, the symmetry under diffeomoprhisms and thus deformations of the triangulation, the recursion relation satisfied by the 6j-symbol, the Hamiltonian constraint operators.

\subsection{3d gravity as a BF theory}

Three-dimensional gravity has been shown to be exactly solvable \cite{Witten:1988hc}, since it is a topological field theory and has no local degrees of freedom (i.e. no gravitational waves) but only global ones, which depend on the topology and boundary of the manifold under consideration. As we have seen before, BF theory is also a topological theory, making it a natural candidate to model three-dimensional quantum gravity. For zero cosmological constant, the $\SU(2)$ BF theory approach  (for Euclidean signature) leads to the famous Ponzano Regge model upon discretization \cite{PR1968}\footnote{For non-zero cosmological constant a spinfoam model is given by the Turaev-Viro model \cite{Turaev:1992hq}.} (for Lorentzian signature, the BF theory under consideration is based on $\SU(1,1)$). 3 dimensional gravity can also be map to a Chern-Simon theory, based on a Lie group $G$ which depend on the signature and cosmological constant, see \cite{Witten:1988hc}\cite{Alexandrov:2011ab}. Analogously to the 2d case, we will start with a BF theory, given by $\su(2)$-valued $1$-forms $A$ and $B$, such that the BF Lagrangian reads
\begin{equation}
    \cL_{BF}[A,B] = \tr(B \wedge F[A]),
\end{equation}
and arrive at the spin-foam partition function still given by \eqref{2dSFpartfunc}. The main difference is the definition of the dual elements, since in three dimensions edges of the spinfoam are dual to triangles of the discretization, and faces of the spinfoam are dual to edges. Below, we detail how to construct the spinfoam amplitude starting from the three-dimensional path integral formulation of gravity. \\

Let us consider the first-order formulation of 3d gravity, \textit{à la Cartan}. The action for Euclidean 3d gravity is written in terms of triads and a spin connection. We denote by $\tau_i=-\frac{i}{2}\sigma_i$ ($\sigma_i$ being Pauli matrices) the basis elements of $\su(2)$ Lie algebra, with $i = 1,2,3$ and commutation relations given by $[\tau_i,\tau_j] = 2 \epsilon_{ij}{}^{k} \tau_k$. A triad is a triple of $\su(2)$-valued $1$-forms $e_\mu=e^{i}_\mu(x)\tau_i$ such that the metric $g_{\mu\nu}$ on the manifold $\cM$ is given by
\be
    g_{\mu \nu}(x) = e^{i}_\mu(x) e^{j}_\nu(x) \delta_{ij}.
\ee
This provides an isomorphism between the tangent bundle of the manifold, $T(\cM)$, and the Euclidean principal bundle $(M,SO(3))$, meaning that locally one can always consider a flat metric $\delta_{ij}$. Here, the choice of $\delta_{ij}$ as an internal metric corresponds to Euclidean geometry with $\SO(3)\cong\SU(2)/\Z_2$ as an internal rotation group. In order to describe a Lorentzian manifold, we would need to consider an $\SO(2,1)$ internal gauge group (double-covered by $\SU(1,1)$) with the Minkowski internal metric, $\eta_{ij}$.

One can define on this bundle a spin connection $A = A^i \tau_i$. The action of Euclidean 3d gravity\footnote{Note that the two formulations, Einstein-Hilbert and Cartan formalism, are not equivalent, and one must impose a relation between the triad and the spin connection in order to retrieve the Einstein-Hilbert action.} is then given by
\be
    S[e,A] = \int_{\cM_3} \langle e \wedge F[A] \rangle\equiv \int_{\cM_3}  e^i \wedge F_i[A],
    \label{3dgravityBF}
\ee
where the Killing form of the Lie algebra $\su(2)$ is taken to be $\delta_{ij}$ and the curvature two form reads:
\begin{equation}
    F^i_{\mu\nu}[A] = \partial_\mu A^i_\nu + \epsilon^{i}_{jk} A_\mu^j A_\nu^k
\end{equation}.

Variation of the action with respect to $A$ and $e$ gives equations of motion, which impose that 
\begin{equation}
    \begin{split}
        F[A] &= 0, \quad \text{the connection $A$ is flat} \\
        T[e, A] & = d e  + A \wedge e = 0 
    \end{split} 
    \label{constraints}
\end{equation}
Canonical analyse of the phase space is done in \cite{Witten:1988hc}\cite{Alexandrov:2011ab} and which can be summarized below. The Lagrangian in the action \ref{3dgravityBF} can be explicitly written:
\begin{equation}
    \mathcal{L}_{3d}[e,A] = \epsilon_{ab} e^i_a \partial_0 A^i_b + e^i_0 \epsilon_{ab}( \partial_a A^i_b - \partial_b A^i_a +  \epsilon^{i}_{jk} A_a^j A_b^k) + A_0^i \epsilon_{ab}(\partial_a e^i_b - \partial_b e^i_a + \epsilon^i_{jk}(A^j_a e^k_b + e^j_a A^k_b) )
\end{equation}
For $A_0^i$ and $e_0^i$ there is no time derivative in the action. They played the role of Lagrangian multiplier and gives the two constraints equations (\ref{constraints}). Spatial components, $A_a^j$ and $e_a^j$ are canonical variables, with Poisson bracket given by:
\begin{equation}
    \{ A_a^j(x),e_b^i(y)\} = \frac{1}{2} \epsilon_{ab} \eta^{ij} \delta^{2}(x-y)
\end{equation}

A counting of freedom degrees can be done here. The kinematic phase space is define with the Poisson bracket and $(A_a^i,e_b^j)$, which is of dimension $2\times2\times3=6$. The number of physical degree of freedom is given by the dimension of kinematics space minus the number of constraint equations (\ref{constraints}), which are $2\times 2 \times 3=0$. Therefore, there are no degrees of freedom in 3-dimensional gravity. \\

On-shell of the torsionlessness condition, which uniquely defines the connection in terms of a triad, the flatness constraint is equivalent to the Einstein equation. The action \ref{3dgravityBF} is invariant under diffeomorphisms and $\SU(2)$ gauge symmetries,
\be
A\rightarrow A^{g} = g A g^{-1} + g dg^{-1}, \quad e\rightarrow e^{g} = g e g^{-1} , \quad g \in \SU(2),
\ee
as well as under shifts,
\be
A \rightarrow A, \quad e\rightarrow e+ d \eta  + A \wedge \eta, \quad \eta \in \su(2),
\ee
which are a truly topological symmetry. \\

Starting from BF formulation of 3d gravity, on can go to Chern-Simon formulation by taking a combination of the triad and the spin-connection. Consider the generator of $ISO(3)$ (or $ISO(2,1)$, $\{J_i,P_i\}_{i=0,1,2}$ such that 
\begin{equation}
    [J_i,J_j] = \epsilon_{ijk} J^k, \quad [J_i,P_j]=\epsilon_{ijk}P^k,\quad [P_a,P_b] = 0
\end{equation}
with Killing form  $\langle J_i,P_j\rangle = \delta_{ij},\quad \langle J_i,J_j\rangle = \langle P_i,P_j\rangle = 0$. Let us define 
\begin{equation}
    \omega_a = e_a^i P_i + A_a^i J_i\,.
\end{equation}
Then, the action \ref{3dgravityBF} can be written as 
\begin{equation}
    S_{CS}[A] = \int_{\cM_3} \langle \omega \wedge d \omega + \frac{2}{3} \omega \wedge \omega \wedge \omega \rangle
    \,.
\end{equation}

\subsection{Discretizing the 3d BF path integral}

Since the local fluctuations of the fields $A$ and $e$ are pure gauge, the theory is topological and does not depend on the choice of discretization. Replace the manifold $\cM$ by an oriented triangulation and consider its dual. The duality rules can be summarized by the following table,
\def\arraystretch{1.2}
\setlength{\tabcolsep}{2em}
\noindent
\begin{center}
\begin{tabular}{| c |c|c|}
\hline
3d triangulation $\Delta$& spinfoam 2-complex $\cC$ & algebraic data
\\\hline\hline
tetrahedron $T$ & vertex $v$  & vertex amplitude $\cA_{v}$
\\\hline
triangle $t$ &edge $e$ & intertwiner $I_{e}$
\\\hline
edge $e$ &face $f$ & spin $j_{f}$
\\\hline 
\end{tabular}
\end{center}

The relevant degrees of freedom of the connection $A$ are encoded in holonomies $g_{e}\in\SU(2)$ associated to the dual edges $e\in\Delta^*$. They define the finite parallel transport along each edge. Then, curvature is associated to dual faces and given by the oriented product of holonomies of edges at the boundary of a face,
\be
    G_f = \overset{\rightarrow}{\prod_{e \in \partial f}}g_{e}.
\ee
Now, consider the path integral formulation of $3d$ gravity and formally integrate out the $e$ field. This leads to a projection onto the flat space-time states,
\begin{equation}
    Z_{BF}^{3d}(M_3)=\int \cD A \cD B \; e^{i S_{BF}^{3d}} = \int \cD A \cD B \; e^{i \int_{\cM_{3}} B \wedge F[A]} \sim \int \cD A \; \delta(F[A]),
\end{equation}
which upon discretization gives exactly the same spin-foam partition function as in two dimensions \eqref{2dSFpartfunc},
\begin{equation}
    Z_{BF}^{3d}(\Delta_3)=\int\prod_e dg_e\prod_f\delta_{SU(2)}(\overset{\rightarrow}{\prod_{e\in f}}g_e)\equiv \int\prod_e dg_e\prod_f\delta_{SU(2)}(G_f).\label{3dZdiscret}
\end{equation}
Let us use the Peter-Weyl decomposition (see Section \ref{secLQG}) to re-write this partition function in the spin basis. Recall that,
in particular, a $\delta$-function, which appears in the state-sum above, can be expanded in terms of the $\SU(2)$ characters $\chi_j$.

We can now come back to the evaluation of the partition function \eqref{3dZdiscret}. Despite their partition functions being apparently identical, the combinatorial properties of the 2- and 3-dimensional triangulations are different. Specifically, in the 3d partition function, each $g_{e}$ appears three times. This is because each edge belongs to the boundary of three faces in the spin foam, which is, in turn, because on the dual (triangulation) side this corresponds to each triangle having three edges. To compute this partition function, we will use $\SU(2)$ representation theory which is recall in \ref{secLQG}. An intertwiner at a node $n$, is a $\SU(2)$-invariant map $I_{n}:\cV^{j_1}\otimes\cV^{j_2}\otimes\cV^{j_3} \to \mathbb{C}$, denoted $|j_1,j_2,j_3\rangle $, where $j_1,j_2,j_3$ are the spin attached to incoming edges. Trivalent intertwiners between three spins exists if and only if those spins satisfy triangular inequalities, $|j_{1}-j_{2}|\le j_{3}\le(j_{1}+j_{2}) $ and are uniquely given by the Clebsh-Gordan coefficients, denoted $\begin{pmatrix}
j_1 & j_2 & j_3\\
m_1 & m_2 & m_3
\end{pmatrix}$, where $m_i$ are matrices coefficients. Since spin representations are isomorphic to their complex conjugate, $(\cV^{j})^{*}\sim\cV^{j}$, one can also define interwiners between two incoming edges and one outgoing, or one incoming and two outgoing or three outgoing. \\

Back to \ref{3dZdiscret}, since each dual face carries its own representation $j$, we will need to compute quantities like
\be
    \cP^{j_1,j_2,j_3} = \int dg \, \cD^{j_1}(g) \cD^{j_2}(g) \cD^{j_3}(g) = |I,\;j_1,j_2,j_3\rangle \langle I,\;j_1,j_2,j_3 |
    \label{3dwigner},
\ee
which is a projector onto the invariant subspace $\operatorname{Inv}_{\SU(2)}[V^{j_1} \otimes\cV^{j_2} \otimes\cV^{j_3}]$ of the kinematical space $V^{j_1} \otimes\cV^{j_2} \otimes\cV^{j_3}$ at the shared edge. Matrix elements $\cP^{j_1,j_2,j_3}_{m_1, m_2, m_3, n_1 n_2 n_3} $ of this projector are given by $3j$ symbols (or intertwiners):
\be
\begin{split}
   [\cP^{j_1,j_2,j_3}]_{m_1, m_2, m_3, n_1 n_2 n_3} & = \int dg \, \cD^{j_1}_{m_1,n_1}(g) \cD^{j_2}_{m_2,n_2}(g) \cD^{j_3}_{m_3,n_3}(g) \\
   &= |I,\;(j_1,m_1),(j_2,m_2),(j_3,m_3)\rangle \langle I,\; (j_1,m_1),(j_2,m_2),(j_3,m_3) | \\
   & =\begin{pmatrix}
    j_1 & j_2 & j_3\\
    m_1 & m_2 & m_3
    \end{pmatrix} \overline{\begin{pmatrix}
    j_1 & j_2 & j_3\\
    n_1 & n_2 & n_3
    \end{pmatrix}} 
\end{split}
\ee
Such objects are graphically represented below, \ref{fig:first}, where the three legs represent a half-link with a spin $j_i$ and a magnetic index $m_i$ each. The fact that these three links terminate at a vertex stands for the intertwiner combining them into a trivial representation. When edges labeled by $g^{-1}$ are under consideration, the formula for the projector \ref{3dwigner} need to by slightly modified, one need to consider $\cD^{j}_{m,n}(g^{-1}) = (-1)^{2j -m_n} \cD^{j}_{-n,-m}(g)$. This will gave extra minus signs in the final expression. A $3j$ symbol decorates a face of the triangulation, as it possesses three edges carrying the three spins of the $3j$ symbol. Furthermore, gluing two tetrahedra, as in \ref{fig:second}, produces (after integration over the Lie group element $g_f$ associated to the face/triangle shared by the tetrahedra) two intertwiners, as described in equation \ref{3dwigner}. Contracting the four intertwiners (of the four faces) of a tetrahedron gives an object called $6j$ symbol:
\be
\begin{Bmatrix}
j_1 & j_2 & j_3\\
j_4 & j_5 & j_6
\end{Bmatrix} =  \sum_{m_1,m_2,..m_6} (-1)^{\sum_{k=1}^{6} (j_k-m_k)} \begin{pmatrix}
    j_1 & j_2 & j_3 \\
    m_1 & m_2 & m_3
\end{pmatrix} \begin{pmatrix}
    j_1 & j_5 & j_6 \\
    -m_1 & -m_5 & m_6
\end{pmatrix} \begin{pmatrix}
    j_4 & j_2 & j_6 \\
    m_4 & -m_2 & -m_6
\end{pmatrix}\begin{pmatrix}
    j_4 & j_5 & j_3 \\
    -m_4 & m_5 & -m_3
\end{pmatrix}\label{6jas3j}
\ee
This object is composed of four $3j$ symbols, with a summation over the six $m_i$ indices. It is invariant under any permutations of the columns or an exchange of upper and lower spins in two columns. However, since each column contains spins associated to the opposite edges of a tetrahedron, 6-$j$ symbols are not invariant under exchanging spins between different columns since this would correspond to a graph with different connectivity. For more properties, see a nice review \cite{Makinen:2019rou}.

\begin{figure}[h!]
\centering
\begin{subfigure}[t]{0.25\textwidth}
    \includegraphics[scale=0.6]{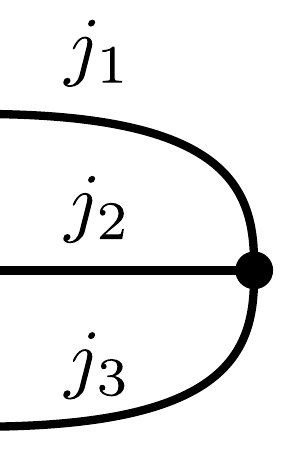}
    \caption{A pictorial representation of a three-valent intertwiner }
    \label{fig:first}
\end{subfigure}
\hspace{2.5ex}
\begin{subfigure}[t]{0.36\textwidth}
    \includegraphics[scale=0.7]{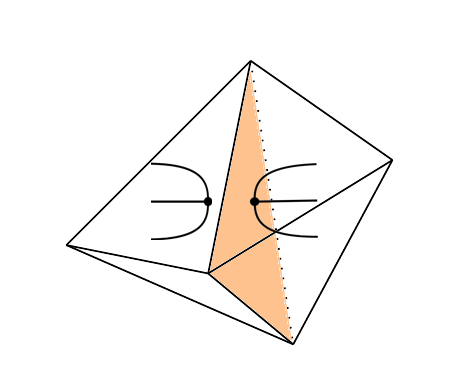}
    \caption{Two tetrahedra sharing one face with the associated three-valent intertwiner on each side of the shared face.  }
    \label{fig:second}
\end{subfigure}
\hspace{2.5ex}
\begin{subfigure}[t]{0.30\textwidth}
    \includegraphics[scale=0.2]{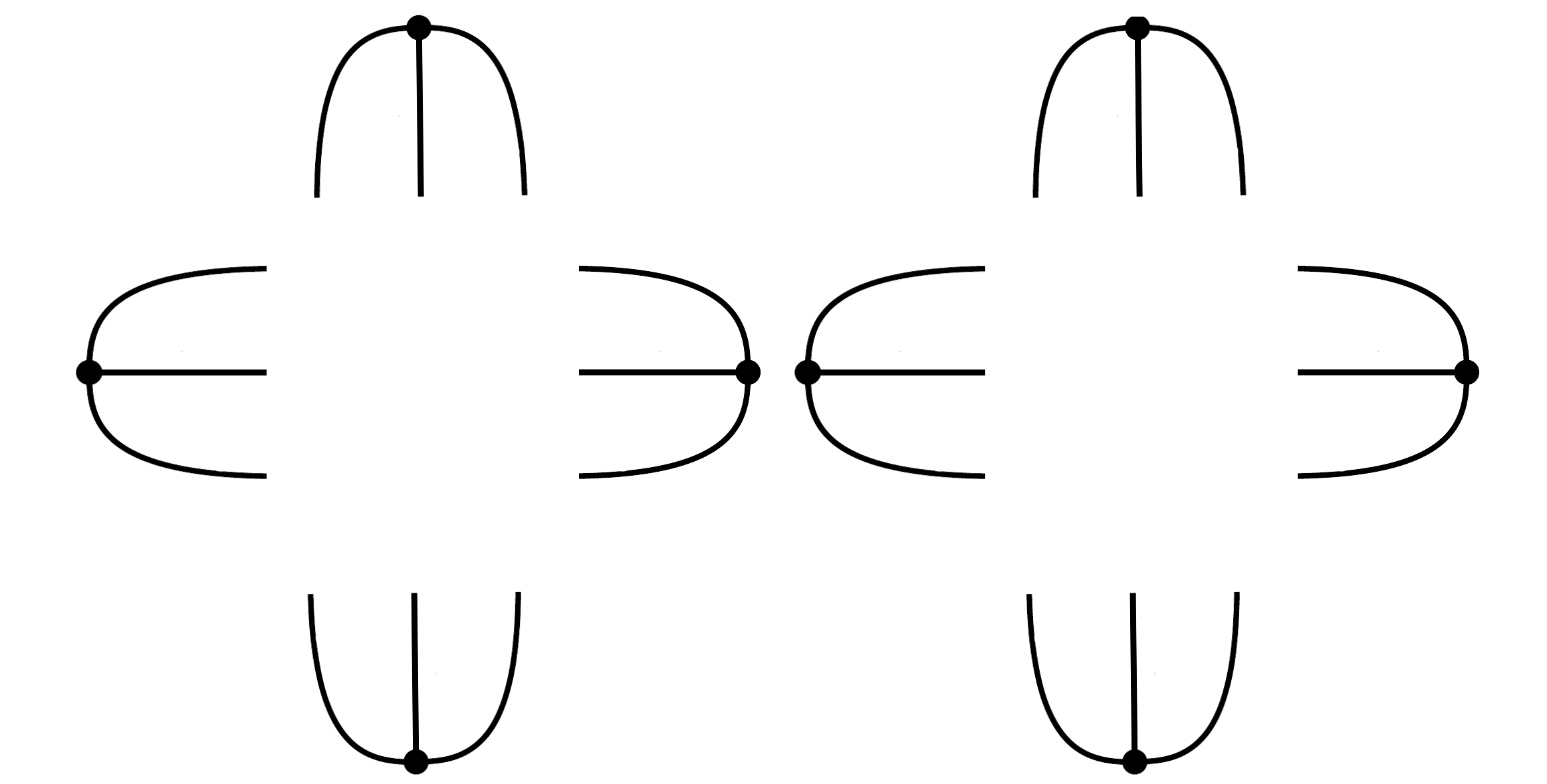}
    \caption{Recoupling of four intertwiners in one tetrahedron}
    \label{fig:third}
\end{subfigure} 
\caption{A quantum triangle is a 3-valent intertwiner. A quantum tetrahedron is made from putting together four such intertwiners. Gluing two tetrahedra together is done by matching the two intertwiners across the shared triangle.}
\label{intertwiners}
\end{figure}

Integration over Lie group elements associated with discretized faces of the triangulation, allows the discretized BF path integral to be rewritten in the so-called spinorial basis. This results in the famous Ponzano--Regge state sum \cite{PR1968},
\be
\cZ_{PR} = \sum_{\{j_e\}}\prod_{e} (-1)^{2j_{e}} d_{j_e} 
 \prod_T
 (-1)^{\sum_{k=1}^{6}j^{(t)}_{k}}
\begin{Bmatrix}
j_{1}^{(T)} &j_{2}^{(T)} & j_{3}^{(T)}\\
j_{4}^{(T)} &j_{5}^{(T)} & j_{6}^{(T)}
\end{Bmatrix},
\ee
where labels $t$ and $T$ run through all triangles and tetrahedra in a triangulation, respectively.\footnote{The $(-1)$ factors come from the presence of $\cD^{j}_{m,n}(g^{-1}) = (-1)^{2j -m_n} \cD^{j}_{-n,-m}(g)$.}
Thus, the spin-foam amplitude amounts to a product of $6j$-symbols assigned to each tetrahedron of the discretization. They are built of the $6$ spins associated to the edges of the tetrahedron. One can think of the $6j$ symbol as the probability amplitude of its boundary spin-network formed by links carrying spins $j_l$ dual to the edges of the tetrahedron and nodes dual to its faces, see \ref{tetraspinnet}. \\
 This path integral is invariant under Pachner moves (see below), corresponding to adding or removing a vertex or an edge in the triangulation. This is due to the topological invariance of the Ponzano--Regge model.

\begin{figure}[h!]
\centering

\begin{tikzpicture}[
    scale=1,
    every node/.style={font=\small},
    dot/.style={circle,fill,inner sep=1.4pt}
]

\coordinate (a) at (0,0);
\coordinate (b) at (2.7,0);
\coordinate (c) at (1.0,3.0);
\coordinate (d) at (3.1,1.0);

\coordinate (ab) at ($(a)!0.50!(b)$);
\coordinate (ac) at ($(a)!0.50!(c)$);
\coordinate (bc) at ($(b)!0.50!(c)$);
\coordinate (ad) at ($(a)!0.50!(d)$);
\coordinate (bd) at ($(b)!0.50!(d)$);
\coordinate (cd) at ($(c)!0.50!(d)$);

\coordinate (A) at (2.35,1.40);   
\coordinate (B) at (1.12,1.63);   
\coordinate (C) at (2.02,0.34);   
\coordinate (D) at (1.05,0.97);   

\draw[thick] (A)--(bc);
\draw[thick] (A)--(bd);
\draw[thick] (A)--(cd);

\draw[thick,densely dotted] (B)--(ac);
\draw[thick,densely dotted] (B)--(ad);
\draw[thick,densely dotted] (B)--(cd);

\draw[thick] (C)--(ab);
\draw[thick] (C)--(ad);
\draw[thick] (C)--(bd);

\draw[thick] (D)--(ab);
\draw[thick] (D)--(ac);
\draw[thick] (D)--(bc);

\draw (a)--(b)--(c)--cycle;
\draw[densely dotted] (a)--(d);
\draw (b)--(d)--(c);

\node[dot] at (A) {};
\node[xshift=6pt,yshift=-2pt] at (A) {$A$};

\node[dot] at (C) {};
\node[xshift=4pt,yshift=-5pt] at (C) {$C$};

\node[dot,label={above:$B$}] at (B) {};
\node[dot,label={left:$D$}] at (D) {};

\node[below=3pt]      at (ab) {$j_1$};
\node[left=4pt]       at (ac) {$j_2$};
\node[below=2pt]      at (bc) {$j_3$};

\node[xshift=-25pt,yshift=-2pt] at (ad) {$j_6$};
\node[right=3pt]      at (bd) {$j_5$};
\node[above=3pt]      at (cd) {$j_4$};

\node at (3.9,1.6) {$\longrightarrow$};

\coordinate (DD) at (4.9,1.5);
\coordinate (CC) at (5.9,0.5);
\coordinate (BB) at (5.9,2.5);
\coordinate (AA) at (6.9,1.5);

\node[dot,label={right:$A$}] at (AA) {};
\node[dot,label={above:$B$}] at (BB) {};
\node[dot,label={below right:$C$}] at (CC) {};
\node[dot,label={left:$D$}] at (DD) {};

\draw (AA)--node[midway,above right] {$j_4$} (BB);
\draw (AA)--node[midway,below right] {$j_5$} (CC);
\draw (AA)--node[midway,xshift=6pt,yshift=12pt] {$j_3$} (DD);

\draw (BB)--node[midway,left]        {$j_2$} (DD);
\draw (BB)--node[midway,xshift=-4pt,yshift=-10pt]  {$j_6$} (CC);

\draw (CC)--node[midway,below]       {$j_1$} (DD);

\end{tikzpicture}

\caption{A tetrahedron with its dual boundary.}
\label{tetraspinnet}
\end{figure}
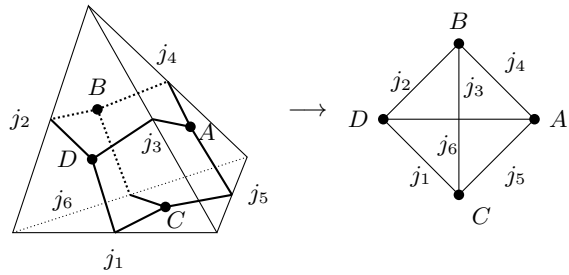

\subsection{The Elementary block of quantum geometry: the $\{6j\}$-symbol}
Properties of the $\{6j\}$-symbol have been widely studied. It represents a quantum tetrahedron. Indeed, the six $\SU(2)$-representations, given by the six spins $j_{k=1.6}$, define the six edges of a quantum tetrahedron. The spins actually give the edge length in Planck unit. Each of the four quantum triangles of the quantum tetrahedron is then defined as an intertwiner between the three spins living on its edges, i.e. a singlet state in the tensor product of the corresponding three $\SU(2)$-representations. The $\{6j\}$-symbol is finally the contraction of those four intertwiners, giving the probability amplitude of gluing those four quantum triangles together into a quantum tetrahedron. Piling up those quantum tetrahedra by gluing them along shared quantum triangles creates the 3d quantum space-time.

\medskip

Those quantum tetrahedra are not mere conceptual constructs, and can actually be understood in a precise way as semi-classical tetrahedra.
Since the spins give the edge lengths in Planck unit, the semi-classical regime is naturally obtained for large spins $j_{k}$'s.
Indeed, their asymptotic expression for large spin $j \gg1$ (i.e. $j \geq 6,7,..$) has been shown to be related to exponential of $i$ times the Regge action (see e.g. \cite{Schulten:1971yv,Schulten:1975yu, Roberts:1998zka,Freidel:2002mj, Dupuis:2010kqh}), which is the action of discrete classical gravity\footnotemark, thus validating its interpretation as a path integral probability amplitude for the geometry.
\footnotetext{The Regge action is actually the evaluation of the Einstein-Hilbert gravitational action on the tetrahedron considered as a region of space-time. Then, assuming that the metric is flat inside the tetrahedron, the action reduces to its Gibbons-Hawking-York boundary term, given by the integral of the extrinsic curvature on the 2d boundary.}
There is nevertheless a crucial subtlety: given six spins $j_{k}$, satisfying the triangular inequality dictating by spin recoupling, and interpreting them as the edge lengths (in Planck unit), a corresponding classical tetrahedron does not necessarily exist, although the $\{6j\}$-symbol  does not vanish. As underlined in several studies (see for instance the recent work \cite{Don__2024}), this can be understood as the manifestation of the tunnelling effect, which allows for an evanescent non-vanishing probability for classically-forbidden configurations. These reveal the deep quantum nature of the spinfoam geometry defined by $\{6j\}$-symbol. 

Mathematically, we look at the squared volume of the tetrahedron with edge lengths $j_{k=1..6}$, given by the Cayley-Menger determinant:
    \begin{equation}
        V^2 = \frac{1}{288}
        \begin{vmatrix}
0 & 1 & 1 & 1 & 1 \\
1 & 0 & j_1^2 & j_2^2 & j_3^2 \\
1 & j_1^2 & 0 & j_4^2 & j_5^2 \\
1 & j_2^2 & j_4^2 & 0 & j_6^2 \\
1 & j_3^2 & j_5^2 & j_6^2 & 0 \\
\end{vmatrix},
    \end{equation}
The triangle inequalities do not imply that this determinant is positive. When it is indeed positive, a classical Euclidean tetrahedron with those edge length exists. When it turns out to be negative, a corresponding classical Euclidean tetrahedron with those edge length does not exist. Instead, there exists a corresponding Lorentzian tetrahedron, implying that the Euclidean geometry is tunnelling through a Lorentzian signature \cite{Barrett:1993db,Davids:1998bp}. Actually, if one goes through half-integer spins, increasing them step by step from 0,  the number of positive and negative squared volume configuration is more or less 50\%-50\%. Nevertheless, the contribution of negative squared volume configurations is exponentially suppressed (at large spins), while the  $\{6j\}$-symbol for positive squared volume configurations oscillate as expected for semi-classical probability amplitudes.

Keeping this in mind, assuming that the spins $j_{k}$'s give a positive squared volume, the $\{6j\}$-symbols behave asymptotically as
\be
\{ 6j\}_T
\underset{\small j \gg 1}{\sim}
\frac{1}{\sqrt{12 \pi V}} \cos\big( S_{\text{Regge}}[j] + \frac{\pi}{4}\big),
\quad \text{where}\quad
S_{\text{Regge}}[j] = \sum_{k=1}^{6} \ell_k \theta_k(\{ \ell_e \})
\quad \text{with}\quad
\ell_e=j_{e}+\f12
\,,
\label{6jasymp}
\ee
where $\theta_k$ is the dihedral angle associated to the edge carrying the spin $j_{k}$ (i.e. the angle between the two triangles sharing this edge).
More precisely, the Regge action $S_{\text{Regge}}$ and dihedral angles $\theta_k$ are defined for the tetrahedron with edge lengths $\ell_{e}=j_{e}+\f12$, equal to the spins up to a shift $\f12$. This shift is sub-leading at large spins, but still produces a phase of order 1 in the asymptotic formula above. 
So at leading order, the edge lengths $\ell_{e}$ are indeed the spins $j_{e}$. Then the $+\f12$ shift can be seen as coming from the metaplectic correction in geometric quantization, when imposing the integrality condition on co-adjoint orbits in (the dual of) the Lie algebra $\su(2)$. It is similar to  the shift in the harmonic oscillator energy spectrum from Maslov's term in Bohr-Sommerfeld quantization (see e.g. \cite{Baez:1999tk,taylor20066}).

Negative and vanishing squared volume configurations are approximated at leading order by Airy functions, which have an exponential decay at large spins.

\medskip

Now, considering a triangulation $\Delta_{3d}$ made of tetrahedra $T$ glued together, the Ponzano-Regge amplitude for given spin assignments $j_{e}$ on the triangulation edges $e$ is simply the product of the $\{6j\}$-symbols for all tetrahedra, which can in turn be approximated by the exponential of the Regge action summed over all possible in/out orientations of each tetrahedron, assuming that all spin $j_{e}$ are large and that all tetrahedra are in positive squared volume configurations:
\be
\cA^{PR}_{\Delta_{3d}}[\{j_{e}\}]
=
\prod_{T}\{6j\}_{T}
\underset{j_{e}\gg 1}{\propto}
\prod_{T}\cos S_{R}[\{j_{e}\}_{e\in T}]
\sim
\sum_{\{\eps_{T}=\pm\}}
e^{i\sum_{T} \eps_{T }S_{R}[\{j_{e}\}_{e\in T}]}
\,,
\ee
where the orientation sign $\eps_{T}$ corresponds to switching the inside/outside of the tetrahedron $T$. We thus see the Regge action appearing in the semi-classical regime of the Ponzano-Regge amplitude at large spins\footnotemark.
\footnotetext{
Now, if one sums over bulk spin assignments, this logic becomes more subtle and one has to be careful showing that the path integral is dominated (or not) by large spins and positive squared volume configurations. 
}

The link with 3d gravity can actually be made more precise in the non-perturbative regime (without going to the semi-classical regime and the large spin approximation) by realizing that, first, the Ponzano-Regge amplitude defines a canonical projector on the moduli space of flat connections \cite{Ooguri:1991ib,Freidel:2005bb,Goeller:2019zpz,Livine:2021sbf}, and, second, it is topologically invariant with the same gauge symmetry as classical 3d gravity \cite{Freidel:2004vi,Freidel:2004nb}.
Let's explore those non-perturbative aspects in the next section.

\subsection{Topological Invariance, Recursion Relations and Wheeler-De Witt Equation}

A key feature of the Ponzano-Regge spinfoam path integral is indeed its topological invariance, that is that the amplitudes do not depend on the details of the bulk triangulation. This topological invariance naturally results from the symmetry of the model under deformation of the triangulations and, thus, it is intimately related to the dynamics of the theory, which is directly inherited from its invariance under space-time diffeomorphisms.
This story is made concrete through precise bridges between:
\begin{itemize}
\item the invariance of the Ponzano-Regge amplitude under Pachner moves, and their relation to the translational  symmetry generated by the action of the Hamiltonian constraints expressed in terms of holonomy operators;
\item the recursion relations satisfied by the $\{6j\}$-symbols, directly resulting from the Biedenharn-Elliott identity responsible for the invariance under Pachner moves, which become 2nd order differential equations in the large-spin limit, recognized as Wheeler-De Witt equations and directly leading to the $\{6j\}$ asymptotic formula in terms of the Regge action;
\item the realization that the Ponzano-Regge amplitude defines a projector onto canonical physical states, i.e. on the moduli space of flat connections on the canonical 2d boundary.
\end{itemize}
Together, this body of results describes the various facets of the dynamics of 3d quantum gravity formulated as a spinfoam path integral.
Let's elaborate on these features below.

\medskip

\begin{figure}[h!]
    \centering
\begin{tikzpicture}[scale=1.5]

\coordinate(x) at (0,0) ;
\coordinate(y) at (2,0);
\coordinate(z) at (1,2);
\coordinate(v) at (2.3,.7);
\coordinate(w) at (1,.8);

\draw (x)--(y);
\draw (y)--(z)--(x);
\draw[dotted] (x)--(v);
\draw (y)--(v)--(z);

\draw (w) node[RoyalBlue]{$\bullet$};
\draw[RoyalBlue,thick] (x)--(w)--(y);
\draw[RoyalBlue,thick] (z)--(w)--(v);

\draw (0.6,1.7) node {$j_1$};
\draw (1.1,1.5) node {$j_2$};
\draw (1.5,1.7) node {$j_3$};

\draw (0.7,-0.2) node {$j_6$};
\draw (0.9,0.4) node {$j_5$};
\draw (2,0.4) node {$j_4$};

\draw[RoyalBlue] (1.3,.9) node {$k_{3}$};
\draw[RoyalBlue] (0.6,0.7) node {$k_{1}$};
\draw[RoyalBlue] (0.9,1) node {$j$};
\draw[RoyalBlue] (1.1,0.6) node {$k_{2}$};

\coordinate(a) at (5.1,2.5) ;
\coordinate(c) at (5.3,-0.3);
\coordinate(b) at (4.7,.5);
\coordinate(d) at (3.9,0.7);
\coordinate(e) at (6,1.3);

\draw (a)--(b)--(c);
\draw[thick, RoyalBlue] (a)--(c);
\draw (d)--(a);
\draw (d)--(b);
\draw (d)--(c);
\draw (e)--(a);
\draw(e)--(b);
\draw (e)--(c);

\draw (4.5,2) node {$j_1$};
\draw (4.9,1.9) node {$j_2$};
\draw (5.3,2) node {$j_3$};

\draw(4.6,1) node {$j_6$};
\draw (4.4,0.7) node {$j_5$};
\draw (5.7,1) node {$j_4$};

\draw (5,-0.2) node {$j_7$};
\draw (5.2,0) node {$j_8$};
\draw (5.6,0) node {$j_9$};

\draw[RoyalBlue] (5.3,1) node {$k$};
\draw[dotted] (e)--(d);

\end{tikzpicture}
    \caption{On the left side $(1-4)$ move, on the right side, $(2-3)$ move: the blue points and edges are the ones that are added or removed during Pachner moves.}
    \label{fig:pachner}
\end{figure}
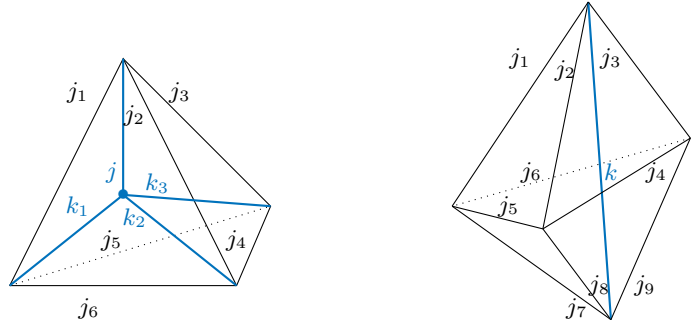

Let's recall the Ponzano-Regge partition function for a 3d triangulation in terms of the $\{6j\}$ symbols, as a sum over spins:
\be
\cZ^{PR}_{\Delta_{3d}}
= 
\sum_{\{j_e\}}\prod_{e} (-1)^{2j_{e}} d_{j_e} 
 \prod_T
  (-1)^{\sum_{k=1}^{6}j^{(t)}_{k}}
 \begin{Bmatrix}
j_{1}^{(T)} &j_{2}^{(T)} & j_{3}^{(T)}\\
j_{4}^{(T)} &j_{5}^{(T)} & j_{6}^{(T)}
\end{Bmatrix}
\,.
\ee
Here we focus on actual 3d triangulations, made of tetrahedra glued together, although the path integral can be written more generally on arbitrary 3d cellular complexes.
The starting point of the proof of topological invariance of this state sum is the remark that two 3d triangulations, with the same topology, can be related by a finite number of Pachner moves \cite{lickorish1999simplicial}. There are two types of Pachner moves here, as illustrated on fig.\ref{fig:pachner}:
\begin{itemize}
\item {\it Adding/erasing an edge or the $(2-3)$ Pachner move}:\\
starting with a double pyramid, made of two tetrahedra glued together by a shared triangle, and adding an edge linking the two opposite apex points, thereby transforming the configuration into three tetrahedra organized around this new edge, or vice-versa;

\item {\it Adding/erasing a point or the $(1-4)$ Pachner move}:\\
starting with a single tetrahedron and adding a point inside and edges linking it to the original points of the tetrahedron, thereby transforming it into four tetrahedra glued together, or vice-versa.

\end{itemize}

The invariance under the $(2-3)$ Pachner move results from  the Biedenharn-Elliott identity, also called the pentagonal identity in category theory:
\be
  \begin{Bmatrix}
    j_4 & j_5 & j_6\\
    j_1 & j_2 & j_3 
  \end{Bmatrix} \begin{Bmatrix}
    j_5 & j_4 & j_6\\
    j_7 & j_9 & j_8 
  \end{Bmatrix}
  =
  \sum_{j} (-1)^{k+\sum_{i=1}^9 j_i }d_k
  \begin{Bmatrix}
    j_1 & j_7 & k \\
    j_9 & j_2 & j_6
  \end{Bmatrix} \begin{Bmatrix}
    j_2 & j_9 & k \\
    j_8 & j_3 & j_4
  \end{Bmatrix} \begin{Bmatrix}
    j_3 & j_8 & k \\
    j_7 & j_1 & j_5
  \end{Bmatrix}
  \,.
  \label{pachner2-3eqn}
\ee 
One can check explicitly that this implies that the Ponzano-Regge partition function is indeed invariant under exchanging a double tetrahedral pyramid with three tetrahedra sharing a common edge (carrying the spin $k$). 

The invariance under  the $(1-4)$ Pachner  similarly results from the following identity, which is actually equivalent to the Biedenharn-Elliott identity (due to the orthonormality of the $\{6j\}$-symbols):
\be
\label{pachner1-4eqn}
d_j 
   \begin{Bmatrix}
    j_1 & j_2 & j_3 \\
    j_5 & j_4 & j_6
  \end{Bmatrix}
  =
  \sum_{k_1,k_2,k_3}(-1)^{j+\sum_{i=1}^6 j_i + \sum_{i=1}^3 k_i} d_{k_1} d_{k_2} d_{k_3} \begin{Bmatrix}
    k_1&  j_4 & j\\
    j_6 & k_3 & j_2
  \end{Bmatrix}\begin{Bmatrix}
    k_2 & j_5& j\\
    j_4 & k_1 & j_3
  \end{Bmatrix} \begin{Bmatrix}
    k_3 & j_6 & j\\
    j_5 & k_2 & j_1
  \end{Bmatrix} \begin{Bmatrix}
    j_1 & j_2 & j_3\\
    k_1 & k_2 & k_3
  \end{Bmatrix} 
  \,.
\ee
This indeed represents, from the left hand side to the right hand side, adding a point within a tetrahedron, thereby splitting it in four tetrahedra glued together. This shows the invariance of the Ponzano-Regge partition function under this $(1-4)$ Pachner move. Up to a crucial subtlety!
Indeed, one is only summing over three spins, $k_{1},k_{2},k_{3}$ out of the four new edges. If one were also summing over the extra spin $j$, with a factor $d_{j}$, as one is supposed to in the definition of the partition function, one would get a divergent factor $\sum_{j}d_{j}^{2}$ on the left-hand side. This means two things:
\begin{enumerate}
\item The original definition of the Ponzano-Regge partition function is typically divergent, as soon as there is a point in the bulk triangulation. There is actually one divergent factor for each bulk point.
\item These divergences can be removed by fixing one spin around each bulk vertex, and more precisely by fixing the spins along a maximal tree on the triangulation. The identities above ensure that the final finite result does not depend on the chosen values for the spins or the choice of a maximal tree. 
\end{enumerate}
This procedure is actually a true gauge theory of the translational gauge symmetry.

Indeed, on the one hand, the $(1-4)$ identity above in equation \eqref{pachner1-4eqn} can be derived as an action of the holonomy operator $\widehat{\chi_{j}}$ around the summit of the tetrahedron (where the spins $j_{1},j_{2},j_{3}$ meet) \cite{Bonzom:2009zd}. This generates a translation of the summit along a new edge carrying the spin $j$. This is also called a tent move \cite{Bonzom:2013tna}. The holonomy operators are actually the Hamiltonian constraint operators of the theory. Since the partition function is invariant under arbitrary translations of the bulk points, one needs to gauge-fix them in order to cleanly define finite transition amplitudes. This is achieved by fixing bulk spins appropriately. 

On the other hand, this gauge-symmetry is easily seen as originating from a redundance of the $\delta$-functions over $\SU(2)$  in the definition of the Ponzano-Regge partition function in terms of group elements, as in equation \eqref{3dZdiscret},
\be
\cZ^{PR}_{\Delta_{3d}}
= 
\int_{\SU(2)^{\# t}} \prod_{t}\rd g_{t}\,\prod_{e}\delta\Big{(}\overrightarrow{\prod_{t \ni e}} g_{t}\Big{)}
\,,
\ee
where one has one group element $g_{t}$ per each triangle $t$ and one $\delta$-function around each edge $e$.
As first shown in \cite{Freidel:2004vi}, if one imposes a trivial holonomy around every edge (i.e. every face of the dual triangulation) in order to get a flat connection, one gets one redundant $\delta$-function around each vertex (where the edges meet) by the Bianchi identity. Removing one such $\delta$-function around an edge $e$ exactly amounts to fixing the spin $j_{e}$. The validity and (topological) limitations of this gauge fixing procedure are discussed in details in \cite{Barrett:2008wh,Bonzom:2010ar,Bonzom:2010zh}.
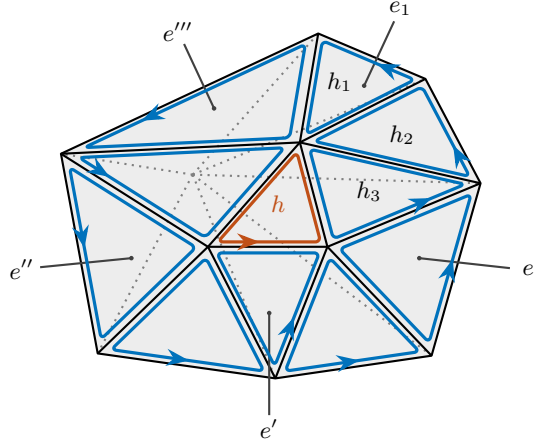
\begin{figure}
    \centering
\begin{tikzpicture}[line join=round,scale=0.6]

\newcommand{\sx}{1.35} 
\newcommand{\circarrow}[4]{
  \coordinate (tmpM) at ($(#2)!0.5!(#3)$);
  \coordinate (tmpG) at ($(tmpM)!0.33333!(#4)$);   
  \coordinate (tmpA) at ($(#2)!0.14!(tmpG)$);
  \coordinate (tmpB) at ($(#3)!0.14!(tmpG)$);
  \coordinate (tmpC) at ($(#4)!0.14!(tmpG)$);
  \draw[#1,very thick,rounded corners=2pt,
        decoration={markings,
          mark=at position 0.5 with {\arrow{Stealth[length=3mm]}}},
        postaction={decorate}]
        (tmpA)--(tmpB)--(tmpC)--cycle;
}
\coordinate (T)  at ({\sx*8.25},10.66);
\coordinate (UR) at ({\sx*10.00}, 9.66);
\coordinate (RR) at ({\sx*10.90}, 7.36);
\coordinate (LR) at ({\sx*10.10}, 3.56);
\coordinate (B)  at ({\sx*7.55},  3.06);
\coordinate (LL) at ({\sx*4.65},  3.61);
\coordinate (UL) at ({\sx*4.05},  8.01);
 
\coordinate (C) at ({\sx*7.95}, 8.26); 
\coordinate (L) at ({\sx*6.45}, 5.96);
\coordinate (R) at ({\sx*8.40}, 5.96);
\coordinate (H) at ({\sx*6.20}, 7.55);

\fill[black!7] (T)--(UR)--(RR)--(LR)--(B)--(LL)--(UL)--cycle;

\foreach \v in {T,UR,RR,LR,B,LL,UL}{\draw[gray,dotted,thick] (H)--(\v);}
 
\draw[thick] (C)--(T);  \draw[thick] (C)--(UR); \draw[thick] (C)--(RR);
\draw[thick] (C)--(UL); \draw[thick] (C)--(L);  \draw[thick] (C)--(R);
\draw[thick] (L)--(UL); \draw[thick] (L)--(LL); \draw[thick] (L)--(B);
\draw[thick] (L)--(R);
\draw[thick] (R)--(RR); \draw[thick] (R)--(LR); \draw[thick] (R)--(B);
 
\draw[thick] (T)--(UR)--(RR)--(LR)--(B)--(LL)--(UL)--cycle;

\draw[darkgray,thick] ({\sx*6.55},9.01)--({\sx*6.06},10.43);
\draw[darkgray,thick] ({\sx*9.00},9.51)--({\sx*9.49},10.93);
\draw[darkgray,thick] ({\sx*5.20},5.71)--({\sx*3.71}, 5.51);
\draw[darkgray,thick] ({\sx*9.90},5.71)--({\sx*11.38},5.49);
\draw[darkgray,thick] ({\sx*7.45},4.50)--({\sx*7.45}, 2.25);

\fill[darkgray] ({\sx*6.55},9.01) circle (0.5mm);
\fill[darkgray] ({\sx*9.00},9.51) circle (0.5mm);
\fill[darkgray] ({\sx*5.20},5.71) circle (0.5mm);
\fill[darkgray] ({\sx*9.90},5.71) circle (0.5mm);
\fill[darkgray] ({\sx*7.45},4.50) circle (0.5mm);

\circarrow{Bittersweet}{C}{L}{R}    
 
\circarrow{RoyalBlue}{C}{UR}{T}
\circarrow{RoyalBlue}{C}{T}{UL}
\circarrow{RoyalBlue}{C}{RR}{UR}
\circarrow{RoyalBlue}{C}{R}{RR}
\circarrow{RoyalBlue}{C}{UL}{L}
\circarrow{RoyalBlue}{L}{UL}{LL}
\circarrow{RoyalBlue}{L}{LL}{B}
\circarrow{RoyalBlue}{L}{B}{R}
\circarrow{RoyalBlue}{R}{B}{LR}
\circarrow{RoyalBlue}{R}{LR}{RR}

\node at ({\sx*5.96},10.71) {$e'''$};
\node at ({\sx*9.59},11.21) {$e_1$};
\node at ({\sx*3.41}, 5.47) {$e''$};
\node at ({\sx*11.68},5.45) {$e$};
\node at ({\sx*7.45}, 1.95) {$e'$};
 
\node[Bittersweet] at ({\sx*7.60},6.90) {$h$};
 
\node at ({\sx*8.6},9.60) {$h_1$};
 
\node at ({\sx*9.62},8.43) {$h_2$};
 
\node at ({\sx*9.08},7.19) {$h_3$};

\end{tikzpicture}
\caption{
We illustrate how the Bianchi identity implies a redundancy in the $\delta$-functions in the definition of the Ponzano-Regge model. Considering a triangulation point and all the edges meeting at that point, the dual bubble around that point is made of faces -- plaquettes -- dual to each edge. The integral formulation of the Ponzano-Regge imposes a $\delta$-function for each face, enforcing that the $\SU(2)$ holonomy around it be trivial. Assuming that the bubble is simply a 2-sphere (i.e. that there is no conical singularity at the point), the flatness of the last holonomy $h$ (in {{\color{Bittersweet} red}}) is automatically implied by the flatness of the holonomies around all the other faces (in {{\color{RoyalBlue} blue}}), thereby making the corresponding $\delta$-function redundant. This creates a divergence, which is gauge-fixed by simply removing this redundancy.}
    \label{fig:BianchiIdentity}
\end{figure}

\medskip

\begin{figure}[h!]
    \centering

\begin{tikzpicture}[line cap=round,line join=round,scale=1.0]

\def\a{2}    
\def\b{0.45} 
\def\H{2}    
\def\gap{1}
\def\shift{7}

\tikzset{
pics/diskmesh/.style={
code={

\draw (0,0) ellipse[x radius=\a,y radius=\b];

\begin{scope}
\clip (0,0) ellipse[x radius=\a,y radius=\b];

\coordinate (A) at (-2.2, 0.10);
\coordinate (B) at (-1.5, 0.45);
\coordinate (C) at (-0.7, 0.20);
\coordinate (D) at ( 0.2, 0.55);
\coordinate (E) at ( 1.2, 0.25);
\coordinate (F) at ( 2.1, 0.05);

\coordinate (G) at (-2.0,-0.35);
\coordinate (H) at (-1.0,-0.20);
\coordinate (I) at ( 0.0,-0.10);
\coordinate (J) at ( 0.9,-0.40);
\coordinate (K) at ( 1.9,-0.25);

\foreach \u/\v in {
A/B,A/G,A/C,
B/C,B/H,
C/D,C/H,C/I,
D/E,D/I,D/J,
E/F,E/J,E/K,
F/K,
G/H,
H/I,
I/J,
J/K%
}
\draw (\u)--(\v);

\foreach \p in {A,B,C,D,E,F,G,H,I,J,K}
    \fill (\p) circle (0.9pt);

\end{scope}
}
}
}

\pic at (0,0) {diskmesh};
\pic at (0,\H+\gap) {diskmesh};
\pic at (\shift,0) {diskmesh};

\pic at (0,\H) {diskmesh};
\pic at (0,2*\H+\gap) {diskmesh};
\pic at (\shift,2*\H+\gap) {diskmesh};

\draw (-\a,0)--(-\a,\H);
\draw ( \a,0)--( \a,\H);
\draw (\shift-\a,0)--(\shift-\a,2*\H+\gap);
\draw (\shift+ \a,0)--( \shift+\a,2*\H+\gap);

\draw (-\a,\H+\gap)--(-\a,2*\H+\gap);
\draw ( \a,\H+\gap)--( \a,2*\H+\gap);

\draw[dotted,thick, {Stealth}-] (-\a+0.1,\H+\gap-0.2)--(-\a+0.1,\H+0.2);
\draw[dotted,thick,{Stealth}-] ( \a-0.1,\H+\gap-0.2)--( \a-0.1,\H+0.2);

\draw[very thick,-{Stealth}] ( -\a-0.5,-0.3)--( -\a-0.5,2*\H+\gap+0.3)node[left=4pt] {time};
\draw[very thick,-{Stealth}] (\shift -\a-0.5,-0.3)--( \shift-\a-0.5,2*\H+\gap+0.3);

\node at (0,\H/2) {$\cZ^{PR}_{\Delta_{3d}}$} ;
\node at (0,\H*3/2+\gap) {$\cZ^{PR}_{\Delta_{3d}}$} ;
\node at (\shift,\H+\gap/2) {$\cZ^{PR}_{\Delta_{3d}}$} ;

\node at (\a+\shift*.17,\H+\gap/2) {$=$} ;
\node at (\a+\shift*.17,\H+\gap/2-0.35) {topological} ;
\node at (\a+\shift*.17,\H+\gap/2-0.65) {invariance} ;
\end{tikzpicture}
    \caption{Illustration of topological invariance of the Ponzano-Regge path integral implying that the transition map in a canonical setting is a projector, thus projecting onto the space of physical states of the theory.}
    \label{fig:composition}
\end{figure}
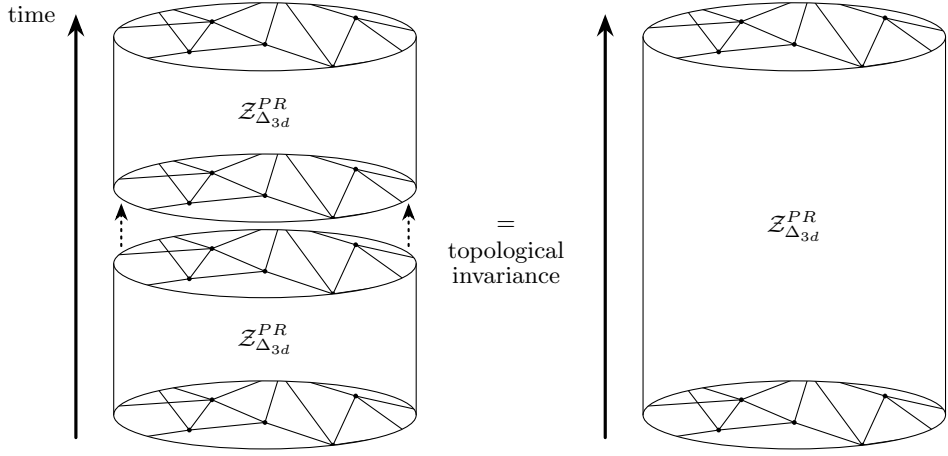

Let us illustrate this topological invariance in application to a canonical setting with a 3d manifold $\cM_{3d}\sim \Sigma_{2d} \times [0,1]$. Here, the canonical slice $\Sigma_{2d}$, assumed to be closed and orientable, evolves along a bounded interval of time. We consider a 3d triangulation  $\Delta_{3d}$ interpolating between initial and final triangulations of $\Sigma_{2d}$, assumed for the sake of simplicity to be the same. As in Atiyah's framework for topological quantum field theories, the Ponzano-Regge path integral defines a projector on the space of spin networks living on (the dual of) the triangulation of $\Sigma_{2d}$. 
Indeed, that Hilbert space of spin networks on the triangulation of $\Sigma_{2d}$ is spanned by assignments of spins to the edges of that boundary triangulation. Then, $\cZ^{PR}_{\Delta_{3d}}$ defines a map from the initial Hilbert space to a final Hilbert space, which is simply an endomorphism. That map only depends on the topology of $\cM_{3d}\sim \Sigma_{2d} \times [0,1]$. As illusrtrated on fig.\ref{fig:composition}, if we compose that map with itself $\cZ^{PR}_{\Delta_{3d}}\circ \cZ^{PR}_{\Delta_{3d}}$, it gives the Ponzano-Regge partition function on $\cM_{3d}$ glued with itself along $\Sigma_{2d}$, which is actually topologically $\cM_{3d}$ again. Thus $\cZ^{PR}_{\Delta_{3d}}\circ \cZ^{PR}_{\Delta_{3d}}=\cZ^{PR}_{\Delta_{3d}}$.
It was shown that this is the expected projector onto the space of physical states of the theory, i.e. on the space of flat $\SU(2)$ connections on $\Sigma_{2d}$ up to gauge transformations, e.g. \cite{Ooguri:1991ib,Freidel:2005bb,Goeller:2019zpz,Livine:2021sbf}. Although this can seem miraculous in the spin representation of the path integral in terms of the $\{6j\}$-symbols, this is natural and straightforward in the group representation in terms of $\delta$-functions on $\SU(2)$ group elements.
This shows that the spinfoam model correctly realizes a quantization of 3d quantum gravity as a discrete topological state sum projecting onto the expected space of physical states.

\medskip

The last panel of this body of results deals with the recursion relation satisfied by $\{6j\}$-symbols. Not only is this a very useful tool to compute numerically the value of the $\{6j\}$'s, but it turns to be an essential piece of the Ponzano-Regge machinery, binding the topological invariance, the dynamics and the semi-classical regime.
First, the recursion relations follow directly from the Biedenharn-Elliott identity realizing the (3-2) Pachner move. Indeed, starting with a $\{6j\}$ with spins $j_{1},..,j_{6}$, you can twice apply the Biedenharn-Elliott identity, attaching a thin tetrahedron to one of the triangles in order to shift a single spin, say $j_{1}$, and obtain, as shown in \cite{Schulten:1975yu,Bonzom:2009zd}, a second-order difference equation:
\be
    \mathcal{A}_{+}[j_1]  \begin{Bmatrix}
    j_1 + 1 & j_2 & j_3 \\
    j_4 & j_5 & j_6
  \end{Bmatrix} + \mathcal{A}_{0}[j_1]  \begin{Bmatrix}
    j_1  & j_2 & j_3 \\
    j_4 & j_5 & j_6
  \end{Bmatrix} + \mathcal{A}_{-}[j_1]  \begin{Bmatrix}
    j_1 - 1 & j_2 & j_3 \\
    j_4 & j_5 & j_6
  \end{Bmatrix} =0\,,
\ee
with the following explicit expressions for the coefficient $A_{0}$:
\beq
A_{0}[j]&=& \bigl(2j_1+1\bigr)\Bigl[2\bigl[j_2(j_2+1)j_{5}(j_{5}+1)+j_3(j_3+1)j_{6}(j_{6}+1)-j_1(j_1+1)j_{4}(j_{4}+1)\bigr] \\
&&- \bigl[j_2(j_2+1)+j_{3}(j_{3}+1)-j_1(j_1+1)\bigr]\bigl[j_{5}(j_{5}+1)+j_6(j_6+1)-j_1(j_1+1)\bigr]\Bigr]
\,,\nn
\eeq
and the two other coefficients given $A_{+}[j_{1}]=j_{1}A[j_{1}+1]$ and $A_{-}[j_{1}]=(j_{1}+1)A[j_{1}]$ with
\beq
A[j_{1}]
&=&\sqrt{(j_1+j_5+j_{6}+1)(j_1-j_5+j_{6})(j_1+j_5-j_{6})(-j_1+j_5+j_{6}+1)} \\
&&\sqrt{(j_1+j_2+j_3+1)(j_1-j_{2}+j_3)(j_1+j_{2}-j_3)(-j_1+j_2+j_3+1)}
\,.\nn
\eeq
As shown in \cite{Schulten:1971yv,Schulten:1975yu}, this recursion relation can be used for fast numerical computations of the $\{6j\}$-symbols, instead of using its more cumbersome expression as a hypergeometric series.
Now those works further showed that this second-order difference equation can be approximated at leading order by a second-order differential equation at large spins.
First, introduce the finite difference operator,
\be
\delta^{2}_{j_{1}} f = f(j_{1}+1) +f(j_{1}-1)-2 f(j_{1})\,,
\ee
for any function of the variable $j_{1}$ and expanding the recursion coefficients at  leading order in terms of the shifted spins  $\ell_{e}=j_{e}+\f12$:
\be
A_{0}\sim32 \ell_{1}\Delta_{123}\Delta_{156}\cos\theta_{1}
\,,\quad
A_{\pm}\sim16 \ell_{1}\Delta_{123}\Delta_{156}\,,
\ee
where $\Delta_{123}$ and $\Delta_{156}$ are respectively the areas of the triangle $(123)$ and $(156)$ in the tetrahedron of edge lenghths $\ell_{e}$ and $\theta_{1}$ is the external dihedral angle on the edge $1$, all given by the Heron formula:
\beq
\Delta_{123}
&=&
\f12\sqrt{(\ell_{1}+\ell_{2}+\ell_{3})(-\ell_{1}+\ell_{2}+\ell_{3})(\ell_{1}-\ell_{2}+\ell_{3})(\ell_{1}+\ell_{2}-\ell_{3})}
\,,\nn\\
\Delta_{156}
&=&
\f12\sqrt{(\ell_{1}+\ell_{5}+\ell_{6})(-\ell_{1}+\ell_{5}+\ell_{6})(\ell_{1}-\ell_{5}+\ell_{6})(\ell_{1}+\ell_{5}-\ell_{6})}
\,,\nn\\
\cos\theta_{1}&=&
- \frac{2\left(\ell_2^2\ell_5^2+\ell_3^2\ell_6^2-\ell_1^2\ell_4^2\right)-\left(\ell_2^2+\ell_3^2-\ell_1^2\right)\left(\ell_5^2+\ell_6^2-\ell_1^2\right)}
{4\Delta_{123}\Delta_{156}}
\,.
\eeq
The recursion relation then reads at leading order as a finite difference equation:
\be
\Big{(}\delta_{j_{1}}^{2} + 2 (1 - \cos\theta_1[\{\ell_{k}\}]) \Big{)}  \begin{Bmatrix}
    j_1  & j_2 & j_3 \\
    j_5 & j_4 & j_6
  \end{Bmatrix}
  \sim 0
  \,.
\ee 
Considering that the physical lengths $x_{e}$ are given by the shifted spins in Planck units, $\ell_{e}=\lambda x_{e}$ with overall factor $\lambda =L_{Planck}^{-1}\gg 1$, the finite difference becomes at leading order in the scale factor $\lambda$ a differential operator, $\delta_{j}^{2}\sim \lambda^{-2}\pp_{x}^{2}$. 
Finally, as proven in details in \cite{Schulten:1971yv,Schulten:1975yu}, solving the resulting second order differential at leading order in $\lambda$  in the WKB approximation  reproduces the large-spin asymptotic formula for the $\{6j\}$-symbols:
\be
\begin{Bmatrix}
    j_1  & j_2 & j_3 \\
    j_5 & j_4 & j_6
  \end{Bmatrix}
\underset{\small j_{k} \gg 1}{\sim} \frac{1}{\sqrt{12 \pi V[\{\ell_{k}\}]}}
\cos\Bigg(\sum_{k=1}^{6} \ell_k \theta_k(\{ \ell_i \}] + \frac{\pi}{4}\Bigg)
\,.
\ee
Since the spins  give the distances and thus represent the metric in this discretized setting, and this second-order differential equation is naturally interpreted as the Wheeler--DeWitt equation for 3d quantum gravity.
Actually, this link can be turned into a stronger, more rigorous, claim.  In fact, the difference equation, defining the exact recursion relations for  $\{6j\}$-symbols, was shown to be explicitly the fact that the $\{6j\}$ symbol is in the kernel of the Hamiltonian constraint operator (generating space-time diffeomorphisms)  quantized as a holonomy operator \cite{Bonzom:2011hm,Bonzom:2011nv}.

This concludes the presentation of the basics of the Ponzano-Regge path integral and of the $\{6j\}$-symbol as the elementary block of 3d quantum geometry.

\subsection{To go beyond: Turaev-Viro, group field theory, quantum group symmetries and holographic duality}

Let us mention and briefly explore three advanced developments of the Ponzano-Regge path integral.

\medskip

\textit{\textbf{Turaev--Viro state sum and generalizations}}
\vspace*{1mm}

\noindent
The main extension of the Ponzano--Regge state sum is the Turaev-Viro model, based on the quantum group $\cU_q(\SU(2))$, obtained by $q$-deforming the gauge group $\SU(2)$.  Their original construction \cite{Turaev:1992hq} defines a topological invariant built from the $q$-deformed $\{6j\}$-symbols, for a deformation parameter root of unity, $q = \exp(\frac{2 \pi i}{r+2})$, for an integer $r$. The  $\{6j\}_{q}$-symbols satisfy the deformed Biedenharn-Elliott identity, thus leading to the invariance under Pacher moves.
Working with a root of unity implies that there are only a finite number of irreducible representations, labeled by bounded half-integer spins  $0\le j \le r/2$. The direct consequence is that the path integral, defined as a sum over spins, is finite, with no divergence and thus no need for gauge fixing. 

This deformation is interpreted as an introduction of a non-vanishing positive cosmological constant $\Lambda = (\frac{2 \pi}{r+2})^{2}>0$. This was first understood in the Chern-Simons reformulation of 3d gravity \cite{Witten:1988hc,Witten:1988hf}, but then also validated by the asymptotics of the $\{6j\}_{q}$-symbols in terms of spherical tetrahedra and their Regge action augmented by the volume term coupled to the cosmological constant \cite{taylor20066}.

The $q$-deformation of $\SU(2)$ and the Turaev-Viro model actually exists for an arbitrary complex deformation parameter $q\in \C^{*}$ (away from the singular values $q=\pm 1$ corresponding to the classical Lie group). However, away from the special case for $q$ root of unity, it has generically the same divergence structure as the Ponzano--Regge model and requires suitable gauge fixing to get finite amplitudes. For instance, a real deformation parameter $q\in\R^{*}_{+}$ corresponds to hyperbolic geometries and a negative cosmological constant  \cite{taylor20066} (see also the recent works \cite{Dupuis:2014fya,Bonzom:2014bua,Bonzom:2021yma,Pan:2022pjh}). Moreover, the $q$-deformation can be understood as being generated by the volume operator acting on spin networks, at least perturbatively, at leading order around the classical value $q=1$ \cite{Freidel:1998ua,Livine:2016vhl}.

More generally, the Turaev-Viro model can be extended to an arbitrary spherical category \cite{Barrett_1999} and can actually be applied to a 3d quantum gravity for an arbitrary cosmological constant and both Riemannian and Lorentzian signature by using the representation theory of the different real forms of $\cU_{q}(\SL(2,\C))$ \cite{Buffenoir:2002tx,Meusburger:2007ad}.

Finally, one can further extend this spinfoam path integral to 3d supergravity and define supersymmetric Ponzano--Regge models, for instance, for $\cN=1$ and $\cN=2$ supergravity \cite{Baccetti:2010xd,Livine:2007dx}. In particular, this allows to see fermionic fields discretized as worldlines of topological defects. The problem in investigating higher supersymmetry is the issue of reducibility in the representation theory of the Lie supergroups $\textrm{OSp}(p|q)$ (in particular, whether the tensor product of irreducible representations are completely reducible or not).

\medskip

\textit{\textbf{Holographic Duality: 3d Quantum Gravity $\leftrightarrow$ 2d Ising}}
\vspace*{1mm}

\noindent

The topological invariance of 3d quantum gravity, and in particular of the Ponzano-Regge path integral, reflects that the theory does not have any local degrees of freedom in the bulk (once the topology is fixed), but it does {\it not} mean that the theory is trivial. Indeed, degrees of freedom will live on topological defects and boundaries and will be encoded in non-local observables. Consider a bounded region of a 3d space-time with trivial topology and no defects in the bulk. Then, all the non-trivial dynamics of geometry will be pushed to the space-time boundary: we have a naturally holographic theory.
\begin{figure}[h]
    \centering
    \begin{tikzpicture}[
  scale=0.8,
  blackdot/.style={circle, fill=black, inner sep=1.5pt},
  orangedot/.style={circle, fill=Bittersweet , inner sep=1.5pt},
  dual/.style={Bittersweet},
  midarrow/.style={dual, decoration={markings,
      mark=at position 0.6 with {\arrow{Stealth}}}, 
      postaction={decorate}},
  verymidarrow/.style={very thick,dual, decoration={markings,
      mark=at position 0.65 with {\arrow{Stealth}}},   
      postaction={decorate}}
]
\node[blackdot] (b1) at (4.45, 5.50) {}; 
\node[blackdot] (b2) at (2.35, 4.55) {}; 
\node[blackdot] (b3) at (6.45, 4.40) {}; 
\node[blackdot] (b4) at (1.40, 3.35) {}; 
\node[blackdot] (b5) at (3.80, 3.05) {}; 
\node[blackdot] (b6) at (5.40, 1.90) {}; 
\node[blackdot] (b7) at (2.55, 1.15) {}; 
\node[blackdot] (b8) at (3.95, 0.60) {}; 
 
\draw[] (b2) -- (b1) -- (b3) -- (b6) -- (b8) -- (b7) -- (b4) -- (b2);
\draw[] (b2) -- (b5);
\draw[very thick] (b1) -- (b5) node[pos=0.22, left=0pt, black] {$Y_{e}$};
\draw[] (b4) -- (b5);
\draw[] (b5) -- (b6);
\draw[] (b5) -- (b8);
\draw[] (b5) -- (b7);
\draw[] (b1) -- (b6);
 
\node[orangedot] (f1) at (3.65, 4.35) {}; 
\node[orangedot] (f2) at (2.25, 3.90) {}; 
\node[orangedot] (f3) at (2.45, 2.30) {}; 
\node[orangedot] (f4) at (3.15, 1.70) {}; 
\node[orangedot] (f5) at (4.45, 1.80) {}; 
\node[orangedot] (f6) at (4.85, 3.10) {}; 
\node[orangedot] (f7) at (5.45, 3.85) {}; 
 
\node[orangedot] (o1) at (3.35, 5.4) {};
\node[orangedot] (o2) at (5.70, 5.5) {};
\node[orangedot] (o3) at (6.5, 2.8) {};
\node[orangedot] (o4) at (5.50, 0.65) {};
\node[orangedot] (o5) at (3.10, 0.45) {};
\node[orangedot] (o6) at (1.20, 1.55) {};
\node[orangedot] (o7) at (1.10, 4.40) {};
 
\draw[midarrow] (f1) -- (o1);   
\draw[midarrow] (f1) -- (f2);   
\draw[midarrow] (f2) -- (o7);   
\draw[midarrow] (f3) -- (f2);   
\draw[midarrow] (f3) -- (o6);   
\draw[midarrow] (f4) -- (f3);   
\draw[midarrow] (f4) -- (f5);   
\draw[midarrow] (f4) -- (o5);   
\draw[midarrow] (f5) -- (o4);   
\draw[midarrow] (f6) -- (f5);   
\draw[verymidarrow] (f1) -- (f6)
    node[pos=0.50, above right=-1pt, Bittersweet
    ] {$y_{e}$};
\draw[midarrow] (f6) -- (f7);   
\draw[midarrow] (f7) -- (o2);   
\draw[midarrow] (f7) -- (o3);   
\end{tikzpicture}

    \caption{A 2d triangulation dressed with coherent state parameters $Y_{e}$ on its edges, and its dual 3-valent graph dressed with the couplings of the dual Ising model $y_{e}$.
    }
    \label{fig:IsingDuality}
\end{figure}
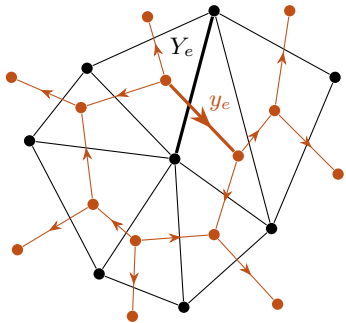

This translates into a concrete finite-distance holographic duality formula between the Ponzano-Regge model for 3d quantum gravity and the 2d Ising model with inhomogeneous couplings. Indeed, working on a triangulated 3-ball with a triangulated 2-sphere as a boundary, one obtains a class of boundary coherent states on the 2d boundary triangulation, controled by a parameter $Y_{e}$ for each boundary edge and exhibiting interesting scale-invariance properties. For those states, the Ponzano-Regge amplitude turns out to be exactly the inverse squared of the Ising partition function on the graph dual to the 2d boundary triangulations. As illustrated on fig.\ref{fig:IsingDuality}, this means that we have a  3-valent graph with a Ising node for each triangle and a Ising link transversal to every triangulation edge. Then the Ising couplings depend simply on the coherent state parameters:
\be
\label{dualityIPR}
\cZ^{PR}_{\Delta_{3d}}[\{Y_e\}_{e\in\Delta_{2d}}]
=
\Big{(}Z^{Ising}_{\Gamma}[\{y_e\}]\Big{)}^{-2}
\quad\textrm{with}\quad
Y_{e}=\tanh y_{e}\,.
\ee
Initially proven by Westbury \cite{Westbury1998} (see also \cite{Costantino2011HDR}), this duality formula relies on the high-temperature loop expansion of the 2d Ising model and reveals a supersymmetry between the bosonic degrees of freedom of the 2d boundary geometry and the fermionic Ising degrees of freedom \cite{Bonzom:2015ova}.
Applied to a single tetrahedron, this formula, combined with the low T/ high T duality of the Ising model, reveals a self-duality of the $\{6j\}$-symbol \cite{Bonzom:2019dpg} reminiscent of self-duality identities for the  $\{6j\}_{q}$-symbol \cite{Freidel:2006qv}.

More generally, it opens an interface between statistical physics and quantum gravity, allowing to export tools and results from one field to the other. In particular, one would like to apply statistical physics methods to study criticality in spinfoam path integrals and loop quantum gravity. On the other hand, one can apply the powerful results on the semi-classical approximation for the large spin asymptotics of spinfoam amplitudes and obtain new geometric formulas for the zeros of the 2d Ising partition function -- Fisher zeroes -- in terms of the geometry of 2d triangulations embedded in the 3d Euclidean space \cite{Livine:2024aix,Garay:2024wvm}. This formula extends results for Ising critical couplings on isoradial graphs and was provided with  a mathematically rigorous proof using the Kac-Ward matrix \cite{Lis:2024diu}.

\medskip

\textit{\textbf{3d Group field theory \& Non-Commutative Symmetries}}
\vspace*{1mm}

\label{sec:3dGFT}

\noindent
Similarly to matrix models generating 2d triangulations, as previously described in section \ref{sec:2dGFT}, one would like a field theory, whose partition function would provide a non-perturbative definition of the sum over 3d quantum space-time geometries. Its Feynman diagrams would be 3d triangulations, or a (subset of a) more general class of 3d cellular complexes, and the evaluation of those Feynman diagrams would give exactly the corresponding Ponzano-Regge amplitude. This would allow to concretely define the sum over 3d triangulations with explicitly weights for each 3d triangulation.
This is realized by  Boulatov's group field theory \cite{Boulatov:1998ze}.

We introduce a field over three copies of $\SU(2)$, invariant under right gauge transformations:
\be
\phi(g_{1},g_{2},g_{3})
=
\phi(g_{1}g,g_{2}g,g_{3}g)
\,,\quad\forall g,g_{i}\in\SU(2)
\,.
\ee
Fourier-transforming into spins, it decomposes over 3-valent intertwiners:
\be
\phi(g_{1},g_{2},g_{3})
=
\sum_{j_{1},j_{2},j_{3}}
\phi^{j_{1},j_{2},j_{3}}_{m_{1},m_{2},m_{3}}
\,
\prod_{i=1}^{3} D^{j_{i}}_{m_{i}\tilde{m}_{i}}(g_{i})
\,
C^{j_{1},j_{2},j_{3}}_{\tilde{m}_{1},\tilde{m}_{2},\tilde{m}_{3}}
\,,
\ee
where the $C$'s  are the $3j$-symbols, equal to the Clebsch-Gordan coefficients up to signs and numerical factors, and $\phi^{j_{1},j_{2},j_{3}}_{m_{1},m_{2},m_{3}}$ are the Fourier components of the field $\phi$.  As such, we realize that $\phi$ represents a field of quantum triangles.

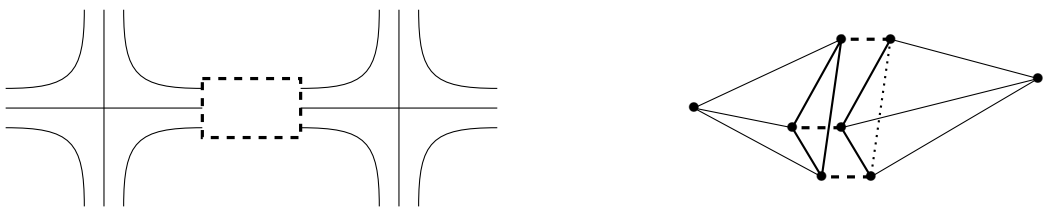
\begin{figure}[h!]
\centering

\begin{tikzpicture}[scale=1.3]

\coordinate (a1) at (-3.5,-0.2);
\coordinate (a2) at (-3.5,0);
\coordinate (a3) at (-3.5,0.2);
\coordinate (b3) at (-2.7,1);
\coordinate (b4) at (-2.5,1);
\coordinate (b5) at (-2.3,1);
\coordinate (a5) at (-1.5,.2);
\coordinate (b2) at (-1.5,0);
\coordinate (a6) at (-1.5,-0.2);
\coordinate (b6) at (-2.3,-1);
\coordinate (a4) at (-2.5,-1);
\coordinate (b1) at (-2.7,-1);

\draw (a2)--(b2) ;
\draw (a4)--(b4) ;
\draw[out=0,in=-90,looseness=1.5] (a3) to (b3);
\draw[out=0,in=90,looseness=1.5] (a1) to (b1);
\draw[out=-90,in=180,looseness=1.5] (b5) to (a5);
\draw[out=90,in=180,looseness=1.5] (b6) to (a6);

\draw[very thick,dashed] (-1.5,0.3)--(-.5,0.3)--(-.5,-0.3)--(-1.5,-0.3)--cycle;

\coordinate (A1) at (-.5,-0.2);
\coordinate (A2) at (-.5,0);
\coordinate (A3) at (-.5,0.2);
\coordinate (B3) at (.3,1);
\coordinate (B4) at (.5,1);
\coordinate (B5) at (.7,1);
\coordinate (A5) at (1.5,.2);
\coordinate (B2) at (1.5,0);
\coordinate (A6) at (1.5,-0.2);
\coordinate (B6) at (.7,-1);
\coordinate (A4) at (.5,-1);
\coordinate (B1) at (.3,-1);

\draw (A2)--(B2) ;
\draw (A4)--(B4) ;
\draw[out=0,in=-90,looseness=1.5] (A3) to (B3);
\draw[out=0,in=90,looseness=1.5] (A1) to (B1);
\draw[out=-90,in=180,looseness=1.5] (B5) to (A5);
\draw[out=90,in=180,looseness=1.5] (B6) to (A6);

\coordinate (a) at (3.5,0);
\coordinate (b) at (5,.7);
\coordinate (c) at (4.5,-.2);
\coordinate (d) at (4.8,-.7);
\draw (a) node {$\bullet$} ;
\draw (b) node {$\bullet$} ;
\draw (c) node {$\bullet$} ;
\draw (d) node {$\bullet$} ;
\draw (a)--(b);
\draw (a)--(c) ;
\draw (a)--(d) ;
\draw[thick] (b)--(c)--(d)--cycle;

\coordinate (A) at (7,+0.3);
\coordinate (B) at (5.5,.7);
\coordinate (C) at (5,-.2);
\coordinate (D) at (5.3,-.7);
\draw (A) node {$\bullet$} ;
\draw (B) node {$\bullet$} ;
\draw (C) node {$\bullet$} ;
\draw (D) node {$\bullet$} ;
\draw[thick,dotted] (B)--(D) ;
\draw (A)--(B);
\draw (A)--(C) ;
\draw (A)--(D) ;
\draw[thick] (B)--(C)--(D);

\draw[very thick,dashed] (B)--(b) ;
\draw[very thick,dashed] (C)--(c) ;
\draw[very thick,dashed] (D)--(d) ;

\end{tikzpicture}

\caption{3d group field theory (GFT): 3d triangulations generated as Feynman diagrams by gluing GFT interaction vertices, representing tetrahedra, together along GFT propagator, representing shared triangles. On the left hand side, a GFT Feynman diagram made of two interaction vertices glued together by one propagator (in bold) with three strands representing the three group elements of $\phi(g_{1},g_{2},g_{3})$. On the right hand side, the corresponding 3d triangulations made of two tetrahedra glued together by one shared triangle (in bold).}
\label{fig:3d GFT}
\end{figure}

We then define the 3d group field theory, through a simple action with a $\phi^{4}$-interaction:
\be
S^{GFT}_{3d}[\phi]=
\f12\int [\rd g_{i}]^{\times 3}
\phi(g_{1},g_{2},g_{3})
\bar{\phi}(g_{1},g_{2},g_{3})
+
\f\lambda{3!}\int\int [\rd g_{i}]^{\times 6}
\phi(g_{1},g_{2},g_{3})
\bar{\phi}(g_{5},g_{4},g_{3})
\phi(g_{5},g_{2},g_{6})
\bar{\phi}(g_{1},g_{4},g_{6})
\,.
\ee
As illustrated in fig.\ref{fig:3d GFT}, Feynman diagrams draw 3d triangulations: the $\phi^{4}$ term creates interaction vertices corresponding to tetrahedra (made of 4 triangles), then the  $\phi^{4}$ term defines a propagator that glues two tetrahedra along a shared triangle. Then the Feynman rules are such that the evaluation of a Feynman diagram gives the Ponzano-Regge amplitude of the corresponding 3d triangulation.

One can add colours to the field $\phi$, and choose various statistics, e.g. \cite{Gurau:2009tw,Gurau:2011xp}, changing the relative (statistical) weights of the induced 3d triangulations. 
One can explore 2d sectors of of this 3d group field theory and realize that it yields matter fields coupled to 3d spinfoams \cite{Fairbairn:2007sv}.
One can investigate the symmetry of that field theory, which are rather unusual and peculiar due to their highly non-local interaction term, and realize that they are invariant under non-commutative deformation of the Poincar\'e group \cite{Girelli:2010ct,Baratin:2011tg}, thereby opening a potential link with quantum group symmetries, $\kappa$-Minkowski space-times and doubly special relativity \cite{Kowalski-Glikman:2004fsz,Kowalski-Glikman:2006ssl}.
Finally, the most important development is the study of renormalization flow of the theory, either by focussing on its group structure, or on its tensorial structure. The first led to the systematic analysis of divergences and cut-off in the evaluation of the Feynman diagrams \cite{Carrozza:2013oiy}. The latter led to the seminal work by Gurau and collaborators on the melonic dominance in the large $N$ limit of tensor models \cite{Gurau:2010ba,Bonzom:2011zz}, implying a dominance of the spherical topology in the group field theory partition function, with notable applications to the SYK model \cite{Witten:2016iux,Gurau:2016lzk}.

\section{4d Spinfoams}

The previous section reviewed the spinfoam quantization of 3d gravity as a topological path integral for discrete 3d geometries. We have explained how this follows from discretizing 3d gravity formulated as a topological BF theory and how the topological invariance is intricately related to the dynamics of the theory and the Wheeler-De Witt equations.

In the present section, we move one space-time dimension up and tackle the case of 4d gravity. Now, general relativity in four space-time dimensions is not a topological field theory. It has local degrees of freedom, whose physical manifestation are propagating gravitational waves. Nonetheless, it turns out that it can be formulated, at the classical level, as a constrained  BF theory, i.e. a topological BF theory with constraints that break the topological invariance and effectively reduce it to the invariance under 4d space-time diffeomorphisms, e.g. \cite{Plebanski,Reisenberger:1998fk,DePietri:1998hnx}. These are called the {\it simplicity constraints}.
This logic can actually be extended to higher dimensions (e.g. \cite{Freidel:1999rr,Montesinos:2021vza}) and even supergravities (e.g. \cite{Ling:2000dk}).

\medskip

The spinfoam strategy is to start with the spinfoam quantization of BF theory as a discrete topological path integral, and then to add a sea of defects in order to implement the constraints raising BF theory to general relativity \cite{Barrett:1997gw,Reisenberger:1997sk}. This led to the Barrett-Crane model for a Riemannian signature \cite{Barrett:1997gw}, and then for a Lorentzian signature \cite{Barrett:1999qw}. Unfortunately, it was realized that the implementation of the constraints was too strong and that the resulting quantum states of geometry could not consistently carry a 3d volume operator \cite{Baez:1999tk,Reisenberger:1998bn}. This further led to a wrong tensorial structure for the graviton propagator  \cite{Alesci:2007tx}.

This issue was tracked down to the strong imposition of the simplicity constraints as first class constraints, i.e. imposing that the corresponding quantum operators vanish on quantum states. Instead, simplicity constraints were appropriately recognized as second class constraints. They are various ways to deal with them at the quantum level. The main two approaches are either to take  into account how they affect the phase space and compute the classical Dirac bracket, then quantize that resulting Dirac bracket (e.g \cite{Alexandrov:2001wt,Alexandrov:2002xc,Alexandrov:2002br}), or to impose them weakly \`a la Gupta-Bleuler using coherent states (for instance performing a holomorphic split of the constraints).

Following the latter path, a weaker imposition of the constraint using coherent states was proposed \cite{Livine:2007vk} and pushed forward in \cite{Freidel:2007py,Livine:2007ya}. This allowed to implement the simplicity constraints on every geometric element similarly to defects relaxing the flatness of the BF theory. This was understood to be consistent with the logic of imposing second class constraints weakly at the quantum level \cite{Engle:2007uq}. This line of investigation finally led to the EPRL model for a path integral over discrete 3+1-d geometries \cite{Engle:2007wy}. It is based on the embedding of $\SU(2)$ algebraic structures into $\SL(2,\mathbb{C})$ structures\footnotemark{}.
\footnotetext{
$\SU(2)$ is the double-cover of the group $\SO(3)$ of three-dimensional spatial rotations, while $\SL(2,\mathbb{C})$ is the double cover of the group $\SO^+(3,1)$ of 4d orthochronous space-time rotations. Keep in mind that $\SO^+(3,1)$ is the subgroup of the Lorentz group $\SO(3,1)$ connected to the identity, and in particular it excludes time-reversing transformations.
}
It effectively defines transition amplitudes for $\SU(2)$ spin network states for 3d quantum geometry by consistently embedding them in 3+1-d space-time structures. A few proposals tried to modulate the imposition of the simplicity constraints around that original proposal, e.g. \cite{Dupuis:2011fz,Banburski:2014cwa}, but the EPRL remains the most widely used spinfoam model. Here, we will present the logic underlying the construction of the EPRL model.

\medskip

It is important to realize how the properties of such spinfoam models differs from the original BF path integral. The simplicity constraints break the shift symmetry, responsible for the topological invariance of BF theory, and in particular its triangulation independence. It is therefore a strong departure from the topological field theory paradigm. It is nevertheless a desirable feature, since the simplicity constraints are meant to re-introduce the local degrees of freedom of general relativity. In some sense, the defects implementing those simplicity constraints by relaxing the flatness constraints of BF theory are supposed to represent those local fluctuations of the gravitational field and of the simplicity constraints geometry. In return, one wishes the resulting spinfoam to stil be invariant under a suitable quantum equivalent of space-time diffeomorphisms. Although such a symmetry has been identified for the Barrett-Crane models \cite{Bonzom:2009zd}, it is still a key open question for the EPRL spinfoam model.

\medskip

Details on the EPRL amplitudes, their algebraic definitions and their properties, were covered by another lecture course during the  \href{https://sites.google.com/cstq.org/2025-lqg-blaumann-school}{\nolinkurl{Les} \nolinkurl{Houches} \nolinkurl{School} \nolinkurl{on} \nolinkurl{Loop} \nolinkurl{Quantum} \nolinkurl{Gravity} \nolinkurl{2025}}. These were taught by Qiaoyin Pan and videos can be found here, \href{https://videos.univ-grenoble-alpes.fr/luniversite/2025-doctoral-training-blaumanns-loop-quantum-gravity-school/video/33878-2025_09_09_11_10_pan-sf-adv-1/}{\nolinkurl{1st} \nolinkurl{lecture}} and \href{https://videos.univ-grenoble-alpes.fr/luniversite/2025-doctoral-training-blaumanns-loop-quantum-gravity-school/video/33899-2025_09_10_18_09_pan-sf-adv-2/}{\nolinkurl{2nd} \nolinkurl{lecture}}. Other lectures on spinfoams, by Francesco Vidotto and Qiaoyin Pan, were also given at the  \href{https://indico.global/event/1306/overview}{\nolinkurl{Les} \nolinkurl{Loops'24} \nolinkurl{Summer} \nolinkurl{School}}, with slides and videos available on the website.

\subsection{4d BF theory: the Ooguri model}

So we shall start with the discrete path integral for the four-dimensional $\SU(2)$ BF theory, which gives the Ooguri model \cite{Ooguri:1992eb}. Then we'll move up to the four-dimensional  BF theory for the Lorentz group $\SL(2,\C)$. Finally, we'll show how to implement the simplicity constraints on discrete quantum geometries, and get the EPRL spinfoam amplitude. 

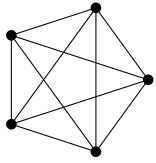
\begin{figure}[h!]
\centering
\begin{tikzpicture}
  s\foreach \i in {1,...,5}
    \fill (\i*360/5:1) coordinate (n\i) circle(2 pt)
      \ifnum \i>1 foreach \j in {\i,...,1}{(n\i) edge (n\j)} \fi;
\end{tikzpicture}
\caption{A 4-simplex $ \sigma$ consists in 5 points connected to each other, forming 5 tetrahedra and 10 triangles.}
\label{4sigma}
\end{figure}

So, let us start with 4d $\SU(2)$ BF theory. It is defined in terms of a 2-form $B$ and a 1-form connection $A$, both valued in the Lie algebra $\su(2)$, thus $B=B_{i}J^{i}$ and $A=A_{i}J^{i}$ in terms of the $\su(2)$ generators $J^{i}$. The BF action then simply reads:
\be
    S\equiv\int_{M_4}B_{i}\wedge F^{i}[A].
\label{4dBF}
\ee
We discretize the path integral using the same spinfoam approach as in lower dimensions, by imposing the flatness of the connection through $\delta$-functions over $\SU(2)$:
\begin{equation}
    Z_{BF}^{4d}(M_4) = \int \cD A \cD B \; e^{i \int_{\cM_{4}} B \wedge F[A]} \sim \int \cD A \; \delta(F[A])\overset{\text{discretization}}{\rightarrow}
    Z_{BF}^{4d}(\Delta_4)=\int_{\SU(2)}\prod_e dg_e\prod_f\delta(\overset{\rightarrow}{\prod_{e\in f}}g_e)\,.
    \label{4dZdiscret}
\end{equation}
Although the expression looks the same  as in three dimensions, the combinatorics of a 4d triangulation (and more generally of a 4d cellular complex) are different.
Now the building block of a 4d triangulated space is the $4$-simplex (see fig. \ref{4sigma}), made of five non-coplanar points. The 3D boundary of each 4-simplex is then made of 5 tetrahedra sharing 10 triangles.

As described earlier in section \ref{duality4drules}, the spinfoam structure is the dual to the triangulation:
\begin{center}
\begin{tabular}{| c |c|c|}
\hline
4d triangulation $\Delta$& spinfoam 2-complex $\cC$ & algebraic data
\\\hline\hline
4-simplex $\sigma$ & vertex $v$  & vertex amplitude $\cA_{v}$
\\\hline
tetrahedron $T$ &edge $e$ & intertwiner $I_{e}$
\\\hline
triangle $t$ &face $f$ & spin $j_{f}$
\\\hline 
\end{tabular}
\end{center}

\noindent
A 4-simplex will be represented by a vertex of the spinfoam cellular complex. A spinfoam edge will link two spinfoam vertices, representing two 4-simplices, and is dual to the tetrahedron shared by those 4-simplices. A spinfoam face -or plaquette- is dual to a triangle. It is formed by going through all the 4-simplices sharing that triangle.

So, we have one $\delta$-function for each spinfoam face, i.e. each triangle. Decomposing it into representations, for instance for a face $f$,
\be
\delta\big{(}\overset{\rightarrow}{\prod_{e\in f}}g_e\big{)}
=
\sum_{j_{f}\in\f\N2}d_{j_{f}}\chi_{j_{f}}\big{(}\overset{\rightarrow}{\prod_{e\in f}}g_e\big{)}
=
\sum_{j_{f}\in\f\N2}d_{j_{f}}
\tr\big{(}
\overset{\rightarrow}{\prod_{e\in f}}D^{j_{f}}(g_e)
\big{)}
\,,
\ee
means that we have one spin $j_{f}$ attached to each spinfoam face $f$, i.e. to each triangle.
Now, each group element $g_{e}$ living on a spinfoam edge $e$, thus going through a tetrahedron, appears in four such $\delta$-functions, one for each triangle of the tetrahedron.
More precisely, a tetrahedron, dual to an edge $e$,  corresponds to one group element $g_{e}$. Let's call $j_{1},..,j_{4}$ the spins associated to its four triangles. We are now integrating the product of the four Wigner matrices $D^{j_{1}}(g_{e)}D^{j_{2}}(g_{e)}D^{j_{3}}(g_{e)}D^{j_{4}}(g_{e)}$ over the group element $g_{e}$. This can be computed in terms of intertwiner states. Indeed the integral $\int \rd g_{e}\,\prod_{f\ni e} D^{j_{f}}(g_{e})$ is an endomorphism of the Hilbert space $\otimes_{f\ni e}\cV^{j_{f}}$, which projects on the space of $\SU(2)$-invariant states, i.e. intertwiner states. In our case with four representations, we are thus considering the projector on the space of $\SU(2)$-invariant states in $\cV^{j_1} \otimes\cV^{j_2} \otimes\cV^{j_3} \otimes\cV^{j_4}$:
\be
\int \rd g_{e}\, \cD^{j_1}(g_{e})\cD^{j_2}(g_{e})\cD^{j_3}(g_{e})\cD^{j_4}(g_{e})
=
\sum_{J} |I^{J}_{j_1j_2j_3j_4}\rangle \langle I^{J}_{j_1j_2j_3j_4}|,
\ee
where $I^{J}_{j_1j_2j_3j_4}$ are a basis of four-valent interwiners, i.e. a basis of the invariant subspace $\text{Inv}_{\SU(2)}[\cV^{j_1} \otimes\cV^{j_2} \otimes\cV^{j_3} \otimes\cV^{j_4}]$.

Contrary to the 3d case, where the space of 3-valent intertwiners at fixed spins $j_{1},j_{2},j_{3}$ is one-dimensional, the intertwiner space in the 4d case typically has a non-trivial dimension. The standard way to build an orthonormal basis is to decompose the 4-valent intertwiner into two 3-valent intertwiners linked by an intermediate  spin $J$, as illustrated on fig. \ref{fig:4valentIntertw}.
This spin labels independent vectors in the four-valent intertwiner basis.
\begin{figure}
    \centering
    \begin{tikzpicture}[scale=0.5]

    \coordinate (A) at (-2,2);
    \coordinate (B) at (2,-2);
    \coordinate (C) at (-2,-2);
    \coordinate (D) at (2,2);
    \node at (0,0) {$\bullet$};

    \draw[thick] (A) -- (B);
    \draw[thick] (C) -- (D);

    \node[left] at (A) {$j_1$};
    \node[left] at (C) {$j_2$};
    \node[right] at (D) {$j_3$};
    \node[right] at (B) {$j_4$};
\end{tikzpicture}
\begin{tikzpicture}[scale=0.67, thick]

  \coordinate (A) at (0,1.5);
  \coordinate (B) at (0,-1.5);
  \coordinate (C) at (4,1.5);
  \coordinate (D) at (4,-1.5);
  \coordinate (L) at (1,0);
  \coordinate (R) at (3,0);

  \draw (A) -- (L);
  \draw (B) -- (L);
  \draw (C) -- (R);
  \draw (D) -- (R);

  \draw (L) -- (R);

  \node[left] at (A) {$j_1$};
  \node[left] at (B) {$j_2$};
  \node[right] at (C) {$j_3$};
  \node[right] at (D) {$j_4$};
  \node[above] at (2,0) {$J$};
  \node[above] at (-2.2,-0.3) {$ = \quad \sum_J c_J $};

\end{tikzpicture}
    \caption{A 4-valent vertex can be stretched into two 3-valent vertices with an intermediate link. This yields a decomposition of 4-valent intertwiner states on a basis of states labeled by the spin $J$ carried by the intermediate link. This is usually called the recoupling basis. There are actually three different basis, whether we recouple $j_{1}$ with $j_{2}$ or $j_{3}$ or $j_{4}$. The matrices of change of basis are given by the $\{6j\}$-symbols.}
    \label{fig:4valentIntertw}
\end{figure}
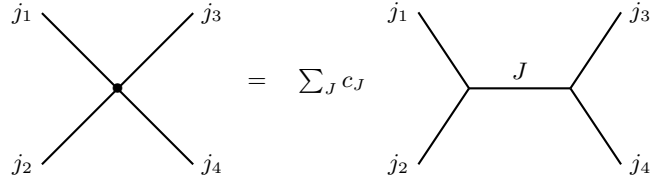
\begin{figure}[h!]
\centering
\includegraphics[scale=1]{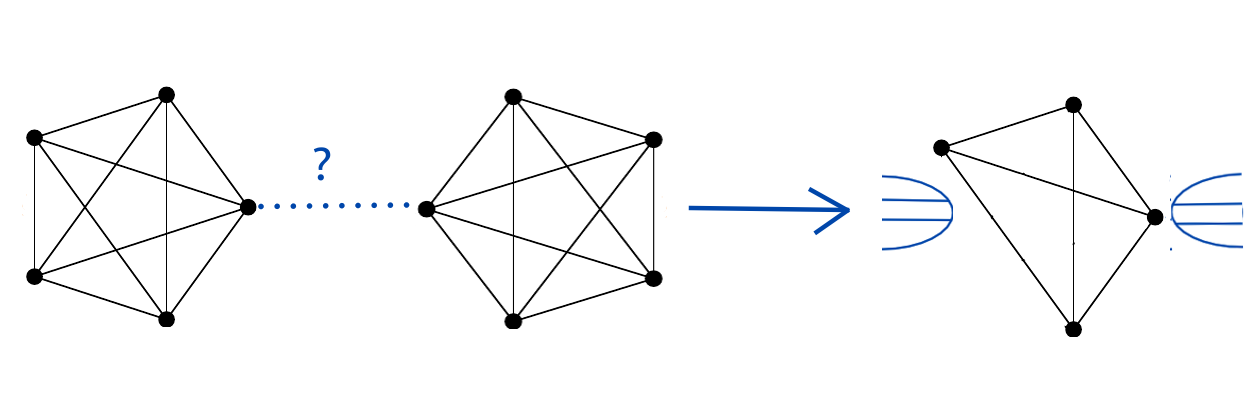}
\caption{Two 4-simplices sharing a boundary tetrahedron. A copy of the associated intertwiner couples to a 4-simplex on each side of the tetrahedron.}
\label{fig:tetraglue}
\end{figure}

Since a tetrahedron $T$, dual to an edge $e$, is shared by two 4-simplices, the corresponding operator $\sum_{J} |I^{J}_{j_1j_2j_3j_4}\rangle \langle I^{J}_{j_1j_2j_3j_4}|$ will glue those 4-simplices together, as represented on fig.\ref{fig:tetraglue}: the $ket$ copy of the intertwiner belongs to one 4-simplex, while the other lives on the other 4-simplex.
Now, we put everything together on a 4-simplex $\sigma$. Its ten triangles $t$ each carry a spin $j_{t}$. Its five tetrahedra $T$ each carry an intertwining spin $J_{T}$. The amplitude of the 4-simplex, depending on the ten $j_{t}$ plus the five $J_{T}$, is then obtained by contracting straightforwardly the five intertwiners, which yields a $15j$ symbol (see e.g. \cite{Makinen:2019rou} for the definition of $3nj$ symbols):
\be
\cA_{\sigma}[\{j_{t},J_{T}\}]=\tr\Big{(}
\bigotimes_{T\in \sigma} I^{J_{T}}_{j_{t_{T,1}}j_{t_{T,2}}j_{t_{T,3}}j_{t_{T,4}}}
\Big{)}
=
\{15j\}_{\sigma}
\quad\in\C\,.
\ee
At the end of the day, the integration of the $\delta$-distribution over the group elements $g_{e}$ gives a sum over spins of the product of $\{15j\}$-symbols, in a structure very similar to the Ponzano-Regge state-sum for 3d quantum gravity:
\be
\cZ_{BF}^{4d}[\Delta_4]
=
\sum_{\{j_t, J_T\}}
\prod_{t\in\Delta_4} d_{j_t} 
\prod_{T\in\Delta_4} \dim J_{T} 
\prod_{\sigma\in\Delta_4} 
\cA_{\sigma}[\{j_{t},J_{T}\}]
\,,
\label{4dZBF}
\ee
This is the Ooguri model \cite{Ooguri:1992eb}. The model has also been $q$-deformed, yielding the Crane-Yetter model \cite{Crane:1993if,Crane:1993cm}, which is the 4d generalization of the Turaev-Viro model (see in appendix \ref{app:CraneYetter} for more details).

\bigskip

Now we can move up to the spin foam construction for the $\SL(2,\C)$ $BF$ theory. We simply replace $\SU(2)$ by $\SL(2,\C)$. We have $\SL(2,\C)$ group elements on the spinfoam edge, i.e. across tetrahedra, and $\delta$-function around every spinfoam face, i.e. around every triangle. We decompose those $\delta$-function in $\SL(2,\C)$-representations using the Plancherel formula, thus replacing the spins by unitary irreducible representations of $\SL(2,\C)$. We then integrate over the group elements, leading to 4-valent intertwiners. In turn, we glue those  4-valent intertwiners together to get the 4-simplex amplitude as a $\{15j\}$ symbol for $\SL(2,\C)$ representations.

Let us give some details on $\SL(2,\C)$ representations.
To start with the $\sl(2,\C)$ Lie algebra is generated by $\su(2)$ generators $L^{3}, L^{\pm}$ and boosts generators $K^{3},K^{\pm}$ satisfying the following commutators:
\beq
&&[L^3,L^\pm]=\pm L^\pm,\quad [L^+, L^-]=2L^3, \nn \\
&&[K^3,K^{\pm}]=\mp L^\pm,\quad [K^+, K^-]=-2L^3. \nn \\
&& [K^3,L^{\pm}]=\pm K^{\pm},\quad [L^3,K^{\pm}]=\pm K^{\pm}, \nn\\
&& [L^+, K^-]=[K^+, L^-]=2K^3, \nn\\
&&  [L^+,K^+]=[L^-, K^-]= [L^3, K^3]=0
\,.
\eeq
Unitary irreducible representations of $\sl(2,\mathbb{C})$ are infinite-dimensional and labeled by a real $p \in \R_+$ and a half-integer $k \in \frac{\mathbb{N}}{2}$, (see e.g. \cite{Barrett:2009mw} for an overview). The Lie algbera has two quadratic Casimir operators and their eigenvalues are given by the representation labels $k$ and $p$ as:
\begin{equation}
\label{casimir}
\vec{K}^2 - \vec{L}^2 = p^2 - k^2 +1\,,\qquad \vec{K}\cdot\vec{L} = kp.
\end{equation}
To write explicitly the action of the Lorentz generators, it is convenient to use a basis simultaneously diagonalizing  these two Casimirs, as well as the $\su(2)$ Casimir $\vec{L}^{2}$ and $L^{3}$, so that basis states are  $|(p,k), j,m \rangle $. In this basis, it can be shown that $\sl(2,\C)$ representations  decompose into a direct sum of $\su(2)$-representations $\cV^j$ as:
\be
 R^{(p,k)} = \bigoplus_{j \in k + \mathbb{N}} \cV^j.
\ee
Then, on the one hand, the rotation generators have the usual action, shifting the magnetic moment $m$ while keeping the label $j$ invariant 
\beq
 L^3 |(p,k),j,m\rangle & = &m |(p,k),j,m\rangle, \nn\\
L^+ |(p,k),j,m\rangle &= &\sqrt{(j+m+1)(j-m)} |(p,k),j,m+1\rangle, \nonumber \\
L^- |(p,k),j,m\rangle  &= &\sqrt{(j+m)(j-m+1)} |(p,k),j,m-1\rangle.
\eeq
And, on the other hand, the boost generators shift the spin $j$:
\beq
 K^3 |(p,k),j,m\rangle
 &=&
 - \alpha_{(j)}\sqrt{j^2-m^2} |(p,k),j-1,m\rangle + \beta_{j} m |(p,k),j,m\rangle +\alpha_{(j+1)}\sqrt{(j+1)^2-m^2} |(p,k),j+1,m\rangle,
\nn \\
 K^+ |(p,k),j,m\rangle
 & =&
 - \alpha_{(j)}\sqrt{(j-m)(j-m-1)}|(p,k),j-1,m+1\rangle + \beta_{j}\sqrt{(j-m)(j+m+1)}|(p,k),j,m+1\rangle \nonumber\\
 & &-\alpha_{(j+1)}\sqrt{(j+m+1)(j+m+2)} |(p,k),j+1,m+1\rangle,\nonumber \\
 K^- |(p,k),j,m\rangle
 &=&
 \alpha_{(j)}\sqrt{(j+m)(j+m-1)}|(p,k),j-1,m-1\rangle  +\beta_{j} \sqrt{(j+m)(j-m+1)} |(p,k),j,m-1\rangle \nonumber  \\ 
 &&+\alpha_{(j+1)}\sqrt{(j-m+1)(j-m+2)}|(p,k),j+1,m-1\rangle,
 \label{boostaction}
\eeq
with the coefficients $\alpha_{j}$ and $\beta_{j}$ depending explicitly on the $\sl(2,\C)$ representation labels:
\begin{equation}
\beta_{j}=\frac{kp}{j(j+1)}
\,,\qquad
\alpha_{(j)}= \f ij\sqrt{\frac{(j^2-k^2)(j^2 +p^2)}{(4j^2-1)}}.
\end{equation}
Then, each triangle $t$ carries such a $\sl(2,\C)$-representation with label $(p_{t},k_{t})$, each tetrahedron carries the space of intertwiners, i.e. $\SL(2,\C)$-invariant states, between the representations attached to its triangles (see e.g. \cite{Maran:2003uc}), and then one can build the $\{15j\}$ as a function of the 10+5 representations, $(p_{t},k_{t})$ and $(p_{T},k_{T})$, dressing the 4-simplex.

We do not give explicitly formulas for those recoupling symbols here, which one can find in standard textbooks about $\SL(2,\C)$ representation theory. We will focus instead on how the simplicity constraints are implemented by imposing geometric conditions on the $\SL(2,\C)$-representations and on the intertwiners, and thereby ``reduce'' the Ooguri model for topological BF theory to a discrete path integral for 4d quantum gravity.

\subsection{4d gravity. Simplicity constraint \& the EPRL spinfoam vertex}

Let us now move up to gravity.
We use the first order Cartan formulation of  general relativity in terms of  a tetrad 1-form field $e$ and a $\sl(2,\C)$-valued connection 1-form $A$.
The metric is a composite field re-constructed from the tetrad,
\be
    g_{\mu \nu}(x) = e^{I}_\mu(x) e^{J}_\nu(x) \eta_{IJ},
\ee
where $\eta_{IJ}$ is the flat metric on the tangent space, $\eta=\operatorname{diag}(-1,1,1,1)$.
The Einstein--Cartan action is expressed as
\be
S_{EC}[e,A]
=
\frac{1}{4\kappa}\int_{\cM_4} \epsilon_{IJKL} e^{I}\wedge e^{J}\wedge F^{KL}[A]
=
\frac{1}{2\kappa}\int_{\cM_4} \star( e\wedge e )_{IJ}\wedge F^{IJ}[A]
\,,
\label{4dgravityEC}
\ee
where the coupling constant is $\kappa=8\pi G$ in terms of Newton's gravity constant $G$. The indices $I,J,K,L$ are coordinate labels on the tangent space, also refered to as the internal Minkowski space. The Levi-Civita tensor (in internal space) $\epsilon_{IJKL}$ defines the internal Hodge dual.

Following the loop quantum gravity approach, we consider an extension of this action at the classical level, by introducing an extra-term,
\be
S_{ECH}[e,A]
=
\frac{1}{2\kappa}\int_{\cM_4} \left[ \star( e\wedge e )_{IJ}\wedge F^{IJ}[A]+\frac1\gamma e_I\wedge e_J\wedge F^{IJ}[A]\right]
\,.
\label{4dgravityECH}
\ee
This is the Einstein-Cartan-Holst action \cite{Holst:1995pc}. The extra-term does not affect the equation of motion: the Einstein equations (for pure gravity) are unchanged. We loosely refer to it as a topological term, although it is not strictly speaking a topological invariant. Nevertheless, it is understood to be related to the Nieh-Yan invariant \cite{Freidel:2005sn,Mercuri:2006wb,Mercuri:2007ki}. From this point of view, the Immirzi parameter appears as a ``mass'' coupling for the squared torsion.

This action can be repackaged as a BF theory,
\be
S_{ECH}[e,A]
=
\frac{1}{2\kappa}\int_{\cM_4}  B_{IJ}\wedge F^{IJ}[A]
\,,\qquad\textrm{with}\quad
B=\star(e\wedge e)+\frac1\gamma e\wedge e
\,.
\ee
A general $\sl(2,\C)$-valued 2-form, such as $B$, would have $6\times 6=36$ components, while the tetrad $e$ only has $4\times 4=16$ components. So there is a big difference in re-packaging $e$ as $B$. At the level of the action, assuming that $B$ is a fully independent tensor, the equation of motion, resulting from the stationarity of the action with respect to variations of $B$, imposes the flatness of the connection $F[A]=0$. Constraining $B$ amounts to relaxing this equation of motion and allowing the connection to fluctuate away from flat connections, leading back to general relativity and its local degrees of freedom.

The goal is now to impose constraints on the field $B$ to effectively reduce  its number of independent components, so that it is of the form $\star(e\wedge e)+\frac1\gamma e\wedge e$. We call this $\gamma$-simplicity of the field $B$. The terminology ``simple'' comes from the fact that a bivector $b$ is called simple if it is simply a wedge product of two vectors, $u\w v$. This is equivalent to requiring that $b\wedge b=0$. The strategy is to adapt this mathematical fact to the bivector field $B$.

Following  this logic, the original set of constraints introduced by Plebanski \cite{Plebanski,Reisenberger:1998fk,DePietri:1998hnx} were indeed of the type $B^{IJ}\w B^{KL}\propto \eps^{IJKL} (\epsilon_{ABCD}B^{AB}\w B^{CD})$ , and were adapted to a non-vanishing Immirzi parameter in \cite{Capovilla:2001zi}.
These are referred to as the quadratic simplicity constraints. The goal is then to hardcode them in the BF path integral \cite{Barrett:1997gw,Freidel:1998pt,Livine:2001jt}. In the discrete setting of a 4d triangulation, the field $B$ is discretized as a bivector living on each triangle $b_{t}\in \w^{2}\R^{3,1}\sim\sl(2,\C)$. These are interpreted as the geometric bivectors to the triangles. Explicitly, considering a triangle with edge vectors $u,v,w$, the bivector is  $b=u\w v= v\w w =w \w u$, its indicates the plane of the triangle and its norm gives the triangle area. Then the quadratic simplicity constraints have a simple geometric interpretation:
\begin{itemize}
\item we have a simple bivector $b_{t}$ for each triangle $t$,
\item then $b_{t}\w b_{t'}=0$ for two triangle $t,t'$ sharing an edge (thus laying in the same hyperplane),
\item and finally, for two non-adjacent triangles $t,t'$ in a given 4-simplex,  $b_{t}\w b_{t'}$ always give the same number (up to a sign), which is actually the 4-volume of the 4-simplex \cite{Baez:1999tk,Livine:2007ya}.
\end{itemize}
However, imposing those quadratic constraints at the quantum level leads to a single intertwiner, the Barrett-Crane intertwiner, not leaving any room for a 3d volume operator and quanta of 3d volume as expected from loop quantum gravity.

\medskip

The issue was recognized as forcing a strong imposition of second class constraints \cite{Engle:2007uq,Livine:2007vk,Livine:2007ya}. Indeed, second class constraints should be imposed weakly, for instance using (holomorphic) annihilation operators and coherent states \`a la Gupta-Bleuler.
The way to solve this conundrum was to suggest linear simplicity constraints taking into account that we need to impose them on canonical states living on a 3d slice, thus allowing to use data about the embedding of that 3d slice into the 4d space-time \cite{Engle:2007uq, Engle:2007qf,Engle:2007wy}.
Now focussing on a tetrahedron (and not a 4-simplex), it lays in a 3d hypersurface. The linear simplicity constraints will simply enforce that the bivectors associated to its four triangles lays in that hypersurface, and are thus orthogonal to the normal vector to that hypersurface. Let's review this in detail, precisely underlining where the Immirzi comes in.

\smallskip

\begin{figure}
    \centering
    \includegraphics[height=40mm]{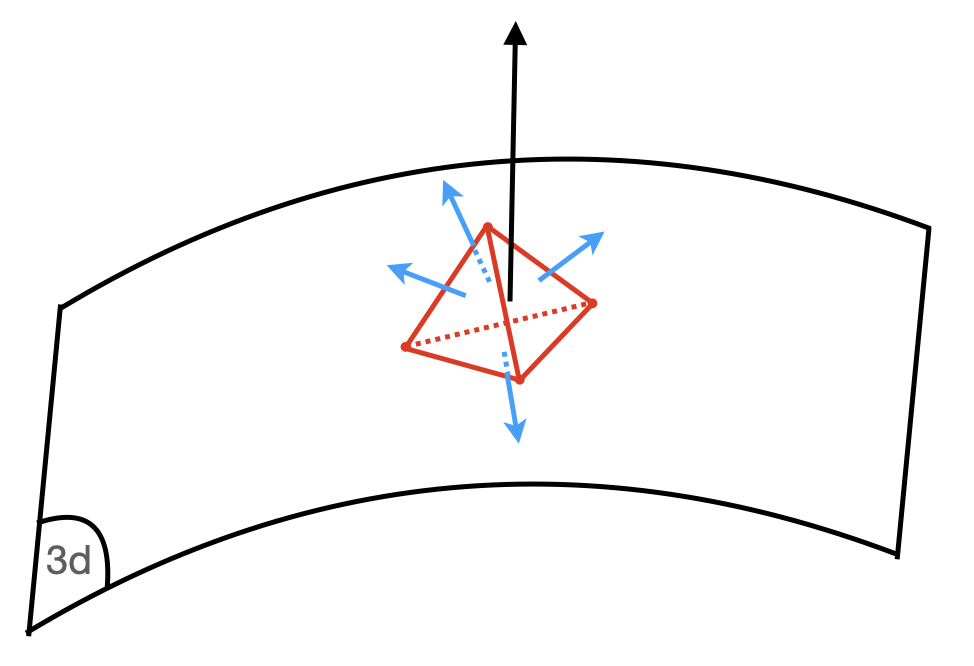}
    \caption{A tetrahedron lies in a 3d hypersurface. Then the vector $n^{I}$ normal to the hypersurface is orthogonal to  each triangle $t$. This is the geometrical interpretation of the discretized linear simplicity constraints, at the heart of the EPRL spinfoam model.}
\label{fig:linearsimplicity}
\end{figure}

Let us introduce the bivector field $E = \star(e \wedge e)$ in terms of the tetrad $e$. Then the $B$ field reads:
\be
B = E- \frac{1}{\gamma} \star E,\quad \text{or} \quad E=\frac{\gamma}{1+\gamma^2}(\gamma B+ \star B)
\,,
\label{EnB}
\ee
keeping in mind that $\star^2=-1$ in Lorentzian signature.
Now let us consider a tetrahedron on the boundary of a 4-simplex. It defines a 3d hypersurface, with normal vector $n$. As we discretize the tetrad field $e$ into  vectors along the edges of the tetrahedron, all those vectors lie in the hypersurface, and are thus  orthogonal to the normal vector $n$, $n^{I}e_{I}=0$. Moving up to the bivector fields $E$ and $B$, this orthogonality condition reads:
\be
n_I (\star E)^{IJ} = 0\,,\qquad
(n_J B^{IJ}) = \gamma \,n_J (\star B)^{IJ}
\,.
\ee
By discretizing the tetrad into edge vectors, we endow each triangle $t$ with a bivector $b_{t}$, which must now satisfy this linear simplicity constraint, as illustrated on fig.\ref{fig:linearsimplicity}:
\be
\forall I\,,\quad
n_J\, b_{t}^{IJ} = \gamma \,n_J\, (\star b_{t})^{IJ}\,.
\ee
Taking the normal vector to be the reference timelike vector, $n^I = (1,0,0,0)$, we decompose the bivectors along their 3d and boost components,
\be
K^I=n_Jb_{t}^{IJ}\quad \Rightarrow \quad K^i=b_{t}^{i0}
\,,\qquad
L^I=n_J(\star b_{t})^{IJ}\quad \Rightarrow\quad L^i=\frac12 \epsilon^{i}{}_{jk}b_{t}^{jk}
\,.
\ee
The simplicity constraint, translated into the orthogonality between the tetrahedron's hypersurface and its normal vector, become a mere proportionality condition between the 3d and boost components of the bivectors:
\begin{equation}
    \vec{K} = \gamma \vec{L}
    \,.
    \label{simplicity}
\end{equation}

\smallskip

At the quantum level, each triangle $t$ carries a $\SL(2,\C)$-representation $(p,k)$. The bivector components $b_{t}^{IJ}$  are quantized as the $\sl(2\C)$ generators in that representation. In particular, the $L^{i}$ become the $\su(2)$ generators, while the $K^{i}$  become the boost generators. The constraints $K^{i}-\gamma L^{i}=0$ do not have closed commutators, they do not define a Lie-subalgebra of $\sl(2,\C)$, and they are indeed second class constraints. We will impose them in a weak sense, that is we look for a Hilbert subspace $\cH\subset R^{(p,k)}$ such that 
\be
    \forall\,\Phi, \Psi \in \cH,\qquad \langle \Phi|\vec{K} - \gamma \vec{L}| \Psi \rangle = 0\,.
\ee
Intuitively, we are looking for coherent (semi-classical) states peaked on $\vec{K} - \gamma \vec{L}=0$.
Plugging this directly in the explicit action of the Lorentz generators for the representation $(p,k)$ given earlier in \eqref{boostaction} imposes that \cite{Engle:2007wy,Dupuis:2010jn,Ding:2010ye}:
\be
\gamma=\beta_{j}=\f{pk}{j(j+1)}\,.
\ee
This is supplemented with conditions on the two Casimir operators $\vK^{2}-\vL^{2}$ and $\vL\cdot\vK$, which must now be weakly equal respectively to $(\gamma^{2}-1)\vL^{2}$ and $\gamma \vL^{2}$, i.e.
\beq
(\gamma^2-1)j(j+1)&\sim&p^2 - k^2 +1
\,,\\
\gamma j(j+1)&\sim&pk\,,
\eeq
which we impose only at leading order for large spins $j\gg 1$, allowing for sub-leading shifts due to operator ordering contributions.
This uniquely fixes the $\sl(2,\C)$ representation labels $(p,k)$ in terms of the $\su(2)$ spin $j$ and the Immirzi parameter $\gamma$\footnotemark{}:
\be
k=j\,,\qquad p=\gamma(j+1)
\,.
\ee
\footnotetext{
Let's point that another commonly used choice of embedding is ($k=j$, $p=\gamma j$), which matches the choice used in the preset lectures at large $j$'s.
}
Note that $j=k$  is actually the lowest value allowed  for the $\su(2)$ spin in the $\sl(2,\C)$-representation $(p,k)$. This is consistent with the intuition from Perelomov coherent states defined as lowest/highest weight states.
Finally, the area of the triangle is given by $E_{0i}E^{0i}$, which  is straightforward to evaluate using the expression of $E$ in terms of $B$:
\be
\cA_{t}^{2}=E_{0i}E^{0i}=\f{\gamma^{2}}{(1+\gamma^{2})^{2}}(\gamma\vK+\vL)^{2}
\underset{\vec{K} = \gamma \vec{L}}{=}
\gamma^{2}\vL^{2}\,.
\ee
Since we are explicitly using  the condition $\vec{K} = \gamma \vec{L}$, this equality only holds  in the weak sense on  states with spins $j$ within the $R^{(p,k)}$ Hilbert space with $k=j$ and $p=\gamma(j+1)$, as explained above:
\be
\forall\,m,m',\qquad
\langle (p=\gamma (j+1),k=j),j,m'|\cA_{t}^{2}| (p=\gamma (j+1),k=j),j,m \rangle
=
\gamma^{2}{j(j+1)} \delta_{m,m'}.
\,,
\ee
leading back to the area spectrum $\gamma\sqrt{j(j+1)}$ expected from loop quantum gravity.

\medskip

This feature  is at heart of the EPRL model. It means that there is a natural automatic embedding of a $\su(2)$ spin $j$ into a $\sl(2,\C)$-representation, if one implements the $\gamma$-simplicity constraints weakly at the quantum level. This is formalized by the $Y_{\gamma}$ map \cite{Rovelli:2014ssa}:
\begin{align}
    Y_\gamma: \quad&\cV^j \hookrightarrow R^{(\gamma (j+1),j)} \nn\\
    &|j,m \rangle \mapsto |(p=\gamma (j+1),k=j),j,m\rangle.
\end{align}
This map extends to embedding whole $\SU(2)$ spin networks into $\SL(2,\C)$ spin networks, effectively realizing the embedding of quantum states of 3d geometries in covariant 4d structures.
To do so, we need to extend the $Y$ from single representations to intertwiners. So let us consider an intertwiner, say 4-valent so as to correspond to a tetrahedron, 
\be
I_{j_{1}j_{2}j_{3}j_{4}}\in \textrm{Inv}_{\SU(2)}[\cV^{j_{1}}\otimes\cV^{j_{2}}\otimes\cV^{j_{3}}\otimes\cV^{j_{4}}]\,.\nn
\ee
One can embed each spin representation $\cV^{j}$ in its corresponding $\SL(2,\C)$ representation $R^{(p=\gamma(j+1),k=j)}$ by the $Y$ map. But this does not yet give a $\SL(2,\C)$ intertwiner, since it is only invariant under $\SU(2)$ transformations. We thus perform a group averaging,
\be 
I_{j_{1}...j_{4}}
\mapsto
\int_{\SL(2,\C)}
\rd G\,\,\bigotimes_{i=1}^{4} D^{(p_{i}=\gamma(j_{i}+1),k_{i}=j_{i})}(G)Y_{\gamma}\,\, I_{j_{1}...j_{4}}.
\ee
The group averaging may seem ad hoc but it has a natural geometrical interpretation, which makes it necessary and essential: it unfreezes the choice of time normal $n^{I}$, which we set to the reference 4-vector $(1,0,0,0)$ to define the embedding map, and integrates over all possible choices of the time normal.

We can then define the vertex amplitude for a 4-simplex $\sigma$. Defining its boundary as a $\SU(2)$ spin network, all the $\SU(2)$ structures are embedded into $\SL(2,\C)$ objects by the $Y$-map. So, as for $\SU(2)$ BF theory,  the ten triangles $t$ of the 4-simplex are dressed with spins $j_{t}$ and the five tetrahedra $T$ are dressed with spins $J_{T}$ encoding the corresponding intertwiner. The resulting spinfoam vertex amplitude of the Engle-Pereira-Rovelli-Livine (EPRL) model is:
\be
\cA^{(EPRL)}_{\sigma}[\{j_{t},J_{T}\}]
=
\int \rd \mu[\{G_{T}\}]\,\,
\tr\Bigg{[}
\bigotimes_{T}I_{\{j_{t}\}_{t\in T}}^{(J_{T})}\bigotimes_{t}Y_{\gamma}D^{(\gamma (j_{t}+1),j_{t})}(G_{\tilde{T}(t)}^{-1}G_{T(t)})Y_{\gamma}
\Bigg{]}
\,,
\ee
where the measure $\rd \mu[\{G_{T}\}]$ integrates over four - out of five- group elements $G_{T}$ with the Haar measure on $\SL(2,\C)$. These group elements associated to the five tetrahedron come from the group averaging defining the coherent intertwiners using the $Y$-map. When gluing those intertwiner together, one gets the integrand above, which is invariant under left group multiplication $G_{T}\mapsto H G_{T}$ for arbitrary $H\in\SL(2,\C)$. This gauge invariance is expected  and physically corresponds to the invariance of the vertex amplitude under Lorentz transformation of the 4-simplex. It is simply gauge-fixed  by dropping one integration over $\SL(2,\C)$, and using $\rd \mu[\{G_{T}\}]$ instead of the mere product of the Haar measures $[\rd G_{T}]^{\times 5}$. The interested reader will find technical details and proofs in \cite{Engle:2008ev}.

To clarify the notations, $T(t)$ and $\tilde{T}(t)$  denote the two tetrahedra sharing the triangle $t$, respectively denote its source and target tetrahedra. This means that we have chosen an orientation of the links of the boundary graph. We need to take into account this orientation appropriately when taking the trace gluing the representation matrices to the intertwiners. 
The full amplitude is given by:
\be
\cA^{(EPRL)}[\Delta_{4d}]=
\sum_{\{j_{t},J_{T}\}}\prod_{t}d_{j_{t}}\prod_{T}d_{J_{T}}\prod_{\sigma}\cA^{EPRL}_{\sigma}[\{j_{t},J_{T}\}_{t,T\in\sigma}]\,.
\ee
Details on the EPRL vertex amplitude, with ready-to-compute formulas in terms of $\{15j\}$-symbols, can be found in \cite{Speziale:2016axj,Dona:2019dkf}. Let us nevertheless explain the choice of  edge and face amplitudes, respectively $d_{J_{T}}$ for tetrahedra (spinfoam edges) and $d_{j_{t}}$ for triangles (spinfoam faces).

The edge amplitude here is taken to be $d_{J_{T}}=2J_{T}+1$. It follows the notation of the seminal work \cite{Speziale:2016axj}, where  the intertwiners $I^{(J_T)}_{\{j_t\}}$ are built by contracting two Wigner $3jm$-symbols through the virtual spin $J_T$, so that $\|I^{(J_T)}\|^2 = 1/d_{J_T}$ with $d_{J_T} = 2J_T+1$. The edge weight $d_{J_T}$ is the inverse of this squared norm. It ensures that  $\sum_{J} d_J \, | I^{(J)}\big\rangle \langle I^{(J)}| = \mathbb{1}_{\mathrm{Inv}}$ is the identity on the space of $\mathrm{SU}(2)$-invariant 4-valent tensors, so that summing over $J_T$ for a tetrahedron shared by two 4-simplices glues their vertex amplitudes consistently.
Let's nevertheless keep in mind that that edge amplitude can always be re-absorbed in the vertex amplitude. For instance, using normalized intertwiners would simply remove this factor and set it to 1.

As for the face amplitude,  $d_{j_t}=2j_t+1$ is the dimension of the $\mathrm{SU}(2)$ irrep carried by the triangle. This is not what one would naively inherit from $\mathrm{SL}(2,\mathbb{C})$ BF theory, where faces carry the Plancherel weight $p^2+k^2$, i.e.\ $\gamma^2(j_t+1)^2+j_t^2$ for the EPRL labels. The $\mathrm{SU}(2)$ dimension is instead fixed by requiring that amplitudes compose correctly under gluing, given that the boundary Hilbert space is spanned by $\mathrm{SU}(2)$ spin networks~\cite{Bianchi:2010fj}. The two choices give different large-spin scalings of the bulk face weight, $\sim 2j$ versus $\sim(1+\gamma^2)\,j^2$. This affects power counting, for example in divergence and renormalization analyses, but not the vertex amplitude itself. As a consequence, changing the face amplitude does change the spinfoam predictions for quantum gravity corrections e.g. \cite{,Livine:2006ab,Bahr:2015gxa}.

\medskip

The asymptotics at large spins of those amplitude can be found in \cite{Barrett:2009mw,Barrett:2010ex} and details on the structure of the resulting semi-classical expansion in terms of saddle points can be found in \cite{Han:2021kll,Han:2023cen,Han:2024lti,Li:2025nog}.
More precisely, in the setting of  \cite{Barrett:2009mw,Barrett:2010ex}, the boundary state of the quantum 4-simplex is a coherent spin network built from Livine--Speziale coherent intertwiners: the triangles are dressed with $\SU(2)$ spins $j_t$ and unit normals $\vec{n}_{t}^{(T)}$ that satisfy closure for each of the five tetrahedra. The spins are rescaled uniformly, $j_t\to\lambda j_t$ with $\lambda\to\infty$, and the amplitude is evaluated by a stationary-phase analysis of its integral representation. The result depends on the geometry of the boundary data. If the data are Regge-like, meaning the five tetrahedra glue consistently into the boundary of a non-degenerate Lorentzian 4-simplex with all tetrahedra and triangles spacelike, then
\begin{equation}
  \mathcal{A}_\sigma \underset{\lambda\to\infty}{\sim} \lambda^{-12}\Big[ N_+\, e^{\,i\lambda S_{\rm R}} + N_-\, e^{-i\lambda S_{\rm R}} \Big],
  \qquad S_{\rm R} = \gamma\sum_{t} j_t\,\Theta_t ,
\end{equation}
where $\Theta_t$ are the Lorentzian dihedral boost angles and $\gamma j_t$ are the triangle areas in Planck units.
Boundary data that are not geometric are suppressed faster than any power of $\lambda$. These results were extended to arbitrary simplicial complexes by Han and Zhang \cite{Han:2011re,Han:2011rf}. They have been checked numerically, and the Regge regime already sets in at moderate spins \cite{Dona:2019dkf}.
This body of works established a clean link between the EPRL spinfoam amplitudes and Regge calculus for discretized gravity, supporting that the EPRL model leads back to general relativity in an appropriate semiclassical regime.

It nevertheless has essential limitations. First, it produces the two terms $e^{\pm i\lambda S_{\rm R}}$, related by a change of orientation, rather than a single exponential (the ``cosine problem''). The proper vertex amplitude \cite{Engle:2011un} was proposed to remove the second term. Second, it requires all spins to be uniformly large.  However, on a fixed simplicial complex with bulk triangles, one also sums over the bulk spins, which are not necessarily asymptotic. Actually, if the bulk spins are scaled together with the boundary spins, stationarity with respect to the bulk spins imposes $\gamma\,\varepsilon_t \in 4\pi\mathbb{Z}$ on the deficit angles $\varepsilon_t$. For generic $\gamma$, this means the real critical points describe only flat geometries, which is recognized as the flatness problem \cite{Bonzom:2009hw,Hellmann:2013gva,Oliveira:2017osu,Engle:2020ffj}.

This led to the more refined semiclassical analysis in terms of saddle points discussed in  \cite{Han:2021kll,Han:2023cen,Han:2024lti,Li:2025nog}.
Those papers showed that this flatness problem critically depended on restricting to real critical points. Analytically continuing the spinfoam action, they found complex critical points associated with curved Regge geometries with small deficit angles. Their contribution is of order $e^{i\mathcal{I}}$, where $\mathcal{I}$ is the Regge action plus higher order  corrections. In a suitable regime of spins, the amplitude then reduces to an integral over Regge geometries weighted by $e^{i\mathcal{I}}$, which also softens the cosine problem. Related evidence comes from numerical studies of corrections beyond leading order \cite{Han:2020fil}.

These results all concern large-spin asymptotics on a fixed spinfoam 2-complex, and this regime should be kept distinct from a true continuum limit. Large spins at fixed combinatorics describe a coarse triangulation with large building blocks. They are a semiclassical limit of the discrete theory, not a limit in which the discretization is removed. Recovering continuum general relativity, with local degrees of freedom propagating on scales much larger than the Planck scale, presumably requires either refining the 2-complex or summing over 2-complexes. Refinement can be studied with coarse-graining and renormalization methods, e.g. \cite{Dittrich:2014ala,Bahr:2016hwc}. Summing over spinfoam 2-complexes can be achieved  by group field theories \cite{Reisenberger:2000fy,Reisenberger:2000zc,Freidel:2005qe,Oriti:2013aqa}, which extends matrix models to sum over random discrete geometries of higher dimensions than 2.  In both approaches, the semi-classical regime may well be dominated by small spins on many vertices rather than by large spins on a few. The flatness problem could then be an artefact of taking the classical limit before the refinement limit. 

\medskip

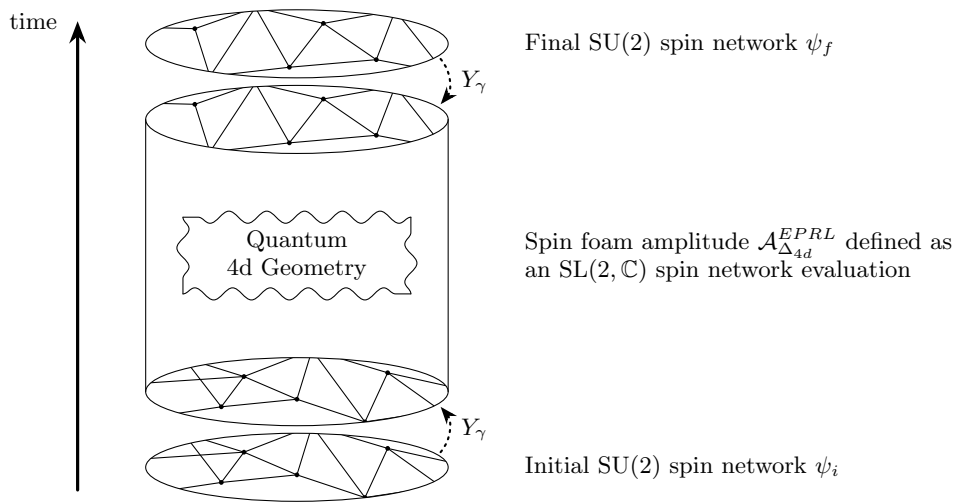
\begin{figure}[t]
\centering
\begin{tikzpicture}[line cap=round,line join=round,scale=1.0]
\def\a{2}    
\def\b{0.45}
\def\H{1.3}    
\def\gap{1}
\tikzset{
pics/diskmesh/.style={
code={
\draw (0,0) ellipse[x radius=\a,y radius=\b];
\begin{scope}
\clip (0,0) ellipse[x radius=\a,y radius=\b];
\coordinate (A) at (-2.2, 0.10);
\coordinate (B) at (-1.5, 0.45);
\coordinate (C) at (-0.7, 0.20);
\coordinate (D) at ( 0.2, 0.55);
\coordinate (E) at ( 1.2, 0.25);
\coordinate (F) at ( 2.1, 0.05);
\coordinate (G) at (-2.0,-0.35);
\coordinate (H) at (-1.0,-0.20);
\coordinate (I) at ( 0.0,-0.10);
\coordinate (J) at ( 0.9,-0.40);
\coordinate (K) at ( 1.9,-0.25);
\foreach \u/\v in {
A/B,A/G,A/C,
B/C,B/H,
C/D,C/H,C/I,
D/E,D/I,D/J,
E/F,E/J,E/K,
F/K,
G/H,
H/I,
I/J,
J/K%
}
\draw (\u)--(\v);
\foreach \p in {A,B,C,D,E,F,G,H,I,J,K}
    \fill (\p) circle (0.9pt);
\end{scope}
}
}
}
\tikzset{
pics/diskmeshtwo/.style={
code={
\draw (0,0) ellipse[x radius=\a,y radius=\b];
\begin{scope}
\clip (0,0) ellipse[x radius=\a,y radius=\b];
\coordinate (A) at (-2.15, 0.22);
\coordinate (B) at (-1.35, 0.2);
\coordinate (C) at (-0.55, 0.48);
\coordinate (D) at ( 0.35, 0.27);
\coordinate (E) at ( 1.30, 0.44);
\coordinate (F) at ( 2.05, 0.16);
\coordinate (G) at (-1.85,-0.29);
\coordinate (H) at (-1.15,-0.43);
\coordinate (I) at (-0.10,-0.31);
\coordinate (J) at ( 1.05,-0.21);
\coordinate (K) at ( 1.95,-0.38);
\foreach \u/\v in {
A/B,A/G,A/C,
B/C,B/H,
C/D,C/H,C/I,
D/E,D/I,D/J,
E/F,E/J,E/K,
F/K,
G/H,
H/I,
I/J,
J/K%
}
\draw (\u)--(\v);
\foreach \p in {A,B,C,D,E,F,G,H,I,J,K}
    \fill (\p) circle (0.9pt);
\end{scope}
}
}
}

\pic at (0,-\gap) {diskmesh};
\pic at (0,0) {diskmesh};
\pic at (0,2*\H+\gap) {diskmeshtwo};
\pic at (0,2*\H+2*\gap) {diskmeshtwo};

\draw (-\a,0)--(-\a,2*\H+\gap);
\draw ( \a,0)--( \a,2*\H+\gap);

\draw[dotted,thick,-{Stealth}] ( \a-0.1,-\gap+0.2)
      to[bend right=40] node[midway,right=1pt] {$Y_\gamma$} ( \a-0.1,-0.2);

\draw[dotted,thick,-{Stealth}] ( \a-0.1,2*\H+2*\gap-0.2)
      to[bend left=40] node[midway,right=1pt] {$Y_\gamma$} ( \a-0.1,2*\H+\gap+0.2);

\draw[very thick,-{Stealth}] (-\a-0.9,-\gap-0.3)--(-\a-0.9,2*\H+2*\gap+0.3)
      node[left=4pt] {time};

\node[inner sep=0pt, minimum width=30mm, minimum height=10mm, align=center]
     (amp) at (0,\H+\gap/2) {Quantum \\ 4d Geometry};
\begin{scope}[decoration={snake, amplitude=0.8mm, segment length=5mm,
              pre length=0mm, post length=0mm}]
  \draw[decorate] (amp.south west) -- (amp.south east);
  \draw[decorate] (amp.south east) -- (amp.north east);
  \draw[decorate] (amp.north east) -- (amp.north west);
  \draw[decorate] (amp.north west) -- (amp.south west);
\end{scope}
\node[anchor=west] at (\a+0.9,-\gap) {Initial $\SU(2)$ spin network $\psi_i$};
\node[anchor=west] at (\a+0.9,2*\H+2*\gap) {Final $\SU(2)$ spin network $\psi_f$};
\node[anchor=west, align=left] at (\a+0.9,\H+\gap/2)
{Spin foam amplitude $\cA^{EPRL}_{\Delta_{4d}}$  defined as \\ an $\SL(2,\C)$ spin network evaluation};

\end{tikzpicture}
\caption{Illustration of the EPRL spinfoam construction, based on embedding $\SU(2)$ spin networks into $\SL(2,\C)$-representations to compute amplitudes for the 4d bulk geometry.}
\label{fig:EPRL}
\end{figure}
This concludes the presentation of the EPRL spinfoam model and of the logic leading to it intertwining geometrical insight and state-sum models for topological quantum field theories. The logic of embedding the $\SU(2)$ structures of loop quantum gravity's spin network states for 3d geometry into the $\SL(2,\C)$ structures of the bulk 4d geometry is illustrated on fig.\ref{fig:EPRL}.

This 4d quantum gravity path integral is thus ready to be studied, computed and analysed, now allowing for deep numerical investigation e.g.\cite{Dona:2019dkf,Dona:2020tvv}. It can for instance be applied to black hole dynamics, and more specifically to black hole to white hole transitions in the Planckian regime \cite{Soltani:2021fno,Han:2024rqb}. Ongoing lines of research  also entail $q$-deforming the model to take into account a non-vanishing cosmological constant \cite{Han:2025mkc}. There are deep issues yet to explore. One of the open research directions is the renormalization flow of the theory under coarse-graining and its fixed points \cite{Dittrich:2014mxa,Asante:2022dnj,Han:2026hyq}. Another open question is the study of symmetries of the model. For instance, an essential question is to identify a symmetry under deformation of the bulk triangulation. This would shed light on the action of space-time diffeomorphisms on spinfoam amplitudes, and how the Wheeler-de Witt equations and the Hamiltonian constraints of loop quantum gravity are represented in this path integral framework.


\section*{Acknowledgments}

This work was supported by the Centre National de la Recherche Scientifique (CNRS) under the International Research Project program (grant awarded to the French–Canadian collaborative project {\it Quantum Geometry - Quantum Symmetry - Quantum Boundary}). O.H. is supported by a NSERC Discovery grant awarded to M.D.

The Loop Quantum Gravity school in Les Houches 2025 was supported by the Blaumann foundation.

Research at Perimeter Institute is supported in part by the Government of Canada through the Department of Innovation, Science and Economic Development and by the Province of Ontario through the Ministry of Colleges, Universities, Research Excellence and Security. 

\appendix
%

\section{$\SU(2)$ representations and Wigner matrices}
\label{app:SU2}

The Lie group $\SU(2)$ consists in the 2$\times$2 unitary matrices,
\be
g=\mat{cc}{\alpha & -\bbeta \\ \beta & \balpha}\in \SU(2)
\,,\qquad
|\alpha|^{2}+|\beta|^{2}=1\,.
\ee
These form a 3-sphere in the four-dimensional real space $\R^{4}$ spanned by the real and imaginary parts of the matrix components $\alpha$ and $\beta$. The obvious measure on $\SU(2)$ is then the  measure on the 3-sphere induced by the Lebesgue measure on $\R^{4}$:
\be
\rd g \,\propto \,\delta(|\alpha|^{2}+|\beta|^{2}-1)\,\rd^{2}\alpha\rd^{2}\beta\,.
\ee
A quick computation allows to check that this measure is indeed invariant under the (left and right) action(s) of $\SU(2)$. To get the Haar measure, we simply need to normalize it by the volume of the unit 3-sphere, which we compute by integrating over the modulus and phases of $\alpha=xe^{i\theta}$ and $\beta=ye^{i\vphi}$ with $x,y\in[0,1]$ and $\theta,\vphi\in[0,2\pi]$:
\be
\int \delta(|\alpha|^{2}+|\beta|^{2}-1)\,\rd^{2}\alpha\rd^{2}\beta
=
(2\pi)^{2}\int_{0}^{1}\int_{0}^{1}
\delta(x^{2}+y^{2}-1)\,xy\,\rd x\rd y
=
2\pi^{2}\int_{0}^{1} x\,\rd x
=
\pi^{2}\,.
\ee

Now, to get representations of $\SU(2)$, we start by considering the obvious action of $\SU(2)$ matrices on complex 2-vectors:
\be
z=\mat{c}{z_{0} \\ z_{1}}\in\C^{2}
\,,\qquad
g\triangleright z=gz\,.
\ee
Since $g$'s act on the vector components $z_{0}$ and $z_{1}$, it is natural to act on polynomials of the complex vector $z$.
There is however a little subtlety when acting on functions, due to the necessary of the matrix transpose to ensure that we define a correct representation of the group:
\be
(g\triangleright f)(z)\equiv
f(g^{t} \triangleright z)
=
f(g^{t} z)
\,,
\ee
\be
(g_{1}\triangleright(g_{2}\triangleright f))(z)
=
(g_{2}\triangleright f)(g_{1}^{t} z)
=
 f(g_{2}^{t}g_{1}^{t}  z)
 =
 f((g_{1}g_{2})^{t}  z)
 =
 (g_{1}g_{2}\triangleright f)(z)
 \,.
\ee
This gives the natural action of $\SU(2)$ group elements on polynomials of two complex variables. That's the approach by Gordan to define irreps of $\SU(2)$, done under the name of ``theory of invariants'' in mathematics well before quantum mechanics arose as a theory of physics:
\be
g^{t} z
=
\mat{cc}{\alpha & \beta \\ -\bbeta & \balpha}\mat{c}{z_{0} \\ z_{1}}
=
\mat{c}{\alpha z_{0} +\beta z_{1} \\ -\bbeta z_{0} +\balpha z_{1}}
\,,\qquad
g\triangleright z_{0}^{a}z_{1}^{b}
=
(\alpha z_{0} +\beta z_{1} )^{a}(-\bbeta z_{0} +\balpha z_{1})^{b}
\,.
\ee
Clearly the action of $\SU(2)$ group elements keeps the total degree of the polynomial $(a+b)$ fixed while it changes the relative degree in $z_{0}$ and $z_{1}$. This allows to recover the irreducible representations of $\SU(2)$. Let us write $(a+b)=2j$ with $\j\in\f\N2$ and $(a-b)=2m$ with $m\in \N+j$ and $-j\le m\le +j$, thus working with the monomials $z_{0}^{j+m}z_{1}^{j-m}$. These form a basis of the irreducible representation of spin $j$. Indeed acting with a $\SU(2)$ group element on one of those monomials at given $j$ simply yields a linear superposition of monomials with the same  $j$:
\beq
g\triangleright z_{0}^{j+m}z_{1}^{j-m}
&=&
(\alpha z_{0} +\beta z_{1} )^{j+m}(-\bbeta z_{0} +\balpha z_{1})^{j-m}
\\
&=&
\sum_{p=0}^{j+m}
\sum_{q=0}^{j-m}
\f{(j+m)!(j-m)!}{p!q!(j+m-p)!(j-m-q)!}
\alpha^{p}(-\bbeta)^{q}\beta^{j+m-p}\balpha^{j-m-q}
z_{0}^{p+q}z_{1}^{2j-(p+q)}
\,.\nn
\eeq
We need to normalize them to consider as states of a Hilbert space. We will use the basic Gaussian measure to define the scalar product for two (holomorphic) functions of $z$:
\be
\la f|\tilde{f} \ra
\equiv
\int \rd^{4}z\,e^{-\la z|z\ra}\, \bar{f}(z)\,\tilde{f}(z)
=
\int \rd^{2} z_{0}\rd^{2} z_{1}
\,e^{-|z_{0}^{2}|-|z_{1}^{2}|}
\,\overline{f(z_{0},z_{1})}\,\tilde{f}(z_{0},z_{1})
\,.
\ee
This gives normalized monomials,
\be
\psi_{j,m}
\equiv
\f{z_{0}^{j+m}z_{1}^{j-m}}{\sqrt{(j+m)!(j-m)!}}
\,,
\qquad
\int_{C} \rd^{2}\xi\, |\xi|^{2n}e^{-|\xi|^{2}}
=
\quad\Longrightarrow\quad
\la \psi_{j,m}|\psi_{j,m'}\ra
=
\delta_{mm'}
\,.
\ee
This allows to get explicitly the Wigner matrices:
\be
g\triangleright \psi_{j,m}
=
\sum_{p=0}^{j+m}
\sum_{q=0}^{j-m}
\f{\sqrt{(j+m)!(j-m)!(p+q)!(2j-p-q)!}}{p!q!(j+m-p)!(j-m-q)!}
\alpha^{p}(-\bbeta)^{q}\beta^{j+m-p}\balpha^{j-m-q}
\psi_{j,p+q-j}
\,,
\ee
getting a transition from $m$ to $\tm=p+q-j$ with
\be
D^{j}_{\tm,m}(g)
=
\sum_{p=m+\tm}^{j+\textrm{min}(\tm,m)}
\f{\sqrt{(j+m)!(j-m)!(j+\tm)!(j-\tm)!}}{p!(p-m-\tm)!(j+\tm-p)!(j+m-p)!}\,
\alpha^{p}\balpha^{p-m-\tm}\,\beta^{j+m-p}(-\bbeta)^{j+\tm-p}
\,.
\ee
First, for $j=\f12$ and $m,\tm=\pm\f12$, it is reassuring to recover the matrix elements $\alpha,\beta,\balpha,-\bbeta$ of the group element $g$ as a 2$\times$2 matrix in the fundamental representation of $\SU(2)$.
Then, the general spin-$j$ matrix elements are mere polynomials in those fundamental matrix elements. It is also obvious to check that those matrices are unitary, $\overline{D^{j}_{ab}(g)}=D^{j}_{ba}(g^{-1})$.

Plugging this directly in the Peter-Weyl integrals $\int \rd g\,D^{j}_{ab}D^{k}_{cd}$ allows a straightforward check of the orthonormality of the Wigner matrix elements. In particular, it is clear that the integration over the phases of $\alpha$ and $\beta$ requires  balancing out the powers of $\alpha$ with $\balpha$, and $\beta$ with $\bbeta$, leading to $a=c$ and $b=d$. A deeper dive in the formula easily leads to $j=k$ and the correct $1/d_{j}$ factor.

\smallskip

Finally, we show to easily compute the action of the $\su(2)$ Lie algebra generators. A $\SU(2)$ group element can be written as the exponentiation of a Lie algebra vector:
\be
g=e^{i\vec{u}\cdot\vec{J}}\,,
\ee
where $\vec{u}\in\R^{3}$ is a 3-vector and $\vec{J}$ consists in the three $\su(2)$ generators satisfying the standard Lie algebra commutator :
\be
[J_{z},J_{x}]=iJ_{y}
\,,\quad
[J_{z},J_{y}]=-iJ_{x}
\,,\quad
[J_{x},J_{y}]=iJ_{z}
\,.
\ee
In the fundamental representation in terms of 2$\times$2 matrices, the $\su(2)$ are simply half the Pauli matrices:
\be
g=e^{\f i2\theta \hat{u}\cdot\vec{\sigma}}=\cos\f\theta2 \,\id +i\sin\f\theta2 \hat{u}\cdot\vec{\sigma}
\,,
\ee
where $\hat{u}$ is a unit 3-vector and $\vec{u}=\theta\hat{u}$.

Thus the action of a generator $J_{a}$ can be derived by computing the action of an infinitesimal group element $g=e^{i\eps J_{a}}$ as $\eps\sim\id$.
For instance, we compute:
\be
e^{\f i2\eps \sigma_{z}}
=
\mat{cc}{e^{\f i2\eps} & 0 \\ 0 & e^{-\f i2\eps}}
\,,
\ee
\be
e^{i\eps J_{z}} \psi_{j,m}
=
\f{e^{\f i2(j+m)\eps}z_{0}^{j+m}\,e^{-\f i2(j-m)\eps}z_{1}^{j-m}}{\sqrt{(j+m)!(j-m)!}} 
=
\psi_{j,m}
+ i\eps m \psi_{j,m}
+\cO(\eps^{2})
\ee
Identifying this expression to $e^{i\eps J_{z}}=\id +i\eps  J_{z}+\dots$ gives simply:
\be
J_{z} \psi_{j,m}=m \psi_{j,m}\,.
\ee
Similarly, we compute
\be
e^{\f i2\eps \sigma_{x}}
=
\id+ \f i2 \eps \mat{cc}{0&1 \\ 1&0}+\dots
\,,
\ee
\beq
e^{i\eps J_{x}} \psi_{j,m}
&=&
\f{(z_{0}+\f i2\eps z_{1})^{j+m}\,(z_{1}+\f i2\eps z_{0})^{j-m}}{\sqrt{(j+m)!(j-m)!}} +\dots
\nn\\
&=&
\psi_{j,m}
+
\f i2\eps
\f{(j+m)z_{0}^{j+m-1}\,z_{1}^{j-m+1}}{\sqrt{(j+m)!(j-m)!}} 
+
\f i2\eps
\f{(j-m)z_{0}^{j+m+1}\,z_{1}^{j-m-1}}{\sqrt{(j+m)!(j-m)!}} 
+\dots
\eeq
leading to the expected formula:
\be
J_{x} \psi_{j,m}
=
\f12\sqrt{(j+m)(j-m+1)} \psi_{j,m-1}
+
\f12\sqrt{(j-m)(j+m+1)} \psi_{j,m+1}
\,.
\ee

\section{Crane-Yetter model}
\label{app:CraneYetter}

Similarly to the Ponzano-Regge model for 3d quantum gravity, the Ooguri model \ref{4dZBF} for 4d $\SU(2)$ BF theory suffers from divergences due to its topological symmetries. One could try to resolve those divergences by a suitable gauge fixing.
However, one can also introduce a $q$-deformation, like for the Turave-Viro model, to truncate the representation theory of $\SU(2)$ to a finite number of representations. Following this approach,  Crane and Yetter defined a regularized version based on the $q$-deformed representations of $\SU_q(2)$ \cite{Crane:1993if,Crane:1993cm}. They derive a state sum of a four-dimensional triangulated manifold with a deformation parameter $q=e^{\frac{2\pi i}{r}}$, where $r=3,4,...$ defines a cut-off for the spin values in a state sum, such that $j=0,\frac12,1,...,\frac{r-2}{2}$. The quantum dimension of a representation $j$ is given by a quantum integer,
\begin{equation}
    \dim_q j\equiv [2j+1]_q=\frac{q^{j+\frac12} - q^{-j-\frac12}}{q^{\frac12} - q^{-\frac12}}\equiv \frac{\sin\left( \frac{(2j+1)\pi}{r} \right)}{\sin\left( \frac{\pi}{r} \right)},
\end{equation}
sometimes defined with an additional sign factor $(-1)^{2j}$. Similarly, generalized $\{15j\}_q$ symbols contain quantum integers in their series expansion. Let us denote a sum of squares of quantum dimensions as
\begin{equation}
    N = \sum_{j=0}^{(r-2)/2} (\dim_q j)^2
\end{equation}
and the number of 3- and 4-simplices as $n_T$ and $n_\sigma$, as before. Then, the Crane--Yetter state sum is given by
\be
\cZ_{q}^{CY} = N^{n_\sigma-n_T}\sum_{\{j_t, i_T\}}\prod_{t\in\Delta_4} \dim_qj_t \prod_{T\in\Delta_4} \dim_q^{-1} i_T
\prod_{\sigma\in\Delta_4} \{15j\}_q.
\ee
Later, Crane, Yetter and Kauffman \cite{Crane:1993cm} showed that this state sum is defined by topological data of a manifold. Namely, if $M$ is a closed 4-dimensional manifold with a signature $\sigma(M)$ and an Euler characteristic $\chi(M)$,
\begin{equation}
    \cZ_{q}^{CY} (M) \sim \kappa^{\,\sigma(M)}N^{\frac{\chi(M)}{2}},
\quad \text{where} \quad \kappa = \kappa(r).
\end{equation}
As a bottom line, the $\SU_q(2)$ state sum with $q$ a root of unity is truncated at representations with finite spin values, which makes it convergent without the need for gauge fixing. Note that a real $q$ corresponds to a negative cosmological constant, but the corresponding state sum requires gauge fixing for convergence.

\bibliographystyle{bib-style}
\bibliography{LQG}

\end{document}